\documentclass[a4paper,fleqn,usenatbib]{mnras}

\usepackage{newtxtext,newtxmath, ulem}
\usepackage[T1]{fontenc}
\usepackage{ae,aecompl}

\usepackage{graphicx}	
\usepackage{amsmath}	

\usepackage{hyperref}
\usepackage[nolist,nohyperlinks]{acronym}
\usepackage{supertabular}
\usepackage{pdflscape}
\usepackage{float}
\usepackage{epstopdf}
\usepackage{multicol}
\usepackage{graphicx}
\usepackage{subcaption}
\usepackage{geometry}
\usepackage{changepage}
\usepackage{siunitx}
\usepackage{dcolumn}
\usepackage{longtable}
\usepackage{afterpage}
\usepackage{array}
\usepackage{varwidth}
\usepackage{comment}

\usepackage{xcolor}
\usepackage{scalerel}

\newcommand\HI{H\protect\scaleto{$I$}{1.2ex}}
\newcommand\HII{H\protect\scaleto{$II$}{1.2ex}}
\newcommand{\SII}{[S\,{\sc II}]}

\newcommand{\OIII}{[O\,{\sc III}]}
\newcommand{\D}{$^\circ$}

\def\arcmin{\hbox{$^\prime$}}
\def\arcsec{\hbox{$^{\prime\prime}$}}

\newcommand{\lmcorc}{J0624--6948}

\newcommand{\kms}{\,km\,s$^{-1}$} 

\definecolor{bv}{rgb}{0.62, 0.17, 0.89}

\newcommand{\lsun}{\ifmmode{{\rm ~L}_\odot}\else{~L$_\odot$}\fi}
\newcommand{\Msun}{\ifmmode{{\rm ~M}_\odot}\else{~M$_\odot$}\fi}
\newcommand{\mjybm}{\,mJy\,beam$^{-1}$}
\newcommand{\ujybm}{\,$\mu$Jy\,beam$^{-1}$}
\newcommand{\ujy}{\,$\mu$Jy}

\def\HII{\hbox{H\,{\sc ii}}}
\def\arcmin{\hbox{$^\prime$}}
\def\arcsec{\hbox{$^{\prime\prime}$}}
\def\kms{km\,s$^{-1}$}

\begin{acronym}[AWGN]
\acro{2MASS}{The Two Micron All-Sky Survey}
\acro{30Dor}[30 Dor]{30~Doradus}
\acro{AGN}{active galactic nucleus}
\acrodefplural{AGN}{Active Galactic Nuclei}
\acro{ATCA}{Australia Telescope Compact Array}
\acro{ATNF}{Australia Telescope National Facility}
\acro{ATOA}{Australia Telescope Online Archive}
\acro{AT20G}{Australia Telescope 20 GHz Survey}
\acro{ASKAP}{Australian Square Kilometre Array Pathfinder}
\acro{BL}[BL-Lac]{BL Lacertae Objects}
\acro{CASS}{CSIRO Astronomy and Space Science}
\acro{CABB}{Compact Array Broadband Back-end}
\acro{CHII}[{\sc CHii}]{Compact \textsc{Hii}}
\acro{CSIRO}{Australian Commonwealth Scientific and Industrial Research Organisation}
\acro{CSS}{Compact Steep Spectrum}
\acro{CME}{Coronal Mass Ejection}
\acro{Dec}{Declination}
\acro{DD}{double-degenerate}
\acro{DS9}[\textsc{DS9}]{\textsc{SAOImage DS9}}
\acro{DSS}{Digital Sky Survey}
\acro{EMU}{Evolutionary Map of the Universe}
\acro{ev}[eV]{electronvolt\acroextra{: 1 eV $\approx 1.6 \times 10^{-19}$ J}}
\acro{FITS}[\textsc{Fits}]{Flexible Image Transport System}
\acro{FRB}{Fast Radio Bursts}
\acro{FRII}{Fanaroff-Riley II}
\acrodefplural{FRII}{Fanaroff-Riley II's}
\acro{FRI}{Fanaroff-Riley I}
\acrodefplural{FRI}{Fanaroff-Riley I's}
\acro{FSRQ}{Flat Spectrum Radio Quasars}
\acro{FWHM}{full width at half-maximum}
\acro{GLEAM}{GaLactic Extragalactic All-sky MWA}
\acro{GMRT}{Giant Meterwave Radio Telescope}
\acro{GRG}{Giant Radio Galaxy}
\acro{GPS}{Gigahertz Peak Spectrum}
\acro{HEMT}{High Electron Mobility Transistor}
\acro{HFP}[HFP]{High Frequency Peaker}
\acro{HzRGs}{High Redshift Radio Galaxies}
\acrodefplural{HVS}{hypervelocity stars}
\acro{HVS}{hypervelocity star}
\acro{pHFP}[pHFP]{Potential High Frequency Peaker}
\acro{HST}{\textit{Hubble Space Telescope}}
\acro{IAU}{International Astronomical Union}
\acro{IFRSs}{Infrared Faint Radio Sources}
\acro{ISM}{interstellar medium}
\acro{IR}{Infrared}
\acro{JY}[Jy]{Jansky\acroextra{, 1 Jy = $10^{-26} \times \mathrm{W~ m}^{-2}~\mathrm{Hz}^{-1}$}} 
\acro{LLS}{Largest Linear Size}
\acro{LFAA}{Low-Frequency Aperture Array}
\acro{LMC}{Large Magellanic Cloud}
\acro{LG}{Local Group}
\acro{LSO}[LSOs]{Large Scale Objects}
\acro{MACHO}{Massive Astrophysical Compact Halo Objects}
\acro{MC}[MCs]{Magellanic Clouds}
\acro{mc}[MC]{Magellanic Cloud}    
\acro{MCELS}{Magellanic Cloud Emission Line Survey}
\acro{MW}{Milky Way}
\acro{MIRIAD}[\textsc{Miriad}]{Multichannel Image Reconstruction, Image Analysis and Display}
\acro{MIT}{Massachusetts Institute of Technology}
\acro{MOST}{Molonglo Observatory Synthesis Telescope}
\acro{MQS}{Magellanic Quasars Survey}
\acro{MRC}{Molonglo Reference Catalogue of Radio Sources}
\acro{MWA}{Murchison Widefield Array}
\acro{NAT}{Narrow-Angle Tail}
\acro{NRAO}{National Radio Astronomy Observatory}
\acro{NVSS}{NRAO VLA Sky Survey}
\acrodefplural{ORC}{Odd Radio Circles}
\acro{ORC}{Odd Radio Circle}
\acro{OPAL}{Online Proposal Applications \& Links}
\acro{OVV}{Optically Violent Variable Quasars}            
\acro{PAF}{Phased Array Feed}
\acro{pc}{parsec\acroextra{: 1 pc $\simeq 3.09 \times 10^{16}$ m}}
\acro{PMN}{Parkes-MIT-NRAO}
\acro{PNe}{Planetary Nebulae}
\acro{PN}{Planetary Nebula}
\acro{PWN}{Pulsar Wind Nebula}
\acro{QSO}{Quasi-Stellar Object}
\acro{RA}{Right Ascension}
\acro{RFI}{Radio-Frequency Interference}
\acro{RMS}[rms]{Root Mean Squared}
\acro{SDSS}{Sloan Digital Sky Survey}
\acro{SED}{spectral energy distribution}
\acro{SI}[$\alpha$]{Spectral Index\acroextra{, $S \propto \nu^\alpha$}}
\acro{SKA}{Square Kilometre Array}
\acro{SMASH}{Survey of the MAgellanic Stellar History}
\acro{SMB}{Super Massive Blackholes}
\acro{SMC}{Small Magellanic Cloud}
\acro{SN}{Supernova}
\acro{SNe}{Supernovae}
\acro{SNR}{Supernova Remnant}
\acrodefplural{SNR}{Supernova Remnants}
\acro{SD}{single-degenerate}
\acro{SUMSS}{Sydney University Molonglo Sky Survey}
\acro{SMBH}{Super Massive Black Hole}
\acro{TOPCAT}[\textsc{Topcat}]{Tool for OPerations on Catalogues And Tables}
\acro{USS}{Ultra Steep Spectrum}
\acro{WBAC}{Wide-Band Analogue Correlator}
\acro{WIFES}[WiFeS]{Wide-Field Spectrograph}
\acro{WISE}{Wide-Field Infrared Survey Explorer}
\acro{VLA}{Very Large Array} 
\acro{VLASS}{VLA Sky Survey}
\acro{VLBI}{Very Long Baseline Interferometry} 
\acro{VLSR}[\textbf{$v_{lsr}$}]{Velocity in the Line of Sight}
\acro{WD}{white dwarf}
\end{acronym}

\title[Mysterious ORC near the LMC -- An Intergalactic SNR?]{Mysterious Odd Radio Circle near the Large Magellanic Cloud -- An Intergalactic Supernova Remnant?}

\author[M. D. Filipovi\'c et al.]{Miroslav D. Filipovi\'c,$^{1}$\thanks{E-mail: m.filipovic@westernsydney.edu.au} 
J. L. Payne,$^{1}$ 
R. Z. E. Alsaberi,$^{1}$ 
R. P. Norris,$^{1,2}$ 
P. J. Macgregor,$^{1}$ 
L. Rudnick,$^{3}$ 
\newauthor
B. S. Koribalski,$^{2,1}$ 
D. Leahy,$^{4}$  
L. Ducci,$^{5,6}$ 
R. Kothes,$^{7}$  
H. Andernach,$^{8}$ 
L. Barnes,$^{1}$ 
I. S. Boji\v ci\'c,$^{1}$ 
\newauthor
L. M. Bozzetto,$^{1}$  
R. Brose,$^{9,10}$
J. D. Collier,$^{11,1}$ 
E. J. Crawford,$^{1}$ 
R. M. Crocker,$^{12}$ 
S. Dai,$^{1}$ 
T. J. Galvin,$^{13}$ 
\newauthor
F. Haberl,$^{14}$ 
U. Heber,$^{15}$ 
T. Hill,$^{1}$ 
A. M. Hopkins,$^{16,1}$  
N. Hurley-Walker,$^{13}$  
A. Ingallinera,$^{17}$ 
T. Jarrett,$^{18,1}$  
\newauthor
P.~J.~Kavanagh,$^{19}$  
E. Lenc,$^{2}$  
K. J. Luken,$^{1,2}$ 
D. Mackey,$^{12}$ 
P. Manojlovi\'c,$^{1}$ 
P. Maggi,$^{20}$ 
C. Maitra,$^{14}$ 
\newauthor
C. M. Pennock,$^{21}$ 
S. Points,$^{22}$ 
S. Riggi,$^{17}$  
G. Rowell,$^{23}$ 
S. Safi-Harb,$^{24}$ 
H. Sano,$^{25}$ 
M. Sasaki,$^{15}$ 
\newauthor
S. Shabala,$^{26}$ 
J. Stevens,$^{2}$ 
J. Th. van Loon,$^{21}$ 
N. F. H. Tothill,$^{1}$ 
G. Umana,$^{17}$ 
D. Uro\v sevi\' c,$^{27,28}$ 
V. Velovi\'c,$^{1}$  
\newauthor
T. Vernstrom,$^{2}$ 
J. L. West,$^{29}$ 
and Z. Wan$^{30}$ 
\\
\\
Affiliations are listed at the end of the paper
}

\date{Accepted 2022 January 20. Received 2022 January 11; in original form 2021 September 21}

\pubyear{2022}

\begin{document}
\label{firstpage}
\pagerange{\pageref{firstpage}--\pageref{lastpage}}
\maketitle


\begin{abstract}
We report the discovery of \lmcorc, a low-surface brightness radio ring, lying between the Galactic Plane and the \ac{LMC}. It was first detected at 888~MHz with the \ac{ASKAP}, and with a diameter of $\sim$196~arcsec. This source has phenomenological similarities to \acp{ORC}. Significant differences to the known \acp{ORC} --- a flatter radio spectral index, the lack of a prominent central galaxy as a possible host, and larger apparent size --- suggest that \lmcorc\ may be a different type of object. We argue that the most plausible explanation for \lmcorc\ is an {\it intergalactic supernova remnant} due to a star that resided in the \ac{LMC} outskirts that had undergone a single-degenerate type~Ia supernova, and we are seeing its remnant expand into a rarefied, intergalactic environment. We also examine if a massive star or a white dwarf binary ejected from either galaxy could be the supernova progenitor. Finally, we consider several other hypotheses for the nature of the object, including the jets of an \ac{AGN} or the remnant of a nearby stellar super-flare.
\end{abstract}

\begin{keywords}
general -- Magellanic Clouds -- radio continuum: general -- galaxies: jets -- stars: flare -- ISM: supernova remnants
\end{keywords}



\section{Introduction}

A new generation of radio telescopes is revealing new features of the radio Universe, due to the combined effects of high-sensitivity, good spatial sampling, and wide area coverage. Rare, low surface brightness features are now more easily detected, as demonstrated by the recent emergence of the class of ring-shaped radio sources known as \acp{ORC} \citep{2021PASA...38....3N,2021MNRAS.505L..11K}. Radio sources often have circular features that can represent nearly spherical objects, including \acp{SNR}, \ac{PNe}, (super)bubbles, \HII\ regions, circumstellar shells or disk-like objects, protoplanetary discs and star forming galaxies, or even ring-shaped imaging artefacts.

While \acp{ORC} resemble \acp{SNR} in some respects, the two populations seem to be distinct. Most known Galactic and \ac{MC} \acp{SNR} are radio emitters. However, detections in the optical and X-ray frequencies have been increasingly noted in recent times \citep{2012SerAJ.184...19M,2017ApJS..230....2B}. \ac{SNR} outliers remind us that they are not a monolithic population. \ac{SN} progenitors, in different \ac{ISM} and at different evolutionary stages, produce unique shapes in individual objects. This includes a population of so-called host-less (intergalactic) \ac{SN}e, discovered beyond our \ac{LG} of galaxies, that cannot be associated to any galaxy \citep{2011A&A...536A.103Z}. However, thus far no such \ac{SN}, nor their remnants, have been seen in our \ac{LG}.

These host-less intergalactic \ac{SN}e may come from hyper or high velocity runaway stars \citep{2011A&A...536A.103Z}, producing a remnant whose velocity will be modulated by a \ac{SN} kick \citep{2017MNRAS.469.2151B,2020MNRAS.497.5344E,2021MNRAS.507.4997E}. This scenario could have some exotic consequences, such as runaway pulsars and \ac{SNe} far from star-forming regions, including the old/outer stellar disk of galaxies such as the \ac{MC} or \ac{MW}. At the same time, galaxy halos and outer discs harbour stars that also explode, though mainly as type~Ia \ac{SN}e \citep{2017MNRAS.471.1390H}.

\begin{figure*}
    \centering
    \includegraphics[width=\textwidth, trim=0 0 0 0,clip]{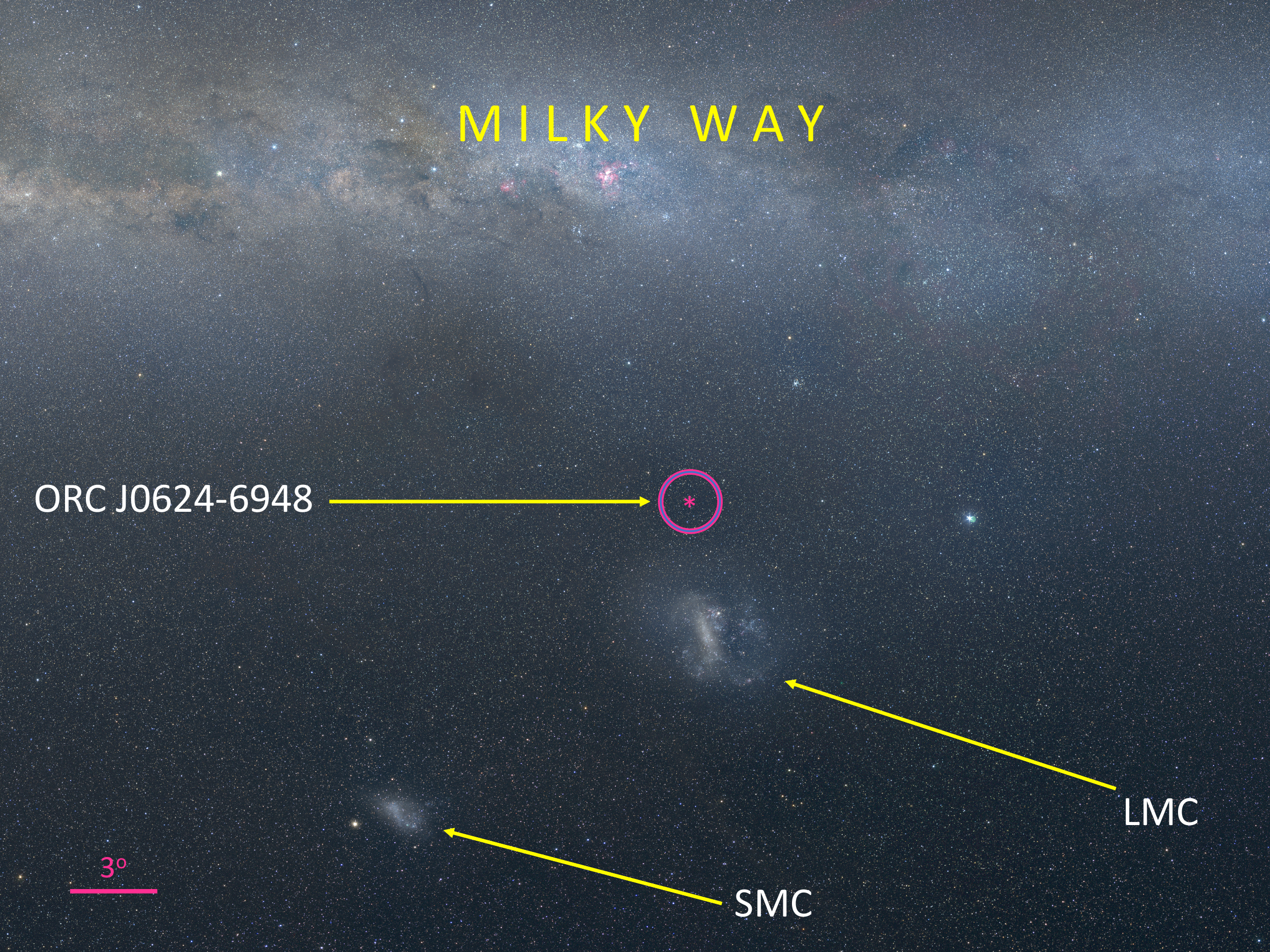}
    \caption{Optical image of the sky around \lmcorc. The position of \lmcorc\ (small object at the centre of the purple ring indicated by *) is indicated with respect to the \ac{MW} galactic plane, the \ac{LMC} and the \ac{SMC}. Optical image credit: Axel Mellinger, Central Michigan University. 
    }
    \label{fig:M1}
\end{figure*}

\begin{figure*}
    \centering
    \includegraphics[width=\textwidth, trim=0 0 0 0,clip]{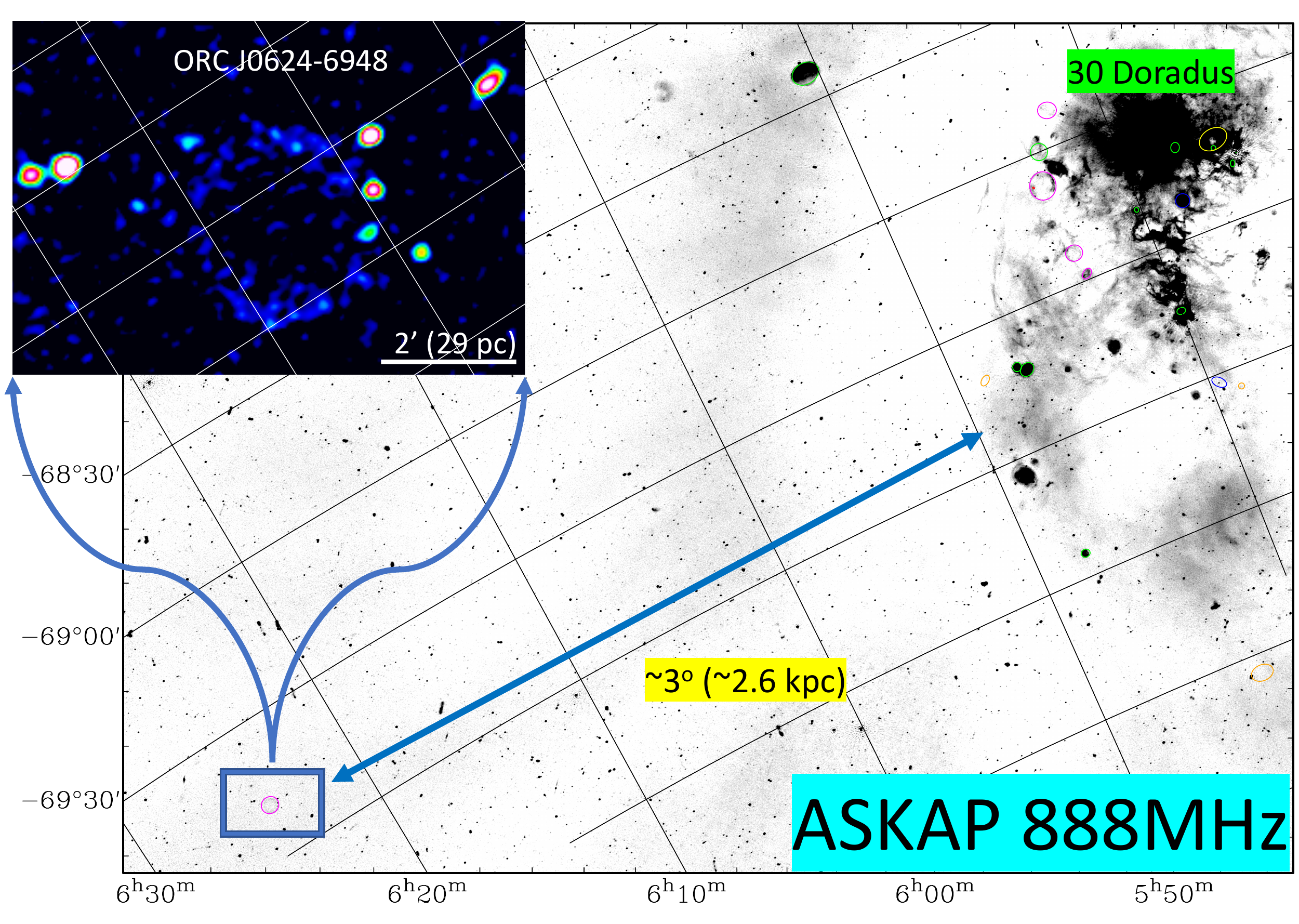}
    \caption{
    \ac{ASKAP} 888~MHz image with \lmcorc\ in the lower left and the \ac{LMC} in the upper right. The various coloured circles/ellipses represent the positions of known \ac{LMC} \acp{SNR} and \ac{SNR} candidates \citep{2017ApJS..230....2B}. The separation of $\sim$2.6~kpc indicated here assumes that \lmcorc\ is at the same distance as the \ac{LMC} (50~kpc).
    The inset in the upper left is zoomed in \lmcorc\ at \ac{ASKAP} native frequency of 888~MHz.
    }
    \label{fig:0}
\end{figure*}

First detected at 888~MHz with the \ac{ASKAP}, \lmcorc\ is located between the \ac{LMC} and the plane of the \ac{MW} (Figs.~\ref{fig:M1} and \ref{fig:0}). We present a multi-frequency study and analysis of \lmcorc. While we argue for an intergalactic \ac{SNR} scenario, we also consider several other hypotheses for the nature of the object, including jets of an \ac{AGN} or the remnant of a nearby stellar super-flare.

\begin{figure*}
\centering
   \includegraphics[width=0.9\textwidth]{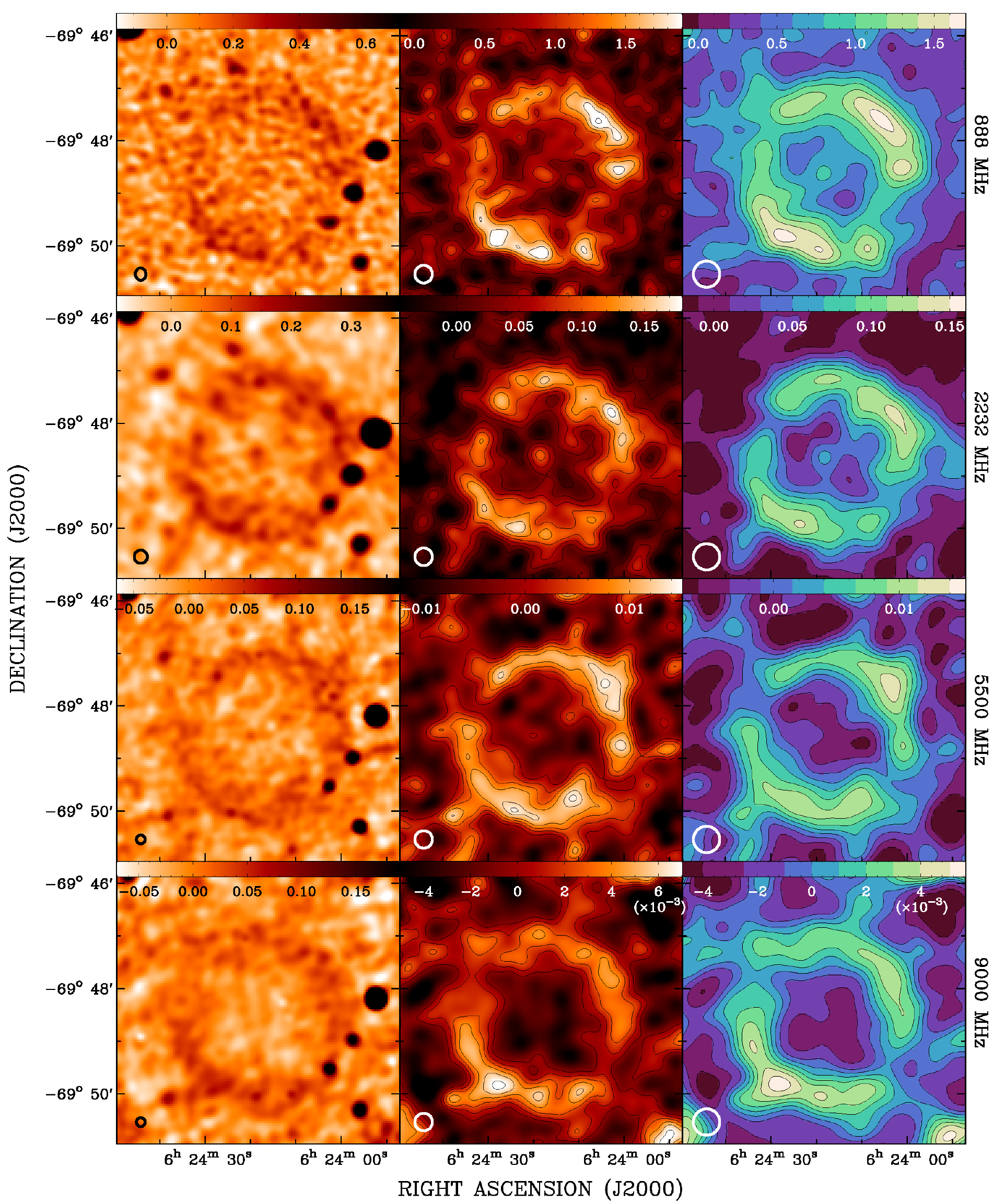}
\caption{Radio continuum images of \lmcorc\ obtained with (first row; left) \ac{ASKAP} at 888~MHz ($\sigma=60$~\ujybm), (second row; left) \ac{ATCA} at 2232~MHz ($\sigma$=20\ujybm), (third row; left) \ac{ATCA} at 5500~MHz ($\sigma$=10\ujybm) and (fourth row; left) \ac{ATCA} at 9000~MHz ($\sigma$=7\ujybm). The images in the left column are with a native synthesised beam (black circle, or in the case of the \ac{ASKAP} image a black ellipse, in the bottom left corner) of 13.87\arcsec$\times$12.11\arcsec, 15\arcsec$\times$15\arcsec, 10\arcsec$\times$10\arcsec\ and 10\arcsec$\times$10\arcsec\ for 888, 2232, 5500 and 9000~MHz, respectively. The intensity scale bars on top of each image are labelled in \mjybm. For the images in the middle and right column we subtracted the four nearby point sources (see Table~\ref{tab2}) and smoothed to 20 and 30~arcsec resolution, respectively (white circle in the bottom left corner). The intensity scale bars on top of these images are in brightness temperature (K$_{\rm Tb}$).}
\label{fig:3}
\end{figure*}

\section{Observations and Data Reduction}
\label{ODR}

\subsection{Radio imaging}

\subsubsection{\ac{ASKAP} data}
\lmcorc\ was originally detected in June 2019 as part of the \ac{ASKAP} Early Science Project \citep{2021MNRAS.506.3540P}. The data were processed using the ASKAPsoft pipeline, including multi-frequency synthesis imaging, multi-scale clean and self-calibration \citep{2021PASA...38....3N}. The 888~MHz \ac{ASKAP} image shown in Figs.~\ref{fig:0} and \ref{fig:3} (top row left) has a sensitivity of $\sigma$=58\ujybm\ and a synthesised beam of 13.87$\times$12.11~arcsec at the beam parallactic angle of --84.4~degrees.

\subsubsection{\ac{ATCA} data}
Follow-up observations were obtained with various \ac{ATCA} arrays at 2232, 5500 and 9000~MHz (Table~\ref{tab1}) and images are shown in Fig.~\ref{fig:3}. We used the source PKS~B1934--638 as primary calibrator (bandpass and flux density) and the source PKS~B0530--727 as phase calibrator for all observing sessions (Table~\ref{tab1}). 

The \textsc{miriad} \citep{1995ASPC...77..433S} and \textsc{karma} \citep{1995ASPC...77..144G} software packages were used to reduce, compare and analyse the data. We combine observations of various interferometer arrays to a common $uv$-plane for each of the three frequency bands. Given that our object of interest is diffuse and of low surface brightness, we experimented with the range of various values for weighting and tapering. To maximise and enhance source diffuse emission, we found that a Briggs weighting robust parameter of 1 (closer to natural weighting) is the most optimal choice. Also, an additional 7~arcsec Gaussian taper is included to further enhance the diffuse emission. This allows us to achieve resolutions of \ac{FWHM} of 15$\times$15~arcsec, 10$\times$10~arcsec, and 10$\times$10~arcsec for the 2232, 5500, and 9000~MHz images, respectively. All images were corrected for the primary beam attenuation. The \ac{RMS} noise (or 1$\sigma$) is 20\ujybm\ for the 2232~MHz image, 5\ujybm\ for the 5500~MHz image and 7\ujybm\ for the 9000~MHz image. The \ac{RMS} numbers quoted in this paper represent noise level at the image centre and because of the primary beam response correction the noise increases towards the edges of the image. Given our observing frequencies and arrays $uv$ coverage, the maximum angular scale that the image is sensitive to ranges from 49~arcmin (for \ac{ASKAP}) to 6 and 3.5~arcmin for \ac{ATCA} 5500 and 9000~MHz images.

\begin{figure*}
    \centering
    \includegraphics[width=\textwidth]{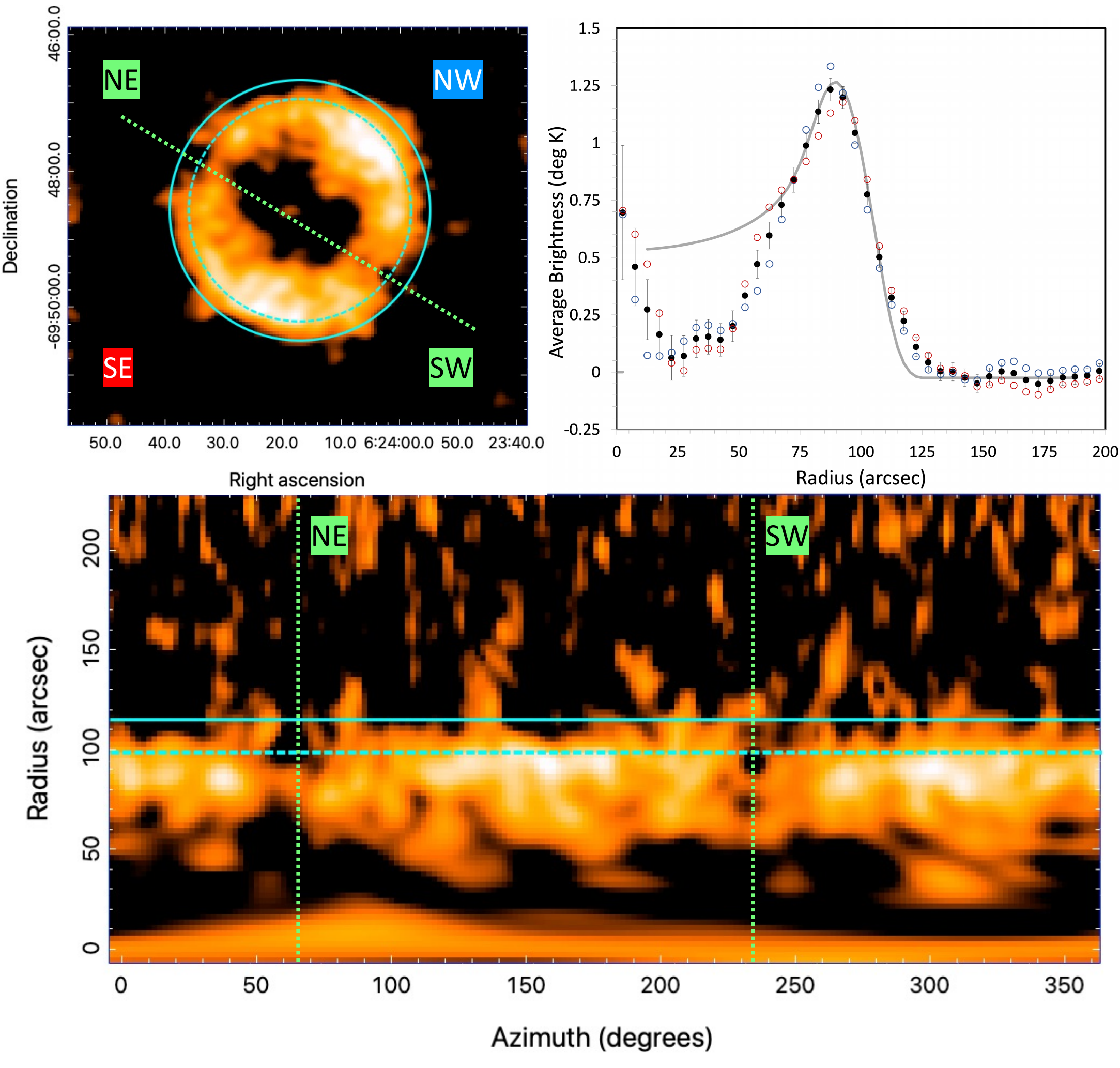}
    \caption{
{\it Top left:} Map emphasising diffuse emission from the \lmcorc, which  was made by combining images at 15~arcsec and 35~arcsec as described in the text. The dotted green line divides \lmcorc\ into two halves. 
{\it Top right:} A radial profile plot of \lmcorc, from the effective 10-cm map is shown for the SE (red), NW (blue) halves, and the total (black). Errors represent the error in the mean brightness of each total ring, based on the image \ac{RMS} and the number of independent beam samples in each ring. The errors for the SE and NW halves are not shown, but are $\sqrt{2}$ higher than the corresponding total ring errors. An approximate fit to the peak in the total ring brightness is shown as a (black) solid line; it represents a uniform emissivity shell with radius of 98~arcsec and width of 10~arcsec, which is seen as 30~arcsec wide in projection. Formal fitting is not possible because the model fails to accurately represent the central brightness. 
{\it Bottom:} The effective 10-cm map at 15~arcsec resolution, seen in polar projection (top left image). The angle on the horizontal axis is azimuth from north (0~degrees)
through east. The dashed horizontal cyan line, at 98~arcsec, is the best fit for the radius of the shell in space, prior to projection, for the total ring (black points in the top right graph). The solid horizontal cyan line, at 115~arcsec, is where the diffuse emission cuts off. The best fit has a width (projected width) of 10~arcsec (30~arcsec), so the inner radius of the shell is 93~arcsec. 
}
    \label{fig:polar}
\end{figure*}

In order to further enhance the signal-to-noise ratio, we averaged the 2232, 5500 and 9000~MHz maps (Fig.~\ref{fig:3}) by extrapolating each of them to 888~MHz using a spectral index (defined by $S \propto \nu^{\alpha}$, where $S$ is flux density, $\nu$ is the frequency and $\alpha$ is the spectral index) of $\alpha=-0.4$ as we determined for the spectral index (Section~\ref{RA}). We convolved all these maps to 15~arcsec, as well as subtracting four point sources from the south-west region (see Table~\ref{tab2}). As a result, the 888~MHz, the extrapolated 2232~MHz, and the extrapolated 5500~MHz maps have all the same noise of $\sim$400~mK, while the 9000~MHz map has a much higher value $\sim$2~K. Thus, we averaged the three lower frequencies together; the combined map's noise is a factor of two lower than that of the \ac{ASKAP} 888~MHz map alone. Hereafter, we refer to this combined frequency image as the `effective 10-cm' (at a weighted frequency of 3623~MHz) map. Finally, a map emphasising the diffuse emission (Fig.~\ref{fig:polar}; top left) was created by adding the effective 10-cm map at 15~arcsec resolution to an image of the large-scale emission alone, at a resolution of 35~arcsec, in the ratio of 0.3 to 0.7. The large-scale emission image was created by filtering out all of the small-scale emission using the multi-resolution filtering technique of \citet{2002PASP..114..427R}, using a box size of 45~arcsec. Note that since this image is at two different resolutions, it is useful for visualisation only, and is not used quantitatively in the paper.

\begin{table}
\centering
\caption{\ac{ATCA} observations centred on \lmcorc\ used here to create images at 2232, 5500 and 9000~MHz. }
\begin{tabular}{ccccc}
\hline
Date       &    Array      &  Project     &  Frequency        & Int. Time \\
           & Config.       &   Code       &   (MHz)           & (minutes)\\
\hline
12-12-2019 &     1.5C      & CX450        &     2232          & 265\\
26-01-2020 &      6A       & C3296        &    2232           & 196\\
23-02-2020 &     367       & C3292        & 2232              & 74   \\
23-02-2020 &     367       & C3292        &       5500, 9000  & 74   \\
15-03-2020 &     6D        & C3295        & 2232              & 119\\
15-03-2020 &     6D        & C3295        &       5500, 9000  & 118\\
28-03-2020 &    H168       & CX310        & 2232              & 87 \\
28-03-2020 &    H168       & CX310        &       5500, 9000  & 87\\
12-04-2020 &    6A         & C3330 	      &   5500, 9000      & 235  \\
10-06-2020 &   1.5C        & CX454 	      &    5500, 9000     & 355 \\
30-10-2020 &   6B          & CX310 	      &  2232             & 177 \\
30-10-2020 &   6B          & CX310 	      &        5500, 9000 & 177 \\
09-02-2021 &   750C        & C3383 	      &  2232             & 98 \\
09-02-2021 &   750C        & C3383 	      &        5500, 9000 & 108 \\
\hline
\end{tabular}
\label{tab1}
\end{table}

\subsection{X-ray imaging}
 \label{sec:xray}
The sky region around \lmcorc\ was observed serendipitously by {\it XMM-Newton} on 14--15$^{\rm th}$~March~2020. Data were reduced and analysed using standard routines in the Science Analysis System (SAS; version 19.0.0) and calibration files. We used only data from the European Photon Imaging Camera (EPIC) instruments which are equipped with charge-coupled devices (CCDs) based on Metal Oxide Semi-conductor technology \citep[MOS1 and MOS2,][]{2001A&A...365L..27T}, as \lmcorc\ was outside the field of view of EPIC-pn \citep{2001A&A...365L..18S}, which was operated in small window read-out mode. After filtering the data for high background contamination, the net exposure times are $\sim$85.62~ks and $\sim$85.60~ks for MOS1 and MOS2, respectively (see {\it XMM-Newton} image in Fig.~\ref{fig:5new}).

\subsection{Other images}
\begin{figure*}
    \centering
    \includegraphics[width=\textwidth, trim=0 0 0 0,clip]{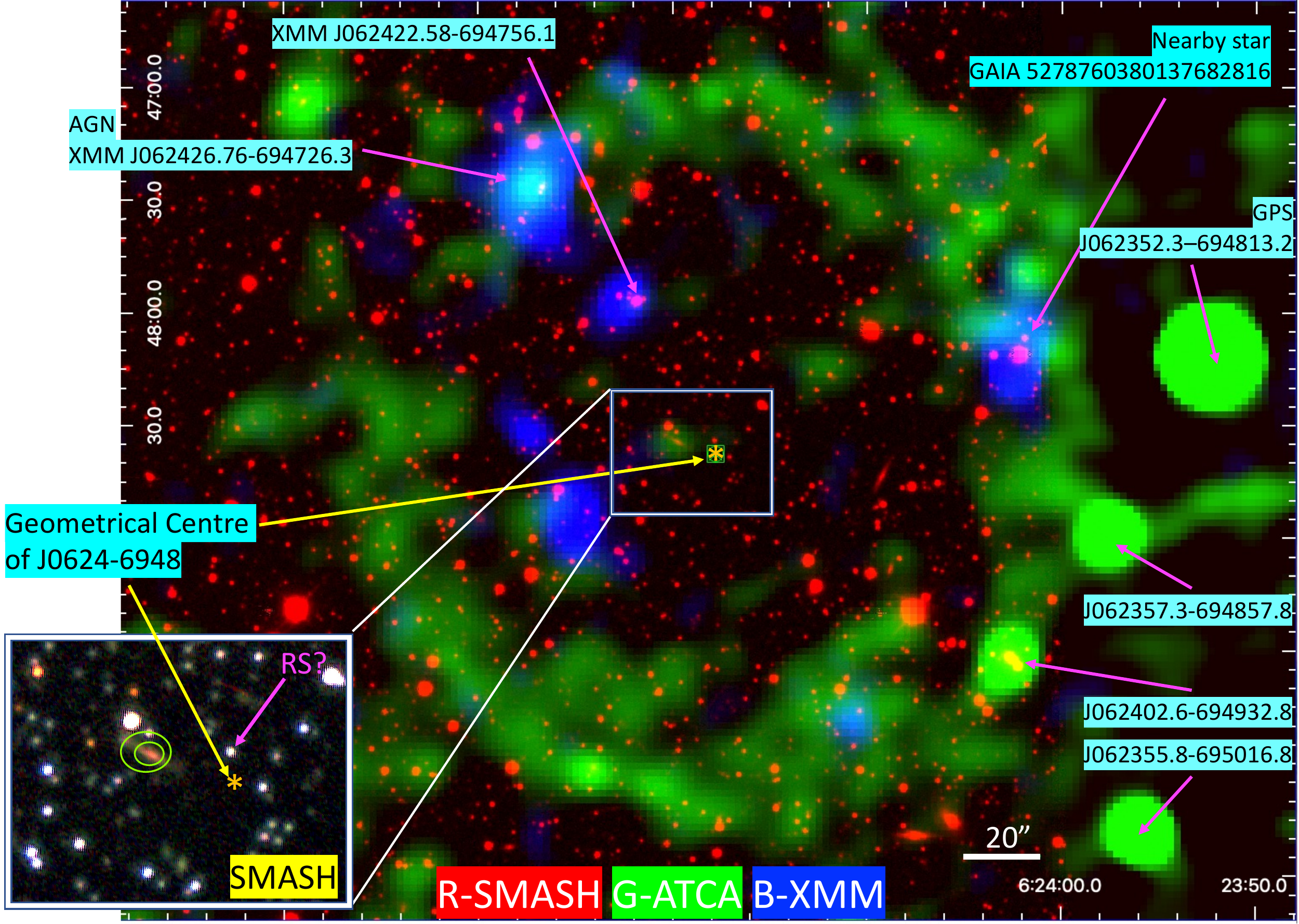}
    \caption{\lmcorc\ RGB image (R-SMASH ($i$-band), G-\ac{ATCA} at 5500~MHz and B-{\it XMM-Newton} (0.3--10~keV) indicating the positions of all relevant objects in the field. The inset RGB image at the bottom left is from SMASH where we used bands R ($r$-band), G ($g$-band) and B ($i$-band). The green ellipses are contours from the \ac{ATCA} 5500~MHz image at 15 and 25~\ujy~beam$^{-1}$. At the centre of these ellipses lies a lenticular galaxy ($r$=22.38~mag) that is located 10.8~arcsec from the \lmcorc\ centre. 
    A possible remnant star (RS?) from a type~Ia SD explosion is indicated by the pink arrow (see Section~\ref{sec:evstar}).}
    \label{fig:5new}
\end{figure*}

We searched for a possible counterpart using Parkes GASS/HI4PI data as well as SkyMapper \citep{2019PASA...36...33O}, \ac{2MASS}, \ac{WISE} and \ac{SMASH} data in optical/IR wavebands. The \ac{WISE} imaging was custom constructed to preserve the native angular resolution \citep[$\sim$6~arcsec in the W1 band;][]{2012AJ....144...68J}. Details about GASS/HI4PI observation can be found in \citet{2016A&A...594A.116H} and \ac{SMASH} data in \citet{2021AJ....161...74N} (see Fig.~\ref{fig:5new}). We also searched other available catalogues and surveys including \ac{MWA} images \citep{2018MNRAS.480.2743F}, the {\it Planck} free-free emission maps \citep{2016ApJS..227...23C} and {\it Fermi-LAT}~4FGL and 3FHL~GeV gamma-ray source catalogues \citep{2020ApJ...892..105A}.

\section{Properties of \lmcorc}
\label{RA}

The apparent location of \lmcorc\ is between the \ac{LMC} and the plane of the \ac{MW}, about 3$^{\circ}$ \citep[$\sim$2.6~kpc at the distance of the \ac{LMC} of 50~kpc;][]{2019Natur.567..200P} east of the \ac{LMC}'s radio continuum eastern edges \citep[RA(J2000) = 05$^\mathrm{h}$50$^\mathrm{m}$00$^\mathrm{s}$ and Dec(J2000) = --70$^{\circ}$00\arcmin00\arcsec; ][]{2021MNRAS.507.2885F} and in the direction towards the \ac{MW} (Figs.~\ref{fig:M1} and \ref{fig:0}). The geometric centre of \lmcorc\ (determined as described in \citet{2016A&A...586A...4K,2021arXiv211100446K}) is located at RA(J2000) = 06$^\mathrm{h}$24$^\mathrm{m}$17.78$^\mathrm{s}$ and Dec(J2000) = --69$^{\circ}$48\arcmin37.8\arcsec\ ($\Delta$RA~\&~$\Delta$Dec $\sim$2~arcsec; Galactic coordinates: $l$=280.168\D\ and $b$=--27.662\D). A red-green-blue (RGB) colour-composite image of \lmcorc, consisting of optical (R), radio (G) and X-ray (B) observations is shown in Fig.~\ref{fig:5new}.

\subsection{Flux density and spectral index}
 \label{sec:fluxspectra}
 
For the \lmcorc\ flux density measurements, we used the method described in \citet[][Section 2.4]{2019PASA...36...48H}. After careful region selection that excludes all obvious point sources, the total radio flux density measured for \lmcorc\ is 11.7$\pm$5.8~mJy at 888~MHz, 9.1$\pm$3.5~mJy at 2232~MHz, 4.5$\pm$1.9~mJy at 5500~MHz and 3.6$\pm$1.8~mJy at 9000~MHz. This gives a spectral index of $\alpha=-0.54\pm0.08$ (Table~\ref{tab2}). The large error in flux density measurements is due to the low surface brightness of \lmcorc. Flux densities of four point sources, in each band, were estimated using the \textsc{Aegean} Source Finding suite of tools \citep{2018PASA...35...11H} with default parameters. They are excluded from the estimate of \lmcorc\ total flux.

\begin{figure}
\centering
\includegraphics[scale=0.315,angle=270]{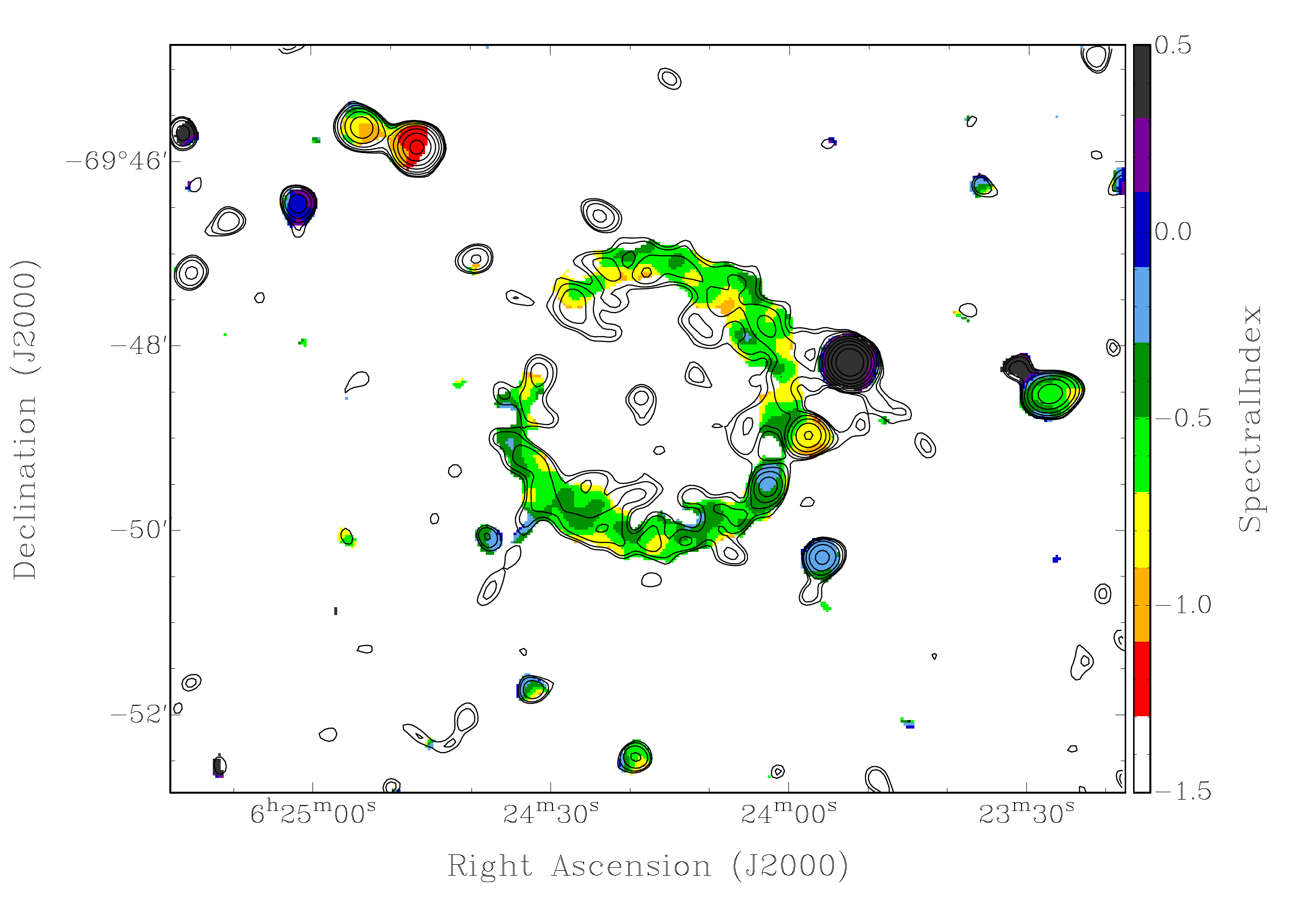}
\caption{Spectral index map of the \ac{LMC} \ac{ORC} J0624-6948 made from \ac{ASKAP} (888~MHz) and \ac{ATCA} (2232, 5500 and 9000~MHz) images. Overlaid contours (black) are from \ac{ATCA} at 2232~MHz and correspond to 0.06 (3$\sigma$), 0.105, 0.165, 0.285, 0.525, 1.01, 1.97 and 3.89~\mjybm.}
\label{fig:6}
\end{figure}

We also produced a spectral index map for \lmcorc\ using 888, 2232, 5500 and 9000 MHz images. First, the images were smoothed to their lowest angular resolution (15~arcsec) using \textsc{miriad} task \textsc{convol}, then they were re-gridded using \textsc{miriad} task \textsc{regrid}. The \textsc{miriad} task \textsc{maths} was used to create the spectral index map from these data as shown in Fig.~\ref{fig:6}. We find that the observed $\alpha$ varies from $-0.40 > \alpha > -0.75$ with a typical uncertainty of $\Delta\alpha$=0.19. Thus, there is no evidence for spectral index variations across \lmcorc.

We detect a weak ($\sim$4$\sigma$) point-like central radio source at RA(J2000) = 06$^\mathrm{h}$24$^\mathrm{m}$18.7$^\mathrm{s}$ and Dec(J2000)=--69$^{\circ}$48\arcmin34.3\arcsec. The integrated flux density of this source is: $\mathrm{S}_{888}=210\pm20~\mu\mathrm{Jy}$, $\mathrm{S}_{2232}=125\pm15~\mu\mathrm{Jy}$ and $\mathrm{S}_{5500}=31\pm5~\mu\mathrm{Jy}$, leading to a spectral index of $\alpha=-1.03\pm0.25$ (Table~\ref{tab2}). Given such a steep spectral index, this central radio source is unlikely a \ac{PWN} \citep{2012SSRv..166..231R}. It is consistent with being a pulsar or central \ac{AGN} engine of \lmcorc\ (Fig.~\ref{fig:5new}). However, it is located 10.8~arcsec away from the geometrical centre of \lmcorc. At the location of the 5500~MHz radio source, we find in \ac{SMASH} (optical) images a faint galaxy ($r$=22.38~mag), which makes this source very unlikely to be a pulsar or \ac{PWN}. 
Based on these optical images, \lmcorc\ could either be an edge-on late-type or lenticular (S0) galaxy (Fig.~\ref{fig:5new}, left inset). The optical faintness of this galaxy, together with an expected small Galactic extinction (A$_{r}\sim$0.17~mag; \cite{2011ApJ...737..103S}), leads us to estimate its redshift as $z \gtrsim 0.7$. This redshift was estimated using the ANNz2 \citep{annz_bib} algorithm and the k-Nearest Neighbours algorithm described in Luken~et~al.~(2022, in prep.) using the SMASH $g$, $r$, $i$, and $z$ magnitudes and AllWISE W1-W4 magnitudes. Given the lack of similar optical data to use as a training set for the models, we hesitate to put a firm estimate on the redshift, and instead provide a cautious lower limit.

Nearby point and point-like radio sources marked in Fig.~\ref{fig:5new} have spectral indices typical for background objects, as can be seen from a small group of four radio sources towards the south-west edges of the \lmcorc\ ring (Table~\ref{tab2} and Fig.~\ref{fig:8}). There is no obvious connection or association of these sources with the extended ring-like structure of \lmcorc. We classify ASKAP~J062352.3--694813.2 as a \ac{GPS} source \citep{2018MNRAS.477..578C} with a turnover frequency at $\sim$4900~MHz. ASKAP~J062357.3--694857.8 is a typical background radio source with $\alpha=-0.79\pm0.03$. The spectral indices of the other two sources (ASKAP~J062402.6--694932.8 and ASKAP~J062355.8--695016.8) suggest a dominantly thermal nature or possible variable radio sources.

\begin{figure}
\centering
\includegraphics[width=0.65\textwidth, trim=130 50 0 0,clip]{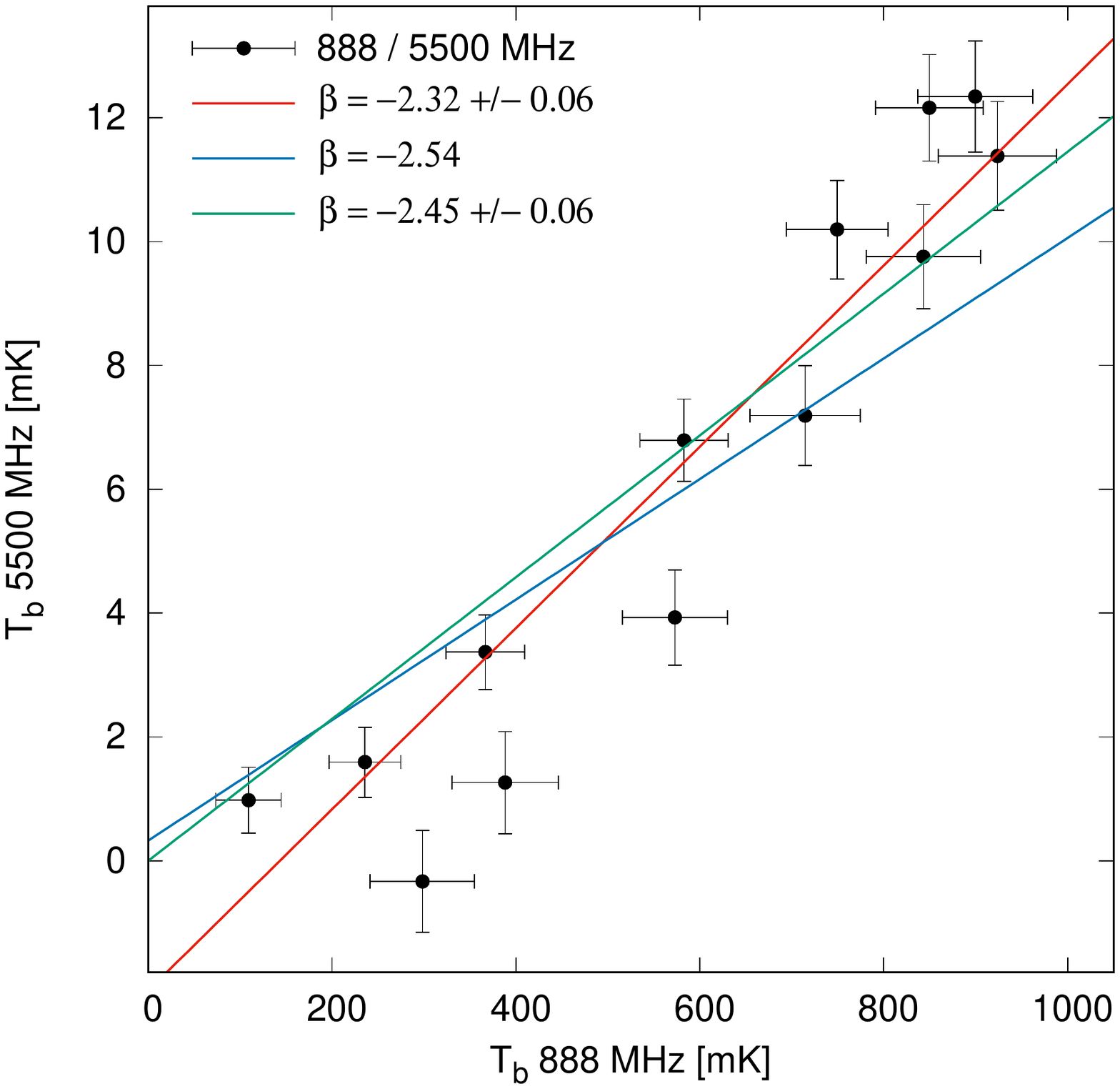}
\caption{Sample TT-plot between the \ac{ASKAP} observation at 888~MHz and the \ac{ATCA} observation at 5500~MHz. The TT-plot is produced from ring-averaged radial profiles that cover the visible shell between 51~arcsec and 123~arcsec radius with a 6~arcsec ring separation. The temperature fitted spectrum ($\beta$) is indicated in red (with free y-intercept) and green (forcing the linear fit to go through the origin of the diagram) and the blue line indicates a spectrum with a spectral index of $-2.54$ for comparison (note that $\beta=\alpha+2$; where $\alpha$ is previously defined flux density spectral index). The fitted red spectrum shows an obvious negative offset from the origin of the diagram for the higher frequency, indicating significant missing short spacings.}
\label{fig:7}
\end{figure}

\begin{figure*}
\centering
\includegraphics[width=0.33\textwidth,scale=0.7, trim=0 0 0 0 0,clip]{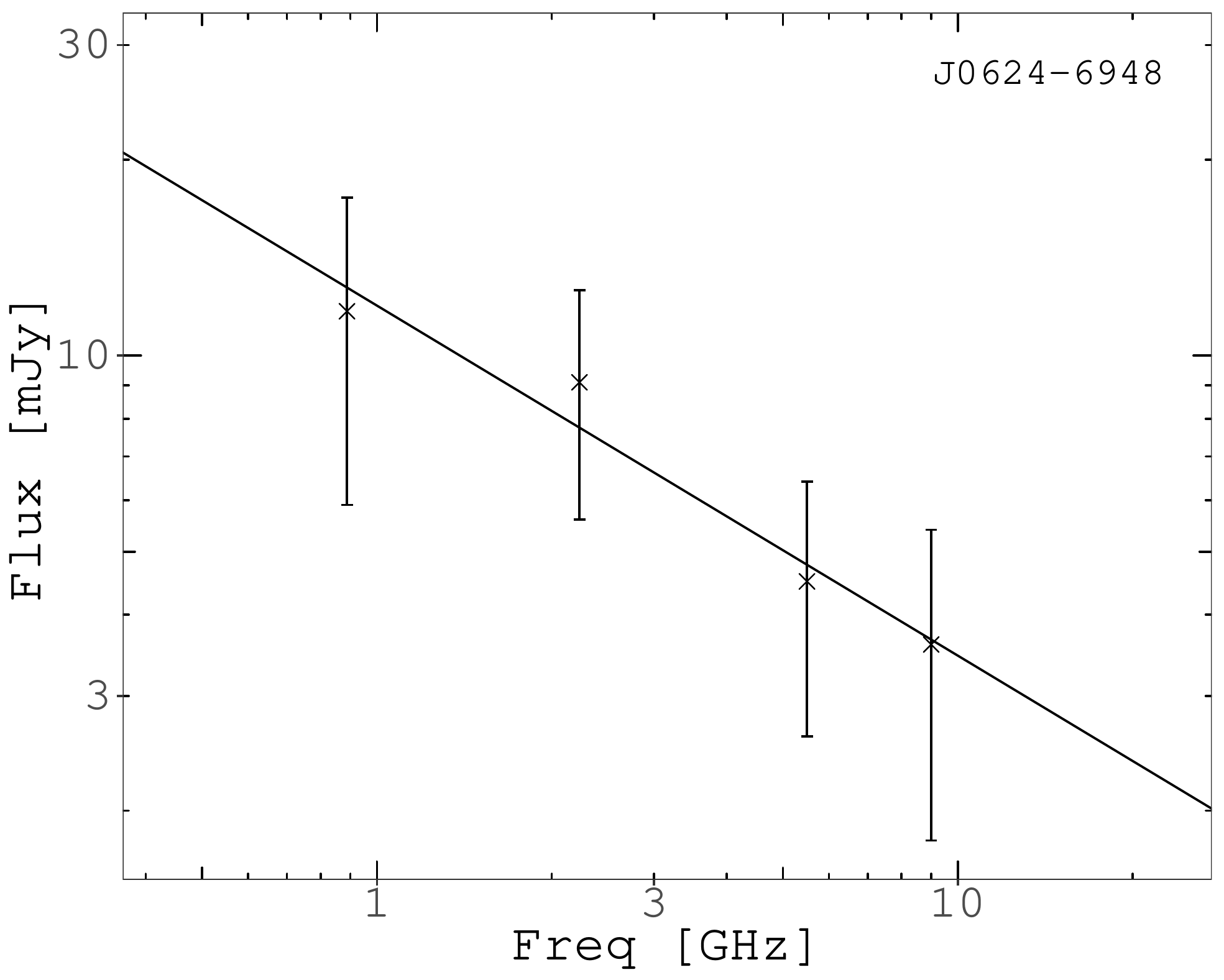}
\includegraphics[width=0.33\textwidth,scale=0.7, trim=0 0 0 0 0,clip]{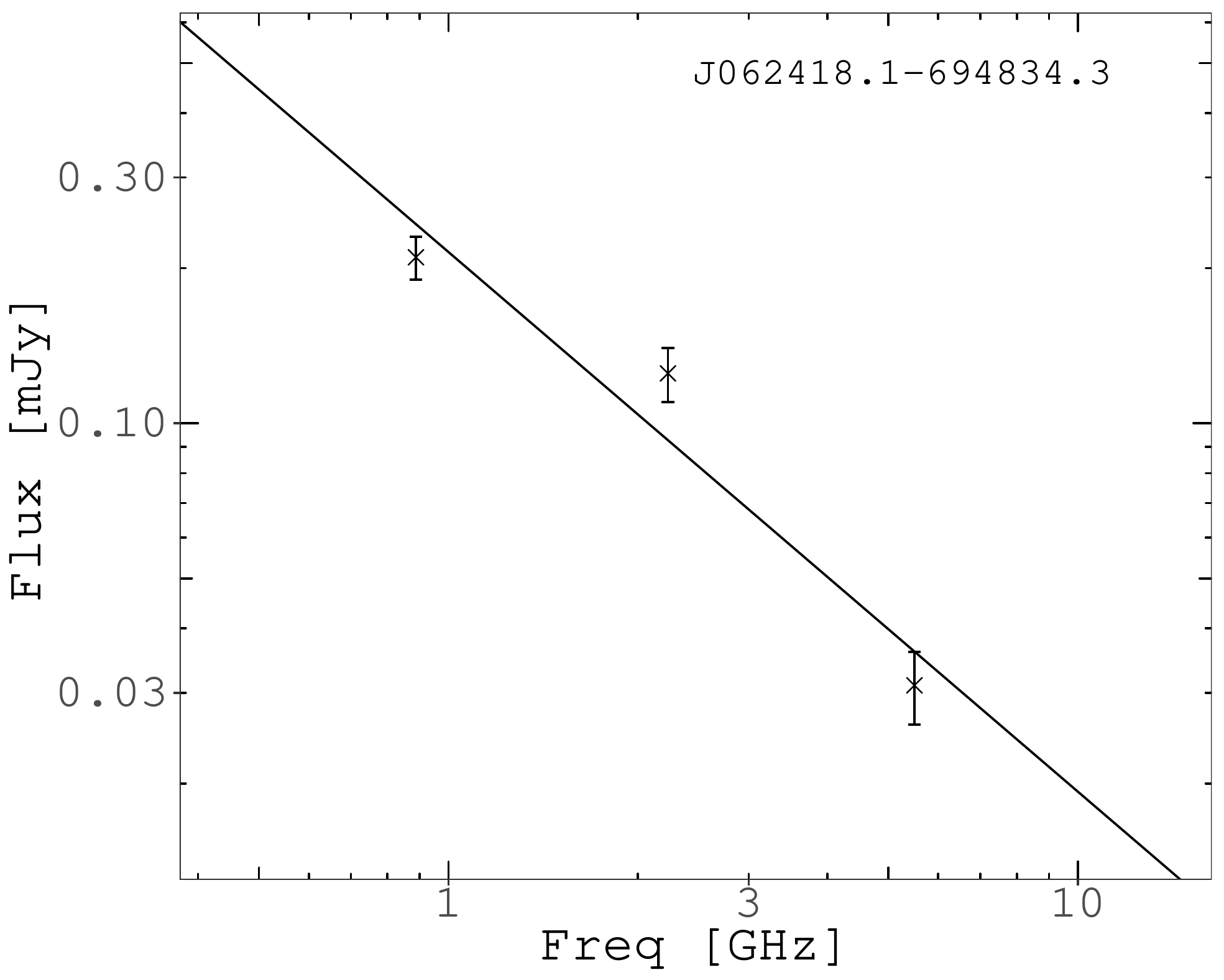}
\includegraphics[width=0.33\textwidth,scale=0.7, trim=0 0 0 0 0,clip]{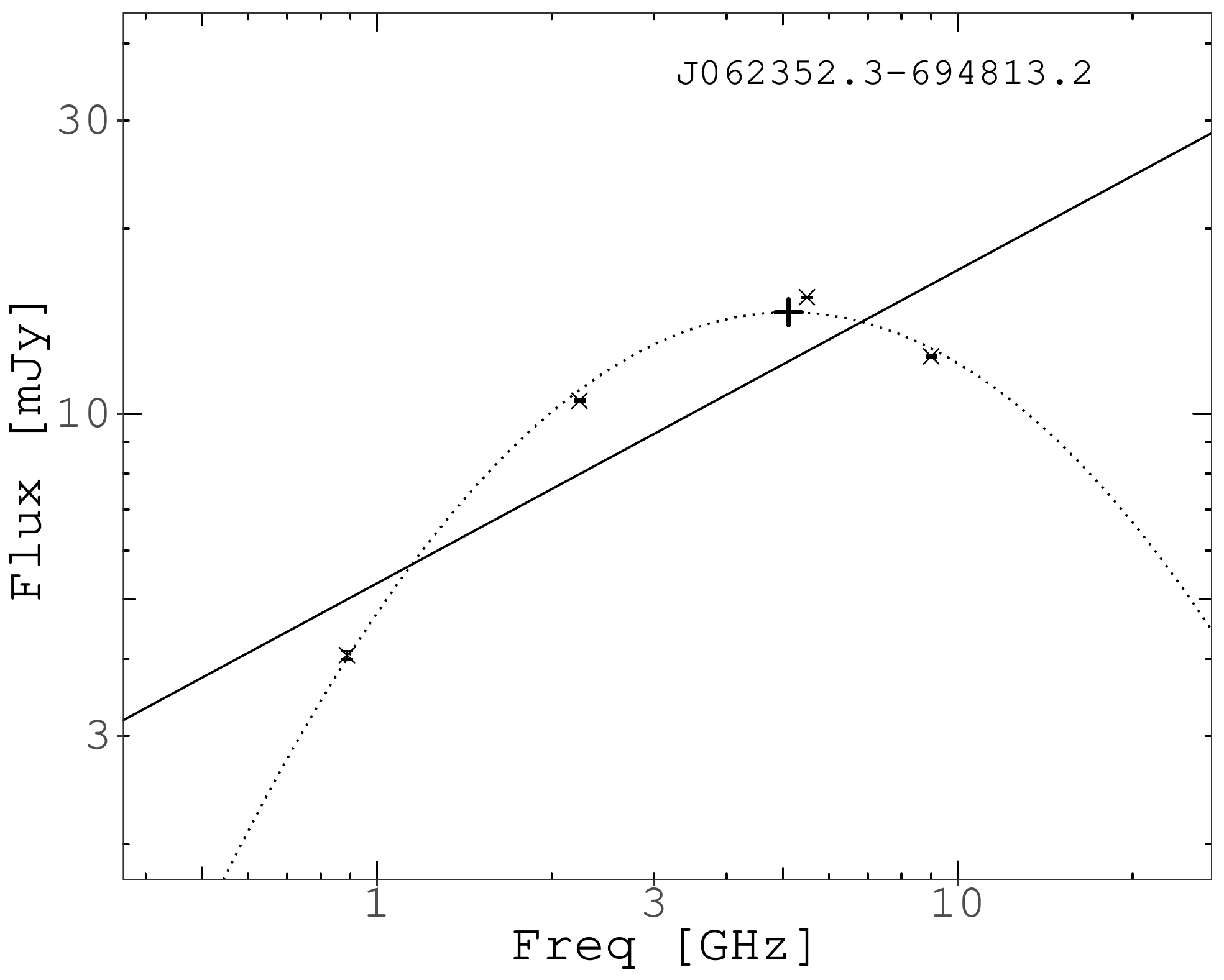}\\
\includegraphics[width=0.33\textwidth,scale=0.7, trim=0 0 0 0 0,clip]{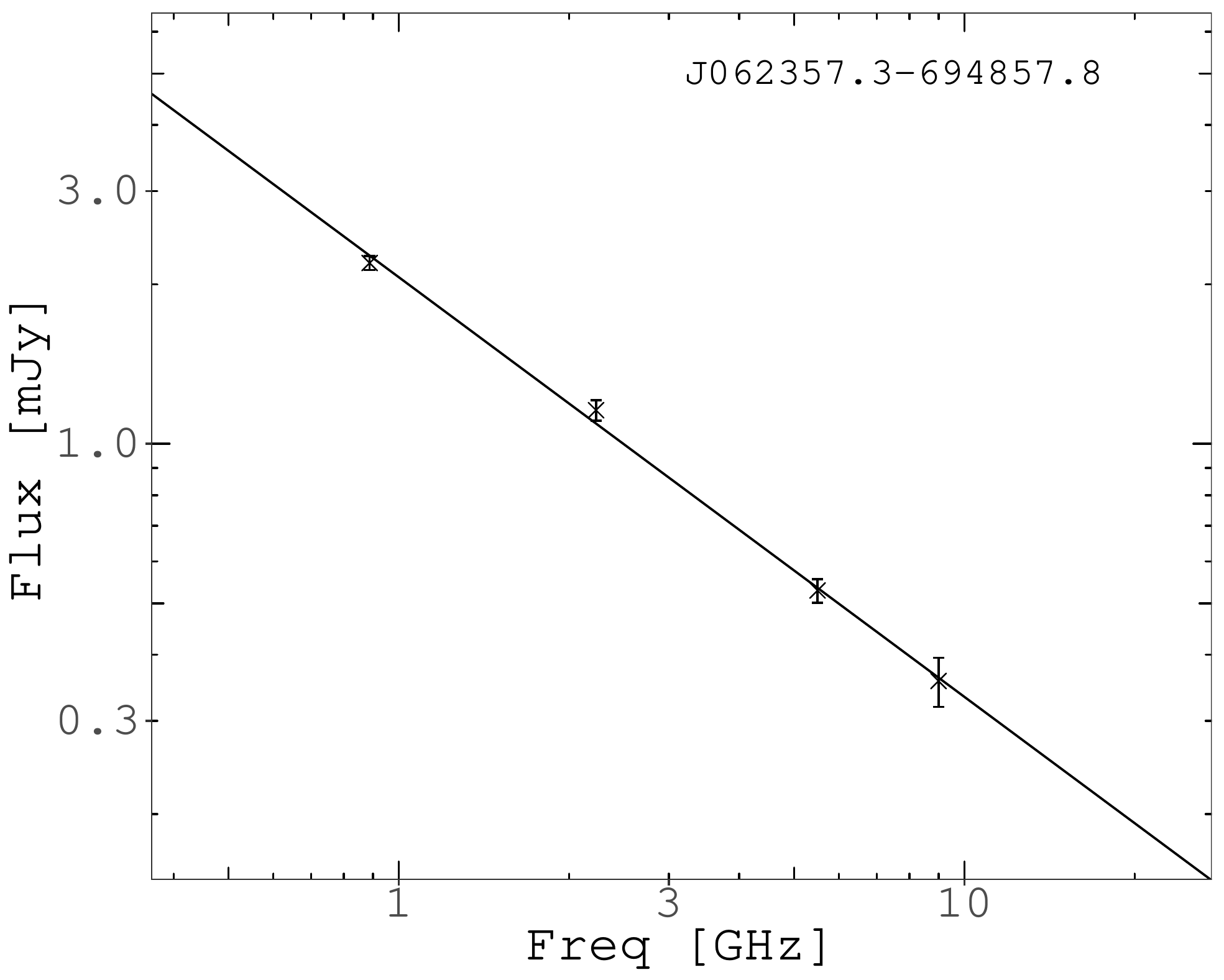}
\includegraphics[width=0.33\textwidth,scale=0.7, trim=0 0 0 0 0,clip]{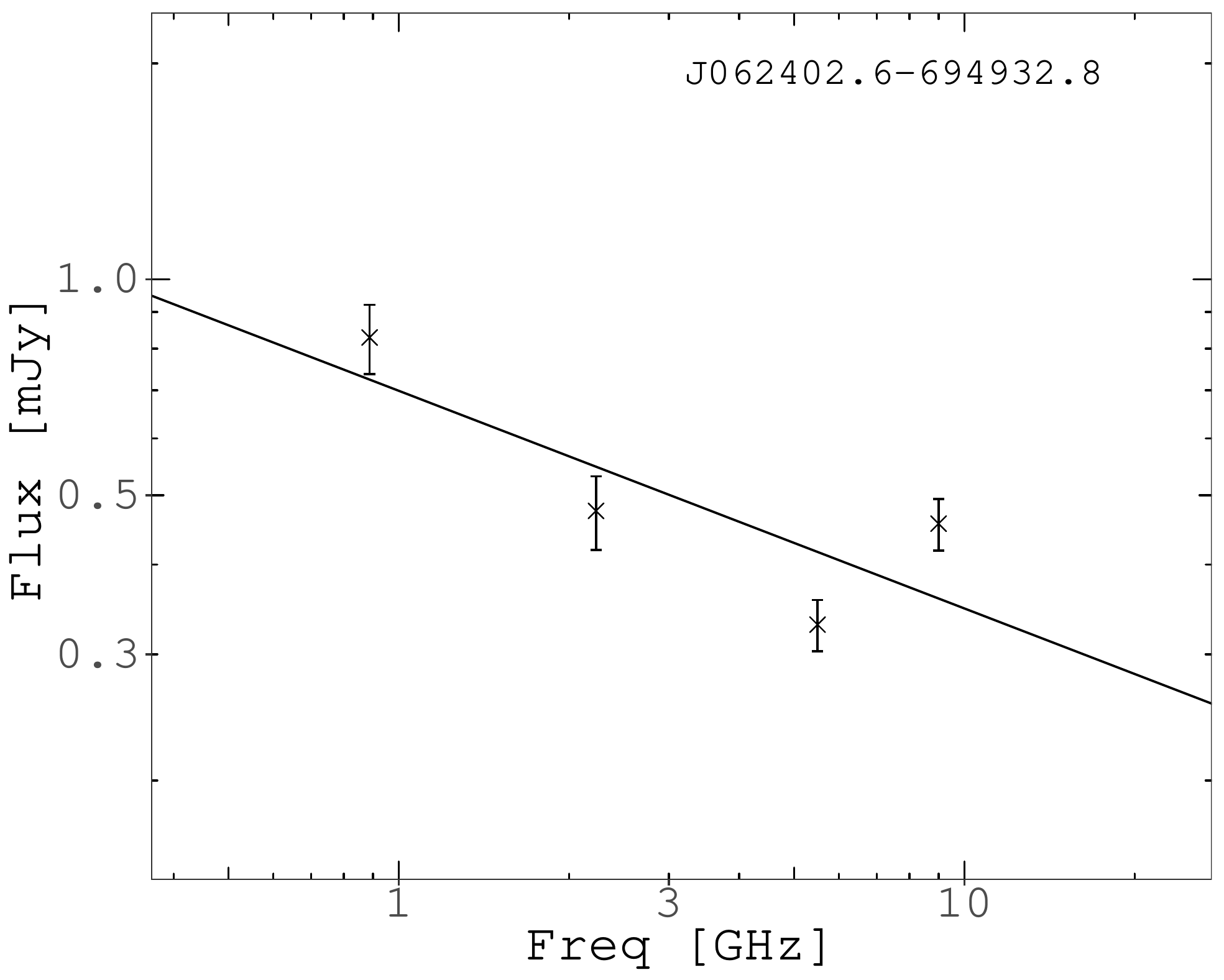}
\includegraphics[width=0.33\textwidth,scale=0.7, trim=0 0 0 0 0,clip]{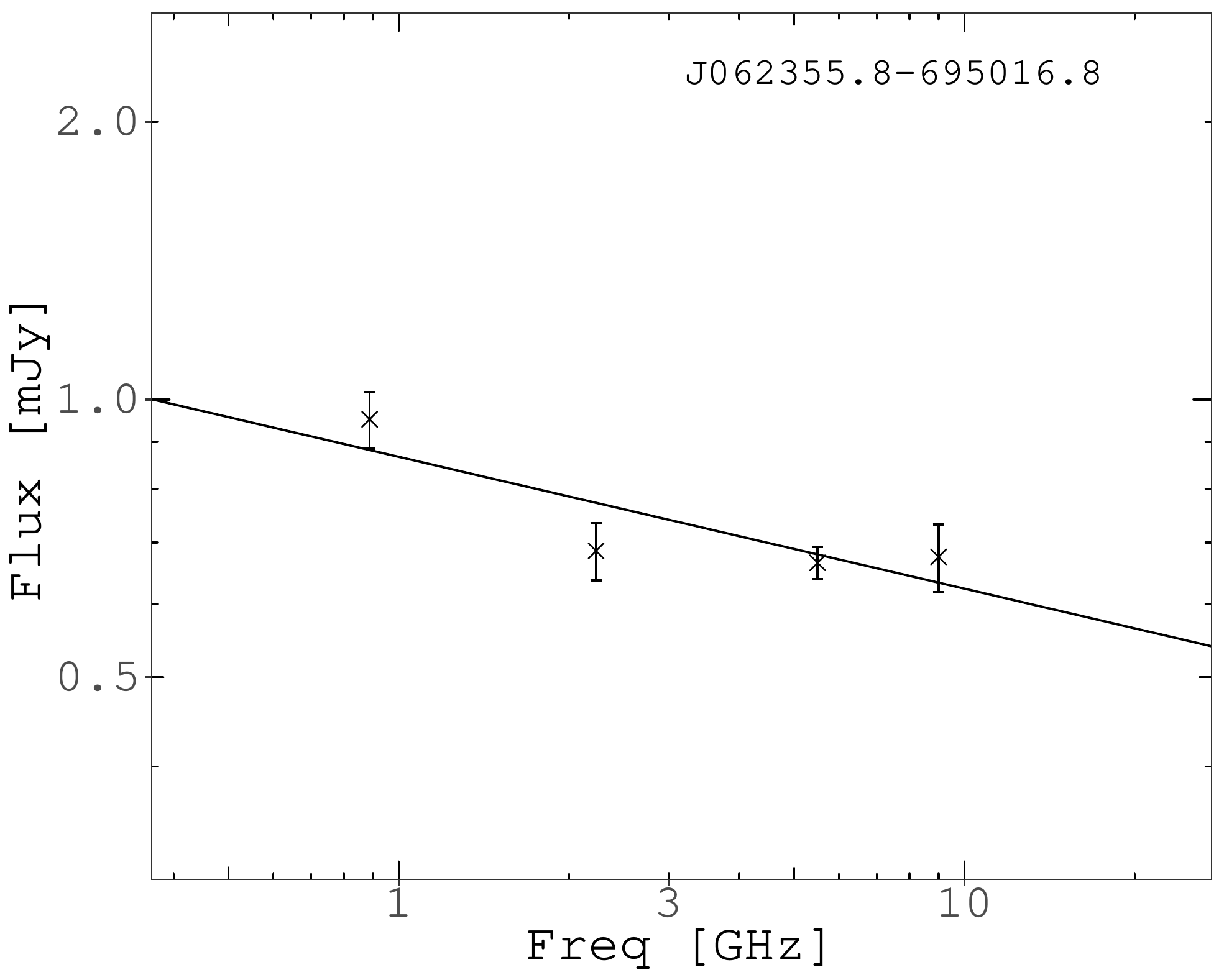}\\
\caption{Spectral index plots for \lmcorc\ (top left), central source (top middle) and four bright sources near south-west from \lmcorc. See Table~\ref{tab2} for more details. 
The plus (+) sign in the top right image indicates the turnover frequency at $\sim$4900~MHz for the \ac{GPS} source ASKAP~J062352.3--694813.2.}
\label{fig:8}
\end{figure*}

We tried to confirm the spectral index of \lmcorc\ by producing TT-plots \citep{1962MNRAS.124..297T} between the radial profiles of the source at different frequencies. This proved to be very effective to properly determine the spectral index of a steep spectrum source located on top of a diffuse flat spectrum source \citep{2020MNRAS.496..723K}. TT-plots should therefore give more reliable results for observations that may suffer from missing short spacings, which in essence cause a diffuse negative source. If the observations suffer from missing short spacings it would then produce an offset from the origin of the diagram. The most reliable results we expect to get from TT-plots are between frequencies with the largest frequency gap: 888 and 5500~MHz (see Fig.~\ref{fig:7}), 888 and 9000~MHz, and 2322 and 9000~MHz. We fitted the TT-plots with a linear function of the form: $a + b\cdot T_b$, with $a$ as a free parameter (red in Fig.~\ref{fig:7}) and with $a = 0$ (green in Fig.~\ref{fig:7}). The negative y-offset of the red fit is obvious and indicates missing short spacings at the higher frequency. If we now imagine the missing short spacings as a diffuse low surface brightness source that peaks in the centre of \lmcorc\ this would change the gradient of the radial profile so that the resulting red fit would be steeper and green fit would be flatter. But since we do not know the amplitude of the missing short spacings, the real spectral index could be anywhere in between. Averaging those fits for the three largest frequency gaps result in a spectral index of $\alpha = -0.4 \pm 0.1$. Since we do not know the proper probability distribution for the spectral index between the two fits, the uncertainty is just a crude estimate. This result is somewhat flatter than the above spectral index estimate of $\alpha=-0.54\pm0.08$ but still overlapping.

\begin{table*}
\centering
\caption{Flux density and spectral index estimates for \lmcorc, central source (J062418.1--694834.3) and four bright sources near the south-west rim. These four sources are possibly background objects that may not be related to \lmcorc. Source J062352.3--694813.2 is classified as a \ac{GPS} and with no power law spectrum as such, we did not estimate the spectral index as for the other sources.}
\begin{tabular}{cccccccc}
\hline
Source Name         & RA (J2000) &  Dec. (J2000)          &  S$_{\rm 888\,MHz}$    & S$_{\rm 2232\,MHz}$   & S$_{\rm 5500\,MHz}$   & S$_{\rm 9000\,MHz}$   & $\alpha \pm \Delta\alpha$ \\
    ASKAP           & (h m s)    & (\D\ \arcmin\ \arcsec) & (mJy)                  & (mJy)                 & (mJy)                 & (mJy)                 &  \\
%
\hline
\lmcorc\            & 06 24 17.78 & --69 48 37.8 & 11.7$\pm$5.8   & 9.1$\pm$3.5    & 4.5$\pm$1.9    & 3.6$\pm$1.8    & --0.54$\pm$0.08   \\
J062418.1--694834.3 & 06 24 18.17 & --69 48 34.3 & 0.21$\pm$0.02  & 0.125$\pm$0.015& 0.031$\pm$0.005& ---            & --1.03$\pm$0.25   \\
J062352.3--694813.2 & 06 23 52.32 & --69 48 13.2 & 4.06$\pm$0.06  & 10.51$\pm$0.05 & 15.49$\pm$0.03 & 12.41$\pm$0.04 & \ac{GPS}   \\
J062357.3--694857.8 & 06 23 57.35 & --69 48 57.8 & 2.19$\pm$0.06  & 1.16$\pm$0.05  & 0.53$\pm$0.03  & 0.36$\pm$0.04  & --0.79$\pm$0.02 \\
J062402.6--694932.8 & 06 24 02.61 & --69 49 32.8 & 0.83$\pm$0.09  & 0.47$\pm$0.06  & 0.33$\pm$0.03  & 0.46$\pm$0.04  & --0.30$\pm$0.12 \\
J062355.8--695016.8 & 06 23 55.83 & --69 50 16.8 & 0.95$\pm$0.07  & 0.69$\pm$0.05  & 0.67$\pm$0.03  & 0.67$\pm$0.06  & --0.14$\pm$0.05 \\
\hline
\end{tabular}
\label{tab2}
\end{table*}

\subsection{Morphology}
 \label{sec:morph}

The very high degree of circularity of \lmcorc\ (compared to the \ac{SNR} sample of \citet{Ranasinghe:2019quc,2011ApJ...732..114L}) is evident in the polar projection of the effective 10-cm map (Fig.~\ref{fig:polar}; bottom). The peak brightness of the ring occurs at an approximate sky-projected radius of 90~arcsec. To explore the 2D geometry of \lmcorc, we have created radial profile plots of the NW and SE halves and the total ring. Attempts to model the total \lmcorc\ ring profile as the projection of a spherical shell with uniform emissivity over-predict the brightness towards the centre of \lmcorc. This implies that the front and back of the shell have lower emissivities than the sides. Despite the use of a wide range of interferometer arrays (where the shortest baseline is 46-m (array: EW367) at all \ac{ATCA} frequencies), the missing short spacing problem, already indicated in the discussion about the TT-plots (Section~\ref{sec:fluxspectra}), is causing a significant reduction of the diffuse emission coming from the shells moving towards us and away from us so that the central region in particular at the higher frequencies is nearly devoid of emission.

Because of the missing short spacing problem, our fitting of \lmcorc\ with a uniform emissivity shell is only illustrative, giving its approximate radius and width. The radial profiles of the NW and SE halves have slightly different shapes, while the best match for the total ring is a shell of 2D radius 98~arcsec and a width of $\sim$10~arcsec. A sharp outer edge is visible when looking at the large-scale emission alone; it cuts off at $\sim$115~arcsec at all azimuths. The 98~arcsec and 115~arcsec fiducial radii are shown in Fig.~\ref{fig:polar} (top right). For all analyses in this paper, we adopt a diameter of 196~arcsec and a shell thickness of 10~arcsec with error of 4~arcsec.

\subsection{Polarisation}
 \label{sec:polarisation1}
We searched for polarisation from \lmcorc\ using the good sensitivity 2232~MHz and 5500~MHz \ac{ATCA} maps by averaging the total and bias-corrected polarised intensities in concentric rings. We detected no significant polarisation at any radius, and determined 3$\sigma$ upper limits at the location of the peak flux of the ring of 7 and 9~per~cent, respectively. There is a marginal (2$\sigma$) hint of polarised flux at one location on the 2232~MHz ring, which would need to be verified at higher sensitivity. The only high significance polarisation that we detected is associated with the \ac{GPS} source J062352.3--694813.2 (Fig.~\ref{fig:8} and Table~\ref{tab2}). While at 2232~MHz this source is depolarised, at 5500~MHz and 9000~MHz we measure mean polarisation fractions of 1.8$\pm$0.3 and 2.1$\pm$0.2~per~cent, respectively.

\subsection{X-rays}
As there is no obvious X-ray emission that correlates with the extended radio emission from \lmcorc, we run the source detection in the 0.3$-$10~keV data of MOS1 and MOS2 simultaneously \citep[for the details of the source detection procedure adopted here, see ][]{2014A&A...566A.115D}. We detect five X-ray point sources within the outer circle of \lmcorc, as shown in Table~\ref{tab3}. The detection likelihoods for the faintest and brightest sources are $\sim$23.4 and $\sim$718.2, respectively. The brightest {\it XMM-Newton} source in the field, XMMU~J062426.76--694726.3 (0.3$-$10~keV rate, $1.46\pm0.08 \times 10^{-2}$~cts~s$^{-1}$, 1$\sigma$ positional error: $\sigma_{\rm pos}=0.34$~arcsec) is detected at various infrared \ac{WISE} and SMASH optical bands -- indicating its \ac{AGN} nature (see Fig.~\ref{fig:5new}). Source XMMU~J062402.35$-$694811.1 ($\sigma_{\rm pos}=0.66$~arcsec) can be associated with the star {\it Gaia}\footnote{The Global Astrometric Interferometer for Astrophysics ({\it Gaia}) is an European Space Agency (ESA) mission designed to chart a 3D map of Galaxy, providing key positional and radial velocity measurement data \citep{2021A&A...649A...1G}.}~EDR3 (early data release 3)~5278760380137682816 (see Section~\ref{super-flare}). Sources XMMU~J062424.84$-$694853.3 ($\sigma_{\rm pos}=0.85$~arcsec) and XMMU~J062410.68$-$694947.2 ($\sigma_{\rm pos}=1.10$~arcsec) can be associated with infrared unWISE \citep{2019ApJS..240...30S} and CatWISE \citep{2021ApJS..253....8M} sources, although the position offsets of the counterparts associated to the {\it XMM-Newton} sources are large: $\sim$2.2~arcsec and $\sim$1.8~arcsec for XMMU~J062424.84$-$694853.3 and XMMU~J062410.68$-$694947.2 respectively. Source XMMU~J062422.58$-$694756.1 ($\sigma_{\rm pos}=0.89$~arcsec) can be associated to the star {\it Gaia}-EDR3~5278760448849205888 source at distance of 317~pc (see Fig.~\ref{fig:5new}).

To determine the upper limit of the X-ray flux we extracted spectra from the area covered by \lmcorc\ and from nearby background regions from both the MOS1 and MOS2 vignetting-corrected event lists, produced using the SAS task {\it evigweight}. The \lmcorc\ extraction region was defined based on the centre and dimensions determined above. The \lmcorc\ spectra comprise the non-X-ray background (NXB), the astrophysical X-ray background (AXB), and any possible contribution from \lmcorc\ itself. The method adopted for constraining the NXB and AXB are described in detail in, e.g. \citet{2016A&A...585A.162M, 2016A&A...586A...4K}, and references therein, whereby the background spectra and filter-wheel closed data are used to constrain the AXB and the NXB, respectively. We account for possible emission from \lmcorc\ using a simple collisional ionisation equilibrium model. While this model is not formally required, including it in our fits allowed us to determine an upper limit to its contribution. This derived upper limit depends on the shape of the spectrum adopted, i.e. the assumed spectral parameters. Since \lmcorc\ is likely an \ac{SNR} in the ED or early Sedov phase, (with age estimates in the range from 2200 to 7100 yrs, see Section~\ref{sec:evolutionarystate}), we approximated its spectrum as a thermal plasma with a temperature of 1~keV. We then fitted all spectra simultaneously in XSPEC \citep{Arnaud1996} v12.11.1 using the abundance table of \citet{Wilms2000}, the photoelectric absorption cross-sections of \citet{Bal1992}, and atomic data from ATOMDB~3.0.9 (\url{http://www.atomdb.org/index.php}). We determined the upper limit to the absorbed X-ray flux in the 0.3--10~keV range from the \lmcorc\ model normalisation upper limit to be 1.2$\times10^{-14}$~erg~cm$^{-2}$~s$^{-1}$, corresponding to a surface brightness of 2.4$\times10^{-15}$~erg~cm$^{-2}$~s$^{-1}$~arcmin$^2$ over the area of \lmcorc. Assuming the \ac{SNR} \lmcorc\ is more evolved with a temperature of 0.5~keV leads to a minor decrease in these upper limit estimates. If it is less evolved with a temperature of 2~keV, the estimated limits increase by a factor of $\sim$2.

\begin{table*}
\centering
\caption{Details of five bright {\it XMM-Newton} X-ray sources (0.3$-$10~keV) near \lmcorc. }
\begin{tabular}{ccccccl}
\hline
Source Name         &RA (J2000)&Dec. (J2000)&  $\sigma_{\rm pos}$   & Likelihood  & Count Rate      \\
XMMU~J              & (degree) &   (degree) &  (arcsec)             &             & (10$^{-2}$~cts~s$^{-1}$) \\
\hline
062426.76--694726.3 & 96.11223 & --69.79074 & 0.34                  & 748.2       & 1.46$\pm$0.08 \\    	
062402.35--694811.1 & 96.00978 & --69.80310 & 0.66                  & 295.3       & 0.72$\pm$0.06 \\
062424.84--694853.3 & 96.10350 & --69.81481 & 0.85                  & 107.7       & 0.50$\pm$0.06 \\
062422.58--694756.1 & 96.09409 & --69.79893 & 0.89                  & 46.6        & 0.28$\pm$0.04 \\
062410.68--694947.2 & 96.04451 & --69.82978 & 1.10                  & 23.4        & 0.22$\pm$0.04 \\
\hline
\end{tabular}
\label{tab3}
\end{table*}

\section{\lmcorc\ as a rogue \ac{SNR}}
 \label{sec:SNR}

The overall appearance and radio spectral properties of \lmcorc\ suggest a possible \ac{SNR} origin; if it were found at low galactic latitudes, it would likely be included in existing \ac{SNR} catalogues. However, its location far from the plane of the \ac{MW} and from the \ac{LMC} create challenges for the \ac{SNR} hypothesis. In this section, we examine the plausibility that \lmcorc\ is a rogue \ac{SNR} associated with either the \ac{MW} or \ac{LMC}. Because this rogue \ac{SNR} hypothesis appears quite speculative, we provide a comprehensive look at all the various issues we considered in its evaluation.

\lmcorc\ could have either be formed well outside the \ac{MW} or the \ac{LMC} bar, or its location could be the result of an explosive kick from an evolved star (post-main sequence) that began in either galaxy and resulted as an \ac{SNR}. We thus start with examining possible progenitors in Section~\ref{sec:progenitor}, followed by a summary of what is known about the presumed local environment in Section~\ref{sec:environment}. We then look at the radio properties (Section~\ref{SNRproperties}) and X-ray limits (Section~\ref{sec:xray+}) to establish whether these are consistent with an \ac{SNR} origin (Section~\ref{sec:evolutionarystate}).

\subsection{Possible progenitors}
 \label{sec:progenitor}
The uniqueness of the possible \ac{SN} progenitor may indicate the nature and evolutionary stage of \lmcorc. Here, we examine various scenarios that are based on the object's shape, size, location, radio properties as well as the fact that we do not see an obvious counterpart in any other waveband. In Section~\ref{sec:fluxspectra} we concluded that there is no associated object such as a pulsar or a \ac{PWN}  in the centre as the nearest pulsar/\ac{PWN} J0540--69.3 \citep{2014ApJ...780...50B} is more than 4~degrees away. Given the circular shape \citep{Ranasinghe:2019quc,2011ApJ...732..114L} and location well outside the \ac{MW} bar \citep{2017MNRAS.471.1390H}, \lmcorc\ would be more likely a type~Ia \ac{SN} explosion from a star that was formed (and lived) in the outskirts of the \ac{LMC}. Somewhat less favourably, \ac{SN} of either type~Ia or core-collapse progenitor that came from the \ac{LMC} as a \ac{HVS} which travelled across the \ac{LMC} boundaries and exploded as an \ac{SN} could also be the origin for \lmcorc.

\subsubsection{Hypervelocity runaway from the \ac{LMC} or \ac{MW}} 

\citet{2017MNRAS.469.2151B,2020MNRAS.497.2930B} and \citet{2011MNRAS.414.3501E} suggest that objects like \lmcorc\ could result as the remnant of a \ac{SN} hypervelocity runaway from the \ac{LMC} or a \ac{MW}. \citet{2021MNRAS.507.4997E} notes that the \ac{LMC} \acp{HVS} will outnumber \ac{MW} ones by a factor $\sim$2.5.

So-called hostless core-collapse \ac{SN}e (CCSNe; beyond 10~kpc of the galaxy outskirts) are most likely to come from \ac{HVS} \citep{2011A&A...536A.103Z}. Direct evidence for the existence of \ac{HVS} in the \ac{LG} is shown in \citet{2018AJ....156...98P} who found 10 \ac{LMC} runaways over a wide range of masses that appear to have been ejected from the massive \ac{LMC} star cluster R136 in the tangential plane to distances of up to 98~pc but at only $\upsilon_{max}=120$~\kms. Even more convincingly, \citet{2005ApJ...634L.181E}, \citet{2008A&A...480L..37P}, \citet{2018A&A...620A..48I} and \citet{2019MNRAS.483.2007E} detected the B-star HE~0437--5439 (a.k.a. HVS~3) as a hypervelocity ($\upsilon$=723~\kms) runaway with 9.1~M$_{\odot}$, which travelled $\sim$18~kpc from the \ac{LMC} centre in 20~Myrs. In fact, most observed \ac{HVS} escaping galaxies are O and B type stars \citep{2015ARA&A..53...15B}.   All these examples suggest that \lmcorc\ (if an \ac{SNR} of CCSNe type) could have a \ac{HVS} progenitor.

Another possibility is a type~Ia progenitor. Such an explosion would occur with a delay after binary formation of $t_{delay}=300\times t_{300}$~Myr, where $t_{300}$ is the delay time (the interval between the progenitor formation and the \ac{SN} type~Ia explosion, close to the lifetime of a 3$\Msun$ star) in units of 300~Myr \citep{2012PASA...29..447M,2021MNRAS.502.5882F}.  If the progenitor velocity is zero before formation and constant after formation, the progenitor lifetime is equal to the travel time. If the progenitor is from the \ac{LMC} then its velocity is $\sim$7$/t_{300}$~\kms. If it from the Galaxy its velocity is $\sim$160$/t_{300}$~\kms. The lower velocities required for an \ac{LMC} origin are easier to obtain, thus arguing for that case. However, individual cases could have quite different delay times depending predominantly on the masses and separation of binary components at birth.

The typical \ac{SN} type~Ia would result from the merger of a \ac{WD}s as a \ac{DD} \citep{1984ApJ...277..355W} or a \ac{WD} and main~sequence/giant star as a \ac{SD} \citep{1973ApJ...186.1007W}. For a DD scenario, it is unlikely to find any remnant star because the thermonuclear explosion is believed to completely destroy the white dwarf. An exception to this is the dynamically-driven, double-detonation DD scenario, which leads to a surviving hyper-velocity \ac{WD} companion \citep{2013ApJ...770L...8P,2018ApJ...865...15S}. On the other hand, for the \ac{SD} scenario we expect to find a main sequence (MS), red-giant (RG) or subdwarf~B (sdB) star that would produce a surviving helium donor such as the star US708 \citep{2005A&A...444L..61H,2015Sci...347.1126G} which is the second fastest unbound \ac{MW} \ac{HVS} at a current space velocity of 994~\kms\ \citep{2020A&A...641A..52N}. Also, nearby ($<$2~kpc) `zombie' stars as \ac{WD}s that survived the thermonuclear explosion as found by \citet{2019MNRAS.489.1489R} are considered as well as a subclass of the above two that are dynamically driven \ac{DD} double-detonation (D$^6$) scenarios \citep{2018ApJ...865...15S}. While these example progenitors are all of Galactic origin, we expect that similar scenarios are quite possible to exist in the \ac{LMC} as well as in the old stellar disk region.

The above examples would certainly suggest that the progenitor of \ac{SNR} candidate \lmcorc\ could come from a runaway \ac{HVS} of CCSNe origin or a normal velocity type~Ia progenitor from either galaxy, albeit, with roughly 20 times higher velocities required for \ac{MW} origin than for \ac{LMC} origin. For a single star system (CCSNe), massive star lifetimes (10~M$_{\odot}$ and up) are 25~Myr down to $\sim$5~Myr  (vs. delay times for type~Ia which are $\sim$100 Myr to $\sim$1 Gyr). Therefore, the peculiar velocities for an origin in the \ac{LMC} are $\upsilon\sim$80~\kms\ to $\upsilon\sim$400~\kms\ for CCSNe and $\upsilon\sim$2~\kms\ to $\upsilon\sim$20~\kms\ for type~Ia. For an origin in the \ac{MW}, the velocities are much higher: $\upsilon\sim$2000~\kms\ to $\upsilon\sim$10,000~\kms\ for CCSNe and $\upsilon\sim$50~\kms\ to $\upsilon\sim$500~\kms\ for type~Ia. However, the location of \lmcorc\ on the orbit path of the \ac{LMC} \citep[see e.g. fig.~6 of][]{2017MNRAS.469.2151B} favour more the \ac{LMC} scenario than a Galactic one.

In the \ac{SMBH} slingshot ejection scenario \citep{1988Natur.331..687H}, velocities as high as $\upsilon$=4000~\kms\ are possible for tidally disrupted close binaries \citep{2015ARA&A..53...15B}. In that scenario, the star must originate from the centre of our Galaxy. In fact, an A-type star (S5-HVS\,1; Galactic coordinates: $l$=337.4361779510168\D\ $b$=--57.4004150381868\D) at $\sim$9~kpc distance from the Sun has been found recently, travelling at a $\upsilon\sim$1700~\kms\ \citep{2020MNRAS.491.2465K,2021A&A...646L...4I} and {\it Gaia} proper motions showed its origin near the Galactic centre. Hence, there is a possibility that the origin of \lmcorc\ as an \ac{SNR} may be a star that would come from the centre of the \ac{MW}. However, given that only one such star has been found despite intensive searches (and that one is of relatively low mass), we regard the \ac{MW} origin scenario as very unlikely. Moreover, the direction of the ejection from the \ac{SMBH} depends on the orbital plane around the \ac{SMBH} of the incoming stellar binary, which is not random, because those stars are likely injected from disks of stars or stellar clusters near the Galactic centre.

At much more moderate velocities of 70 to 350~\kms\ a dynamical ejection from a \ac{LMC} stellar cluster is very plausible \citep{2007MNRAS.376L..29G}. We note that HVS~3 is more than twice as distant from the \ac{LMC} centre as \lmcorc\ and, accordingly the ejection velocity would need to be as high as $\upsilon$=870~\kms \citep{2018A&A...620A..48I}. This is very hard to achieve in dynamical ejection scenarios unless an intermediate mass stellar black hole is involved \citep{2019MNRAS.489.4543F}. The lower ejection velocities that we derived can be explained without invoking exotic mechanisms.

In summary, a variety of hypervelocity stars are known in both the \ac{LMC} and \ac{MW}, and their properties make them plausible candidates for the progenitor of \lmcorc.

\subsubsection{Evolved star from the old \ac{LMC} stellar disk}
 \label{sec:evstar}
We also examine the scenario that the progenitor of \ac{SNR} candidate \lmcorc\ is a star that has always lived in the old (outer) stellar disk of the \ac{LMC}, but formed more recently. Despite the sparsity of matter in this region, there could be localised new star formation triggered by disequilibrium due to the interactions between the \ac{LMC} and the stellar and dark matter halos of the \ac{MW}. 

\begin{figure*}
\centering
\includegraphics[width=\textwidth, trim=0 0 0 0, clip]{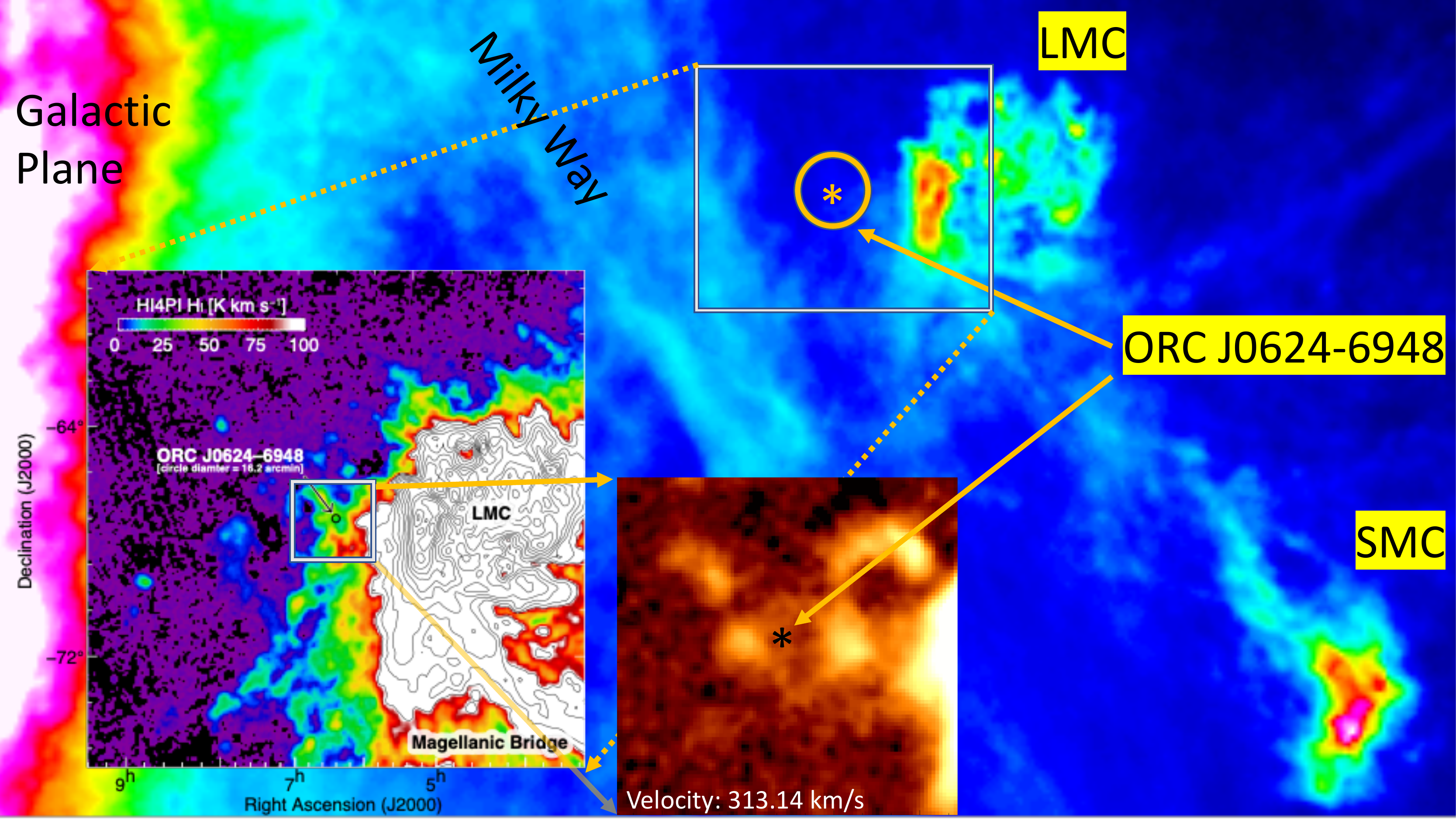}
\caption{The Parkes GASS and HI4PI (insets) \HI\ images indicating the position of \lmcorc\ with respect to the \ac{LMC}, \ac{SMC} and \ac{MW}. The image beam size is 16.2~arcmin. In the leftmost inset the integrated velocity range is from $\upsilon$=189.5 to $\upsilon$=319.6~\kms. The contour levels are 100, 150, 250, 400, 600, 850, 1150, 1500, 1900, 2350 and 2800~K~\kms.}
\label{fig:9}
\end{figure*}

There is clear observational evidence of local outflows emanating from supergiant shells in the \ac{LMC} and a trailing filament of \HI\ gas mainly originating from the 30-Doradus region as well as the Leading Arm connection. Indeed, looking in the Parkes Galactic All-Sky Survey (GASS/HI4PI; angular resolution is $\sim$16.2~arcmin; \cite{2016A&A...594A.116H}) \HI\ image (Fig.~\ref{fig:9}), we can see that the \ac{LMC} \HI\ envelope covers a much larger area than the optical (and radio-continuum) emission, including the location of this new object that is within the \HI\ envelope of the \ac{LMC}. 

Evidence and modelling for the presence of such interaction waves and star formation has been presented by \citet{2016ApJ...825...20B} and \citet{conroy}. \ac{SMASH} \citep{2021AJ....161...74N}, {\sc SkyMapper} and the \ac{WISE} all-sky Mollweide projection images do show an increased density of K giant stars $60-100~\mathrm{kpc}$ from the Galactic centre, especially in the NE and SW regions (in equatorial sense) of the Galactic Halo \citep{conroy}. The latter may be due to the local wake of the passage of the \ac{LMC}. The model of \citet[][their fig.~1]{conroy} predicts some local increased stellar density in our region of interest between the \ac{LMC} and \ac{MW}, albeit relatively small. Nevertheless, several recent studies confirm a large envelope of stars around the \ac{LMC} (and \ac{MW}) \citep{2020MNRAS.497.3055C}. This strongly argues that a range of stellar variety would already exist in these regions for at least the past tens-to-hundreds of million years. We also note that the location of \lmcorc, some 3~degrees from the edge of the \ac{LMC} bar, is just within the outer shell of stars that surround the \ac{LMC}, clearly apparent in a large-area \ac{WISE}~W1 (3.4~$\mu$m) panoramic mosaic of the \ac{LMC} region \citep{2019ApJS..245...25J}, which suggests \lmcorc\ is still within the outer confines of the \ac{LMC} dwarf galaxy. 

The situation with respect to the \ac{MW} is a little more complicated. \lmcorc\ is located at a Galactic longitude of about $280\degr$ at $28\degr$ from the Galactic plane. For the evolutionary models discussed below, we adopt a somewhat arbitrary 5~kpc distance for a \ac{MW} origin. Adopting this distance does cause some problems; at 5~kpc distance it would be about 2.5~kpc below the Galactic plane, away from any potential formation site within the \ac{MW} for its progenitor star. However, if \lmcorc\ were much closer, it would be very small in linear size and very young, so the \ac{SN} should have appeared in the historical record (see APPENDIX~\ref{app:1}). Further away than 5~kpc would increase the distance from the plane and make a relation to the \ac{MW} even less likely. We could, of course, find the location of the spiral arms in the plane of the Galaxy and optimise the potential distance to \lmcorc\ within the \ac{MW}, but that is not constrained by our data so  we adopt an optimal distance of 5~kpc if \lmcorc\ is related to our Galaxy.

\begin{figure*}
\centering
\includegraphics[width=0.49\textwidth, trim=0 0 0 0, clip]{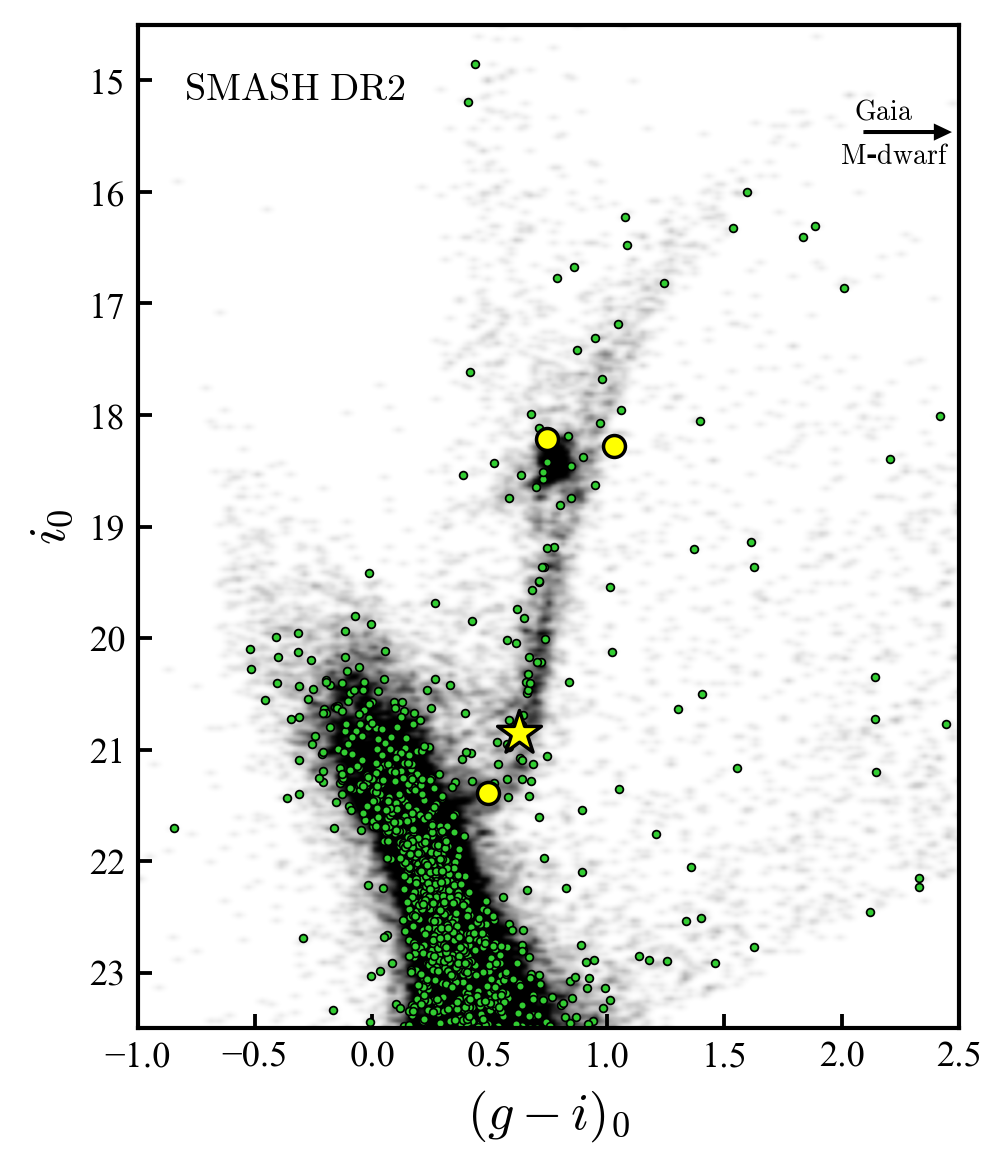}
\includegraphics[width=0.49\textwidth, trim=0 0 0 0, clip]{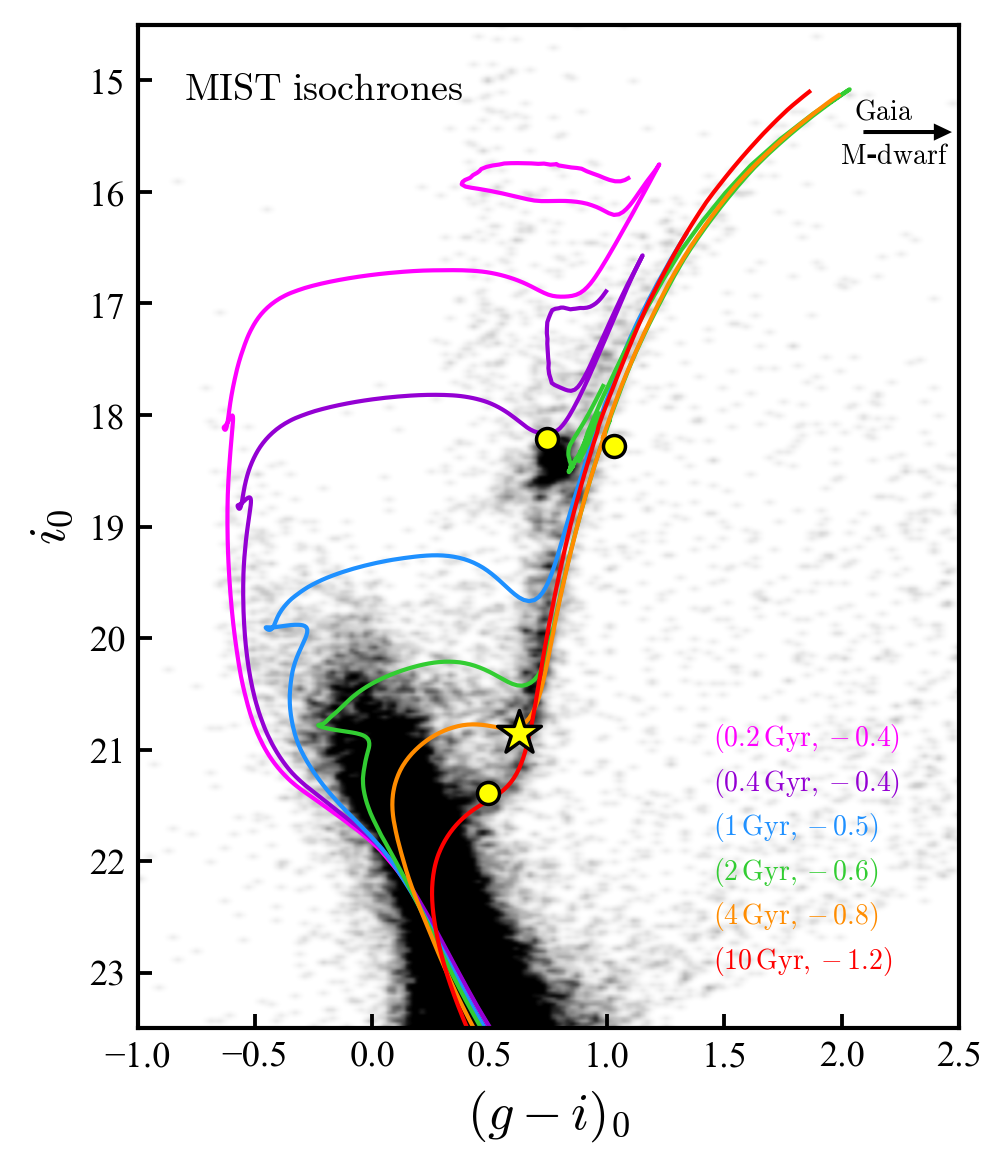}
\caption{Colour-magnitude diagrams for stars in the vicinity of \lmcorc\ from \ac{SMASH} DR2. {\it Left:} Stars within 12~arcmin of the geometric centre of \lmcorc\ indicate the typical local populations and are plotted in grey as a smoothed Hess diagram (projected density distribution). Stars projected inside the ring itself (i.e., located within 100~arcsec of its centre) are marked by green points, and the four evolved stars within 20~arcsec by yellow points. The star on the lower part of the RGB discussed in the text (also see left inset in Fig.~\ref{fig:5new}) is marked with a star symbol. The arrow to the upper right indicates the location of the {\it Gaia} M-dwarf discussed in the text, which has colour $(g-i)_0 = 3.24$. {\it Right:} Isochrones from the MIST library \citep{2016ApJS..222....8D,2016ApJ...823..102C} over-plotted on the same Hess diagram as shown in the left-hand panel. These are appropriate for the observed age-metallicity relationship in the \ac{LMC} field \citep[e.g.][]{2013AJ....145...17P}. For clarity, evolution after the helium flash is shown only for the 200~Myr, 400~Myr, and 2~Gyr isochrones; the loci for the 1~Gyr, 4~Gyr and 10~Gyr tracks are very similar to that of the 2~Gyr isochrone.}
\label{fig:10}
\end{figure*}

We queried the \ac{SMASH} public data release~2 \citep{2021AJ....161...74N}, which covers the region surrounding \lmcorc, and in Fig.~\ref{fig:10} display a colour-magnitude diagram (CMD) showing all stars with robust photometry within 12~arcmin of its geometric centre to sample the local stellar population. LMC members are clearly the dominant population at this location, providing weak statistical evidence that \lmcorc\ is indeed associated with the LMC under the assumption that it has a stellar origin. Isochrones from the MIST library \citep{2016ApJS..222....8D,2016ApJ...823..102C} suggest that most stars in the vicinity of \lmcorc\ have ages older than $\sim$1~Gyr, but there are also low-density younger populations present with ages down to $\sim$200~Myr. However, when considering stars projected within the ring itself (i.e., within a radius of 100~arcsec of its geometric centre) there are no objects obviously younger than $\sim$1~Gyr. Overall, this indicates that the type of stellar populations expected to host a type~Ia \ac{SN} are abundant at this location, while the young massive stars required for a \ac{SN} core-collapse (CC) are exceedingly rare.

There are four stars within 20~arcsec of the centre of \lmcorc\ that have evolved off the main sequence. These are marked with yellow points in Fig.~\ref{fig:10}. The most interesting of these is a star on the lower red giant branch, which sits just $<4$~arcsec from the centre of \lmcorc\ (see Fig.~\ref{fig:5new} left inset) and, according to the stellar tracks, must be at least $\sim$4~Gyr old. This is a plausible candidate for a surviving companion (remnant star) from a \ac{SN} type~Ia \ac{SD} scenario, but is too faint for {\it Gaia} EDR3 to provide any astrometric data.

There are recent discoveries of another two similar objects located in the outskirts of the \ac{LMC}. Only $\sim$11~arcmin (or 160~pc at the distance of the \ac{LMC}) east from \lmcorc\ is the \ac{LMC} classical nova AT~2018bej \citep{2018ATel11610....1C,2020ATel13545....1D}. While also possible, we have no evidence that AT~2018bej resulted from a runaway star. This nearby location of a similar type of object further strengthens an \ac{LMC} \ac{SNR} scenario. A second possible runaway (or isolated) \ac{SNR} candidate (LMC~SNR~J0509-6402) is detected in a recent optical study by \citet{2021MNRAS.500.2336Y} and is positioned some 2$^{\circ}$ (1.75~kpc) north from the outermost (radio-continuum) boundaries of the \ac{LMC} but well inside of the \ac{LMC} \HI\ envelope. While this \ac{LMC} \ac{SNR} candidate is a bit larger (D=107~pc), the most distinctive difference to \lmcorc\ is the lack of radio detection. 

Several Galactic \acp{SNR} are detected at high Galactic latitudes but none is seen at distances further than a few kpc \citep{2019SerAJ.199...23S}. Another two high-latitude Galactic \acp{SNR} are detected in new {\it eROSITA} X-ray images \citep{2021A&A...648A..30B,2021MNRAS.507..971C} and are associated with the \ac{MW} but at distances of up to $\sim$3~kpc. These two are older (40~kyr) Galactic \acp{SNR} and tentatively classified as Type~Ia \ac{SN} explosions. However, none of these sources have a strong radio-continuum signature, which is different from \lmcorc\ and further challenges our understanding of the physical processes that would occur in such objects.

\subsection{The local environment: MW \& LMC alternatives }
 \label{sec:environment}

If \lmcorc\ is associated with the \ac{LMC}, then, with a nominal distance of 50~kpc, it would be 2.5--3~kpc away from any potential progenitor formation site. Similarly, we argued in Section~\ref{sec:evstar} that the most plausible distance from the Galactic Plane is 5~kpc. We have shown above that this is plausible in both the hypervelocity and old, outer disk scenarios. The \ac{LMC} 50~kpc distance is well-constrained, since even a distance of 45 or 55~kpc, e.g., would place \lmcorc\ at an implausibly large 5-6~kpc away from the \ac{LMC}. Therefore, in the absence of good constraining data, we adopt a distance of 50~kpc for \lmcorc\ assuming it is related to the \ac{LMC}.

The only direct information on the local environment comes from \HI\ observations, but only when averaged over large scales which may or not reflect the immediate environment of the presumed \ac{SNR}.  The integrated intensity of \HI\ in the vicinity of \lmcorc\ is $\sim$25~K~\kms\ (over the velocity range $\upsilon\sim$190~\kms\ to $\upsilon\sim$320~\kms, corresponding to the \ac{LMC} component; see inset, Fig.~\ref{fig:9}). This corresponds to a column density $N_{\rm H}$ of $\sim$4.5~$\times10^{19}$~cm$^{-2}$. At the location of the \lmcorc, this cloud is $\sim$1~degree wide, which would translate to $\sim$0.9~kpc at the assumed \ac{LMC} distance of 50~kpc. If we assume that the \HI\ is distributed evenly along a 0.9~kpc line of sight we can get an approximation of the average ambient density of the \ac{SNR}. \citet{1990ARA&A..28..215D} determined that \HI\ column densities derived assuming optically thin emission should be multiplied by 1.1 to 1.3 to account for \HI\ self-absorption. Using 1.2 as a medium value we get an integrated \HI\ column density of 5.4$\times10^{19}$~cm$^{-2}$. Another factor of 1.1 gives an atomic column density of 6$\times10^{19}$~cm$^{-2}$, taking the Helium into account. With the line of sight of 0.9~kpc, we find an average density of 0.017~cm$^{-3}$.

\subsection{Properties of \lmcorc\ as an \ac{SNR} candidate}
 \label{SNRproperties}
In this section, we assume that there is a plausible \ac{SN} progenitor, as discussed above. We then provide a comprehensive examination of the observed radio properties and X-ray limits of \lmcorc\ to evaluate the plausibility that it is an \ac{SNR}.

\subsubsection{Size, structure and radio spectrum}
 \label{sec:ssrs}
The circular appearance of \lmcorc\ and the prominence of its radio emission compared to other wavelengths are consistent with the \ac{SNR} hypothesis. With the above determined potential distances, \lmcorc\ would have a radius of about $R_{\rm LMC}=23.8$~pc if it is related to the \ac{LMC} and a maximum of $R_{\rm MW}=2.4$~pc if related to the \ac{MW}. Both of these are plausible. In addition, \lmcorc's bilateral symmetry along the NW--SE axis (Fig.~\ref{fig:polar}; top left)\footnote{A shell-type \ac{SNR} with two lobes of emission, separated by a symmetry axis \citep{2017A&A...597A.121W}; also called `barrel shape'.}  makes it similar to nearly 45~per~cent of shell-type Galactic \acp{SNR} \citep{2016A&A...587A.148W}.

We note that the \lmcorc\ extreme circularity (see Section~\ref{sec:morph}) is unusual for a \ac{SNR}; we can find only a handful of similar examples, such as the young (under 2000~yr old) \ac{MC} \acp{SNR} such as SN1987A \citep{2018ApJ...867...65C}, MC~SNR~J0509--6731 \citep{2014MNRAS.440.3220B,2018MNRAS.479.1800R} and N\,103B \citep{2019Ap&SS.364..204A}. This high level of circularity in \lmcorc\ might be the result of an explosion in a uniform and low density environment.

The spectral index of $\alpha=-0.4\pm0.1$ (see Section~\ref{sec:fluxspectra}) is somewhat flatter than that of the average  shell type \ac{SNR} but within the observed range of --0.55$\pm$0.20.  \citep{2012SSRv..166..231R, 2014SerAJ.189...15G, 2017ApJS..230....2B,2019A&A...631A.127M,book2}. For example, we note another relatively young shell-type \ac{SNR} at high latitude is the \ac{SNR} G182.4+4.3 discovered by \citet{1998A&A...331..661K} with a similar radio spectral index of $\alpha = -0.42$. 

\acp{SNR} are also characterised by their position in the brightness--to--diameter ($\Sigma$--D) parameter space. With $S_{\rm 1\,GHz}$=11.3~mJy, assuming that the \ac{ASKAP} observations at 888~MHz include all required angular scales and $\alpha = -0.4$, we estimate a radio surface brightness at 1~GHz of $\Sigma=1.6\times10^{-22}$~W~m$^{-2}$~Hz$^{-1}$~sr$^{-1}$ (assuming the emission to be spread smoothly over a circle of 196~arcsec diameter). This value would place \lmcorc\ in the bottom area of the \ac{SNR} $\Sigma$--D diagram \citep[][their fig.~3]{2020NatAs...4..910U,2018ApJ...852...84P}. If it is located in the \ac{MW}, \lmcorc\ would be positioned in the lower left corner of the $\Sigma$--D diagram in \citet[][their fig.~3]{2018ApJ...852...84P} and \citet{2020NatAs...4..910U}, indicating free expansion and an unrealistic low ambient density. For a \ac{LMC} location it would be positioned in the lower right corner, indicating a Sedov phase \ac{SNR} with still very low ambient density. While the low radio surface brightness clearly points to an \ac{LMC} location we note that the transitional phase may last even longer than the well defined free expansion or Sedov phases \citep{2021MNRAS.505..755P}. Finally, as pointed by \citet{2018ApJ...852...84P} the radio brightness in free expansion phase is expected to rise with time as we can also observe in the youngest Galactic \ac{SNR} G1.9+0.3 \citep{2020MNRAS.492.2606L}. But if \ac{SNR} is in transition or Sedov phase its brightness first plateau and then starts decreasing. This might be a good future indicator for the \lmcorc\ evolutionary stage where the next generation of radio observations might help solving this question. 

We looked for further clues to the nature of \lmcorc\ from its bilateral appearance (Fig.~\ref{fig:polar}). The bilateral morphology of \lmcorc\ resembles a number of well-known high-latitude \ac{MW} \acp{SNR}, such as SN\,1006 \citep{2013AJ....145..104R}, G296.5+10.0 \citep{2010ApJ...712.1157H}, G119.5+10.2 \citep{2011A&A...535A..64S} and G93.3+6.9 \citep{2006A&A...457.1081K} as well as young ones like G1.9+0.3 \citep{2014SerAJ.189...41D,2020MNRAS.492.2606L}. A number of studies have looked at the difference angle between the symmetry axis, e.g., and the position angle of the plane of the \ac{MW}. At low latitudes, the symmetry axis is broadly parallel to the \ac{MW} plane \citep{1998ApJ...493..781G, 2016A&A...587A.148W}, with exceptions occurring at high Galactic latitudes, where they  tend to be perpendicular \citep{2016A&A...587A.148W}. \lmcorc's symmetry axis is at position angle of 60(240)~degrees (measured in equatorial coordinates), while lines of constant Galactic latitude have position angles of --4(176)~degrees, so no special relationship is seen. On the other hand, the orientation of the \ac{LMC} bar is 66(246)~degrees which is quite close to the \lmcorc\ value. 

Large-scale magnetic fields in galaxies have geometries that typically follow their spiral arm pattern, and the \ac{LMC} has a geometry that is closer to face-on than edge-on, with an inclination angle of $\sim$35~degrees \citep{2004ApJ...601..260N}. We expect that the large-scale magnetic field of the \ac{LMC} would be weak at the distance of \lmcorc; given this, and the expectation that the magnetic field would follow the spiral pattern, there is no particular reason to expect a relationship between the symmetry axis of \lmcorc\ and the orientation of the \ac{LMC} bar \citep{1998A&AS..130..421F}. At the present time, there are no studies of the large-scale magnetic field of the \ac{LMC} that go out that far to verify whether there is a clearer relationship between the orientation of \lmcorc\ and the large-scale magnetic field of the \ac{LMC}. We also note that a number of \ac{MC} \acp{SNR} --- 1E0102--72.3 (Alsaberi et al., in prep.), SXP\,1062 \citep{2012A&A...537L...1H}, and MC~SNR~J0509--6731 \citep{2014MNRAS.440.3220B,2018MNRAS.479.1800R} --- show bilateral symmetry, but without the clear connection with large-scale galaxy structure. 

To some extent, \lmcorc\ could be similar (on a smaller scale) to Galactic Radio Loops identified in the \ac{MW} \citep{2018Galax...6...56D,2021ApJ...923...58W} but at larger distances. One would expect a handful of such \acp{SNR} (or (super)bubbles) around the \ac{LMC}/\ac{SMC} and several around the \ac{MW} \citep{2021PASA...38....3N}.

\subsubsection{Polarisation of \lmcorc\ as \ac{SNR}}

As shown in Section~\ref{sec:polarisation1}, only an upper limit of 7--9~per~cent can be placed on the polarisation fraction of \lmcorc. Although a detection of polarisation on the ring of \lmcorc\ would help support an \ac{SNR} origin, our upper limits mentioned above are still very much consistent with the \ac{SNR} scenario. Most young \acp{SNR} typically are not strongly polarised because of the highly turbulent magnetic field \citep{1995ApJ...441..300A,2020MNRAS.492.2606L,2014MNRAS.440.3220B}. For example, the \ac{MW} \ac{SNR} Cassiopeia~A at 5~GHz shows a mean degree of polarisation in the diffuse plateau region of $\sim$8--10~per~cent while the ring fractional polarisation drops to $\sim$5~per~cent \citep{1995ApJ...441..300A}. Another striking example is the historical SNR G11.2--0.3 from A.D.~386. This \ac{SNR} has only about 2~per~cent integrated polarisation at 32~GHz, but shows both radial and tangential magnetic fields in its shell, indicating it is transitioning between the free expansion phase and the Sedov-Taylor phase \citep{2001A&A...372..627K}. The low polarisation is also true for old \acp{SNR} in the radiative phase where the thermal electron population causes depolarisation and thus there is very little polarised emission, at lower frequencies \citep{10.1088/2514-3433/ac087e}. Galactic \ac{SNR}~S147 is a good example of this \citep{2008A&A...482..783X}. These older remnants, however, are not a good comparison for \lmcorc, because for these very depolarised cases (like S147) we would also see an optical counterpart, which here we do not detect. And in this case, one would expect to see higher polarised fractions at high frequencies. 

At the other extreme, \citet{2013AJ....145..104R} found in the case of the young \ac{SNR}~SN1006 or G327.6+14.6 (radius$\sim$10~pc and expanding in a denser environment ($\sim$0.05~cm$^{-3}$; \citep{2007A&A...475..883A}) than the \ac{ISM} around \lmcorc), a mean fractional polarisation of 17~per~cent. The recently discovered \ac{MW} SNR~G181.1+9.5 serves as a nice example of what can happen in a low density environment, as it shows 70 to 80~per~cent polarisation at 5~GHz, and is in the late Sedov or early radiative phase \citep{2017A&A...597A.116K}.

From the lack of polarisation and lack of depolarising optical emission, we suggest that \lmcorc\ is most likely a young \ac{SNR} that is still expanding freely or at least in the early Sedov phase, in a low density environment.

\subsection{X-ray and local environment constraints}
 \label{sec:xray+}

\lmcorc, as an \ac{SNR} at the distance of the \ac{LMC} (50~kpc), would have an X-ray luminosity upper limit of $\sim$5~$\times10^{33}$~erg~s$^{-1}$ for a temperature of 1~keV and $\sim$9~$\times10^{33}$~erg~s$^{-1}$ for a temperature of 2~keV. The lowest luminosity of \acp{SNR} in the \ac{LMC} (from \citet{2016A&A...585A.162M,2019A&A...631A.127M}, \citet{2012A&A...537L...1H} and \cite{2021MNRAS.504..326M}) are $\sim$7~$\times10^{33}$~erg~s$^{-1}$ so \lmcorc\ would be among the faintest \ac{LMC} \acp{SNR}, assuming it is indeed an \ac{SNR} and at the distance of the \ac{LMC}. In addition, these upper limits to the X-ray luminosity are significantly lower than ordinary \acp{SNR} of similar radius in the Sedov phase, evolving into densities $\sim$0.1--1~cm$^{-3}$, typical of the \ac{LMC}, which have X-ray luminosities of 10$^{34}$ to 10$^{36}$~erg~s$^{-1}$ \citep{2016A&A...585A.162M}. 

Given its large degree of circularity, the local density at the location of \lmcorc\ at $\sim$3~kpc from the \ac{LMC} and $\sim$50~kpc (Section~\ref{sec:environment}) from the Sun is expected to be as low as $10^{-1} - 10^{-4}$~cm$^{-3}$ \citep{2019PASA...36...45H,2019PASA...36...48H}. As for the \ac{MW} origin, we expect a similar range of environmental densities.

There are 6 \acp{SNR} in \citet{2014SerAJ.189...25P} that have radio surface brightness similar to or lower than the estimated surface brightness of \lmcorc\ (1.6$\times10^{-22}$~W~m$^{-2}$~Hz$^{-1}$~sr$^{-1}$ and similarly large diameters. Those are: G65.1+0.6 (1.8/176), G96.0+2.0 (0.67/30.3), G114.3+0.3 (1.7/14.3), G152.4--2.1 (0.56/31.1), G156.2+5.7 (0.62/32), G190.9--2.2 (0.4/18.8) where surface brightness/diameter in parentheses are given in units of $10^{-22}$~W~ m$^{-2}$~Hz$^{-1}$~sr$^{-1}$ and pc, respectively. The ambient densities are only known for those with X-ray spectra \citep{2020ApJS..248...16L}, which includes only G156.2+5.7, which has \ac{ISM} density of 0.20~cm$^{-3}$. Thus G156.2+5.7 has comparable radio properties to \lmcorc, but is much brighter in X-rays. For G156.2+5.7 to have similar X-ray luminosity to the upper limit of \lmcorc, it would have to have a factor of $\sim$30 lower \ac{ISM} density, i.e. 0.007~cm$^{-3}$.

We do not expect to find strong \ac{SNR} shock tracers in such a rarefied intergalactic environment as in the future optical (\SII\ and/or \OIII\ emission lines) or in deeper X-ray observations. 
The intensity of the thermal X-ray emission of a hot gas scales as the square of the environmental density. An expansion in a very rarefied environment would be faster. Therefore, the reason for faintness in X-rays is the low density in the shock. And similarly, we do not necessarily expect to detect optical emission because of the low \ac{ISM} density. As shown in \citet{2021MNRAS.500.2336Y}, optical emission from \acp{SNR} generally comes from high density clouds and filaments overrun by the \ac{SN} shock. 

Finally, an \ac{SNR} originating from a type~Ia explosion would be quite circular despite a peculiar (with respect to the local \ac{ISM}) progenitor velocity of $\upsilon\sim$10--100~km~s$^{-1}$; the small ratio of peculiar velocity to the shock velocity, which is expected to be in the 1000's of \kms\ and the unperturbed \ac{ISM} ensures this. In contrast, any remnant following from a CCSN of a star moving with $\upsilon\sim$70--350~km~s$^{-1}$ would expand into a strongly elongated stellar wind-bubble created by the progenitor's wind. The resulting \ac{SNR} would be highly non-spherical as shown for instance by \cite{2015MNRAS.450.3080M} and \cite{2021MNRAS.502.5340M}. We note that the stellar progenitor type is unknown, however, any massive star undergoing significant mass-loss would create a sizeable bubble and give rise to a similar non-spherical structure. This renders a type~Ia progenitor scenario more likely than a CCSNR scenario.

\subsection{Evolutionary State}
 \label{sec:evolutionarystate}

The low percentage polarisation of the radio continuum emission, which indicates a lot of turbulence and therefore still strong interaction between the \ac{SN} ejecta and the swept up material, points to an \ac{SNR} that is either still expanding freely or at least in the early Sedov phase. In addition, we find a shell width that is about 10~per~cent of the SNR's radius, pointing to adiabatic expansion and therefore again to the Sedov phase. We therefore assume from now on that our potential \ac{SNR} \lmcorc\ is adiabatically expanding and thus in the early Sedov phase.

We apply the models of \citet{1995PhR...256..157M} to study the possible evolutionary state of \lmcorc\ at the two potential locations. As \lmcorc\ most likely is in the early Sedov phase, we use the time $t_{st}$, the age that marks the transition between free expansion and Sedov phase, as a reference time. According to \citet{1995PhR...256..157M} the well-known young \acp{SNR} Cas~A, Kepler, and Tycho would all likely be younger than $t_{st}$ and SN~1006 a little older. The radius $R_{st}$ at $t_{st}$ depends only on the ambient density and the ejecta mass $M_{ej}$ of the supernova explosion. We assume for a type~Ia explosion $M_{ej} = 1.4~$M$_{\odot}$ and for a core-collapse (CCSNe) explosion $M_{ej} = 10~$M$_\odot$. This would result in ambient densities $n_0$ of $\ge1.2$~cm$^{-3}$ (type~Ia) and $\ge 8.0$~cm$^{-3}$ (CCSNe) for a \ac{MW} location and $n_0 = 0.001$~cm$^{-3}$ (type~Ia) and $n_{0} = 0.008$~cm$^{-3}$ (CCSNe) for a \ac{LMC} location. The low radio surface brightness, the missing thermal X-ray emission, and the location far away from any star forming environment pretty much negate an environment as dense as required for the \ac{MW} location. 

Using the equation by \citet{1995PhR...256..157M} we determine the age $t_{st}$ of \lmcorc\ for the different locations and supernova explosions to be 215~yr (type~Ia) and 712~yr (CCSNe) for a \ac{MW} location and 2200~yr (type~Ia) and 7100~yr (CCSNe) for a \ac{LMC} location for an explosion energy of $1.5\times10^{51}$~erg for the type~Ia explosions \citep{1993A&A...270..223K} and the canonical explosion energy of $10^{51}$~erg for the CCSNe explosions. Further characteristics of these cases are listed in Table~\ref{tab45}.

\begin{table*}
 \centering
 \caption{\lmcorc\ as an \ac{SNR} at LMC and MW distances and with various \ac{ISM} densities. Calculation of characteristics are based on \citet{1995PhR...256..157M}. We show the distance $d$, radius $r$, ejecta mass $M_0$, explosion energy $E_0$, ambient density $n_0$, age $t$, velocity $v_b$ and temperature $T_b$ at the blast wave location, and the swept up mass $M_{sw}$. We used the average explosion energy of $1.5\times 10^{51}$~erg observed for extragalactic type Ia supernova explosions \citep{1993A&A...270..223K}. SNII CCSNe typically have explosion energies between $10^{50}$~erg and $10^{51}$~erg \citep{2015ApJ...806..225P}. We used those lower and upper limits for CCSNe together with a representative ejecta mass of 10~M$_\odot$.}
 \begin{tabular}{cccccccccccc}
\hline\hline
  (1)    & (2)   & (3)  & (4)     & (5)           & (6)             & (7)         & (8)   & (9)            & (10)     & (11)          & (12) \\
Location & $d$   &  $r$ & SN type & $M_0$         & $E_0$           & $n_0$       & $t$   & $v_b$          & $T_b$    & $M_{sw}$      & Comments \\
(MW/LMC) & (kpc) & (pc) &         & ($M_{\odot}$) & ($10^{51}$~erg) & (cm$^{-3}$) & (yr)  & (km\,s$^{-1}$) & $10^7$~K & ($M_{\odot}$) & \\
\hline\hline
  MW     & 5     & 2.4 & SNIa    & 1.4          & 1.5              & 1.2         &  215  & 6780           & 63.8     & 2.3           & $t = t_{st}$ ED/Sedov\\
  MW     & 5     & 2.4 & CCSNe   & 10           & 0.1              & 8.0         & 2250  &  655           & 0.6      & 16            & $t = t_{st}$ ED/Sedov \\
  MW     & 5     & 2.4 & CCSNe   & 10           & 1.0              & 8.0         &  712  & 2070           & 6.0      & 16            & $t = t_{st}$ ED/Sedov \\
\hline
  LMC    & 50    & 23.8 & SNIa    & 1.4          & 1.5              & 0.001       & 2200  & 6780           & 63.8     & 2.3           & $t = t_{st}$ ED/Sedov \\
  LMC    & 50    & 23.8 & CCSNe   & 10           & 0.1              & 0.008       & 22500 & 655            & 0.6      & 16            & $t = t_{st}$ ED/Sedov \\
  LMC    & 50    & 23.8 & CCSNe   & 10           & 1.0              & 0.008       & 7100  & 2070           & 6.0      & 16            & $t = t_{st}$ ED/Sedov \\
\hline
  LMC    & 50    & 23.8 & SNIa    & 1.4          & 1.5              & 0.007       & 4000  & 2600           & 9.3      & 12.4          & Sedov\\
  LMC    & 50    & 23.8 & SNIa    & 1.4          & 1.5              & 0.017       & 5500  & 1800           & 4.4      & 30            & Sedov\\
  LMC    & 50    & 23.8 & SNIa    & 1.4          & 0.5              & 0.007       & 6510  & 1600           & 3.6      & 12.4          & Sedov\\
  LMC    & 50    & 23.8 & SNIa    & 1.4          & 0.5              & 0.017       & 9550  & 1032           & 1.5      & 30            & Sedov\\
  LMC    & 50    & 23.8 & CCSNe   & 10           & 0.1              & 0.007       & 21300 & 712            & 0.7      & 12.4          & late ED\\
  LMC    & 50    & 23.8 & CCSNe   & 10           & 1.0              & 0.007       & 6740  & 2250           & 7.0      & 12.4          & late ED\\
  LMC    & 50    & 23.8 & CCSNe   & 10           & 0.1              & 0.017       & 26400 & 460            & 0.3      & 30            & early Sedov\\
  LMC    & 50    & 23.8 & CCSNe   & 10           & 1.0              & 0.017       & 8350  & 1450           & 2.9      & 30            & early Sedov\\
\hline\hline
\end{tabular}
\label{tab45}
\end{table*}

We used the equations for the hydrodynamic evolution of a young \ac{SNR} by \citet{1995PhR...256..157M} to estimate \ac{SNR} characteristics and list them in Table~\ref{tab45}. In addition to \acp{SNR} that are evolutionary at the transition between ED and Sedov phase, which we already discussed above for the \ac{MW} and the \ac{LMC} location, we added cases for type~Ia and CCSNe explosions expanding into constant ambient densities estimated above from the upper limit of the X-ray emission ($n_0 = 0.007$~cm$^{-3}$) and from the \HI\ observations ($n_0 = 0.017$~cm$^{-3}$). Also, we explore the `unified \ac{SNR} evolutionary models' but only for type~Ia \ac{SN} in Appendix~\ref{app:1}. All of the results we calculate for the \ac{LMC} location look quite reasonable for our observations. 

The low percentage polarisation rules out that the swept up material is dominating the expansion. In that case, we would expect quite homogeneous magnetic fields and therefore high percentage polarisation, at least at the higher frequencies. From this we can certainly rule out the somewhat higher ambient density we get from the \HI\ observations and assume an upper ambient density of about 0.005~cm$^{-3}$ for a type~Ia explosion. We can get another limit from the low radio surface brightness and the width of the shell which makes up about 10~per~cent of the source's radius. This rules out a very early evolutionary phase and gives us a lower limit of about 0.008~cm$^{-3}$ for a type~II CCSN. Therefore, based on this we are not able to definitively say if \lmcorc\ was a type~Ia or a CCSN explosion.

\section{Alternative scenarios for \lmcorc\ origin}
\label{D}

Given the unusual nature of the rogue supernova origin presented above, we also explore a variety of alternative explanations for the origin of \lmcorc.

\subsection{Odd Radio Circle}
    \label{orc}

Odd Radio Circles (\acp{ORC}) are circles of faint diffuse radio emission found in large-area surveys,  \citep{2021PASA...38....3N,2021MNRAS.505L..11K} which are undetected at other wavelengths. Three of the five known \acp{ORC} are single, and each of these has an elliptical galaxy at its centre at a redshift of $z\sim$ 0.2--0.5. Each of these host galaxies is located in an overdensity or is interacting with a companion \citep{2021Galax...9...83N}. These single \acp{ORC} probably represent a spherical shell of emission caused by a shock wave from the central host galaxy (Norris et al., 2022, in prep.). The remaining two \acp{ORC} form a pair, separated by less than one arcmin, which may be associated with a nearby double-lobed radio \ac{AGN} (Macgregor et al., 2022, in prep.).

In the \ac{EMU}~Pilot Survey \citep[PS; area of $\sim$270~sq.~deg.;][]{emups}, we discovered three \acp{ORC} by chance, giving a surface density of \acp{ORC} which is consistent with finding one in the 120~sq~deg \ac{ASKAP} image of the \ac{LMC} \citep{2021MNRAS.506.3540P}. However, \lmcorc\ differs from the known single \acp{ORC} in three respects: (i) the \acp{ORC} have a steeper spectral index than \lmcorc, (ii) \lmcorc's angular size of 196~arcsec is significantly larger than the typical 80~arcsec size of \acp{ORC}, (iii) all single \acp{ORC} have a central elliptical galaxy, whereas the nearest galaxy to the centre of \lmcorc\ is the lenticular galaxy 10.8~arcsec from the centre, discussed in Section~\ref{sec:fluxspectra}.

Because of these differences, we consider it unlikely (but not impossible) that \lmcorc\ is an \ac{ORC}. Below, we consider several possibilities for its \ac{ORC}-like formation.

\subsubsection{Precessing jets}

\citet{2020MNRAS.499.5765H} have shown that precessing jets in an \ac{AGN} can display a number of morphologies. In particular, if the jets are seen end-on, they might be observed as one or two circular discs of emission. These might form rings if the precession timescale is significantly shorter than the source age and the radiative timescale of the hotspot. However, such exact rings remain hypothetical, as they have not yet been observationally confirmed. 

Although, an interesting example of possible precessing jets is seen in the nearby (350~Mpc) ultraluminous infrared galaxy (ULIRG) WISEA~J060253.98-710310.0 \citep{2019PASJ...71...26D}, which is a merging pair of spiral galaxies (Filipovi\'c~et~al., in prep.). These authors show that merging spiral galaxies \ac{SMBH} are most likely responsible for producing spirals of circular (precessing) jets spanning over 600~kpc in length. In this case, one of the two \ac{SMBH}s is `disturbing' the jet emission on a regular basis, and switching it off at the same time, creating rings and jet spirals. If observed directly along the line of sight, one would expect to see nearly-circular rings. We note a number of other examples that would fit into the category of `\ac{AGN} sprinklers' were found by visual inspection of various deep radio survey images, e.g. the \ac{QSO} candidate DES~J061909.04--555843.7 in the \ac{ASKAP} image of the merging galaxy cluster system Abell~3391~--~Abell~3395 \citep{2021A&A...647A...3B}, the \ac{QSO} SDSS~J134545.36+533252.3 as seen in the \ac{VLASS} Sky Survey \citep{2020PASP..132c5001L}, LEDA~783409 (a.k.a. DES~J004506.98--250146.8 and WISEA~J004506.98--250147.0) in a deep \ac{ASKAP} image of the NGC~253 region (Koribalski~et~al.,~in~prep.) and the BL\,Lac object 2MASX~J08120189+0237325 \citep{2020MNRAS.497...94P} as seen in \ac{VLASS} \citep{2020PASP..132c5001L}.

If \lmcorc\ is caused by a precessing \ac{AGN}, it should have a radio-loud \ac{AGN} in the centre. The nearest candidate is the lenticular galaxy discussed above, 10.8~arcsec from the centre (Section~\ref{sec:fluxspectra}). The lack of a bright central radio source might be explained if the jet were currently switched off, although that requires that it has only just switched off, but was previously switched on for at least a precession period to illuminate the ring. However, lenticular galaxies do not often host radio \ac{AGN}. Furthermore, the galaxy appears edge-on, so is unlikely to be hosting an end-on radio \ac{AGN}. Furthermore, the redshift of this galaxy is $z\gtrsim0.7$ (Section~\ref{sec:fluxspectra}), implying a diameter of the ring of $\gtrsim$~1.4~Mpc. If the precessing jets subtend an angle of 10~degrees at the \ac{AGN}, then this implies a distance between the front and back hotspots of $\gtrsim$~8~Mpc, making this one of the largest radio \acp{AGN} ever observed.

We therefore consider it unlikely that \lmcorc\ is caused by precessing jets from an \ac{AGN}.

\subsubsection{Buoyant toroidal structures}

Dynamical evolution of buoyantly rising remnant lobes can produce toroidal structures \citep{2001ApJ...554..261C}, which will be seen as rings when viewed end-on. However, the timescale to evacuate a central cavity is comparable to the sound crossing time, which is of order 200~Myr if the central galaxy is at $z\gtrsim0.7$. This is much longer than the typical remnant fading timescale \citep{2020MNRAS.496.1706S}. While re-acceleration of remnant plasma is possible, the relatively flat radio spectral index of the \lmcorc\ would imply a very strong shock. Therefore, we consider this interpretation less plausible.

\subsubsection{Bent-tailed galaxy}

Another possibility for an \ac{AGN} origin is that it might be caused by a bent-tailed galaxy. For example, the bent-jets from the \ac{NAT} galaxy NGC\,1265 \citep{1986ApJ...301..841O} initially form a semi-circle, but then broaden and merge downstream, creating the large tails characteristic of a narrow-angle tailed galaxy. However, this model suffers from two problems: (a) no tailed galaxy has been observed to form a full circle, (b) no bright radio \ac{AGN} host is found along the \lmcorc\ ring. We examined all compact sources along the \lmcorc\ ring feature and found no evidence that they could be associated with a host galaxy that would be the source of a twin bent-jet structure. There is a bright X-ray source XMMU~J062426.76--694726.3 (Fig.~\ref{fig:5new}) on the north-east rim of the \lmcorc, which is presumably an \ac{AGN}, but although visible at infrared and optical wavelengths, it does not correspond to a radio source, and so appears to be radio-quiet.

\subsection{Nearby remnant of a stellar super-flare}
\label{super-flare}

What if \lmcorc\ is the result of a nearby stellar super-flare(s)? At the NW side of the \lmcorc\ ring (Fig.~\ref{fig:5new}), we found the star {\it Gaia}~EDR3~5278760380137682816 (also known as 2MASS~J06240216--6948116; RA(J2000) = 06$^\mathrm{h}$24$^\mathrm{m}$2.151$^\mathrm{s}$ and Dec(J2000) = --69$^{\circ}$48\arcmin11.64\arcsec). 

This object has a very red colour in \ac{SMASH} DR2 with $(g-i)_0 = 3.24$, which indicates that this star is a foreground M-dwarf \citep[Fig.~3 in][]{2020MNRAS.492..782W}. In addition, this star is much brighter at longer wavelengths. \ac{2MASS} photometry shows that the K band brightness is $11.9$~mag, which is much brighter than the \ac{SMASH} {\it i}-band brightness ($\sim$15.5~mag). From {\it Gaia} EDR3, this star has a proper motion in declination of $\sim$46.086$\pm$0.050~mas~yr$^{-1}$, plus a small proper motion in RA of $\sim$--0.142$\pm0.048$~mas~yr$^{-1}$; it has a $G$-band magnitude of 16.271, and a parallax of 17.078$\pm$0.039~mas, putting it at a distance of $58.5~\mathrm{pc}$. At this distance, the 196~arcsec circular shell could be a giant \ac{CME} that would have a linear diameter of $\sim$0.186~ly (0.057~pc), and with an assumed expansion velocity of $\upsilon$=50--500~km~s$^{-1}$ \citep{2019NatAs...3..742A,2021NatAs.tmp..246N} would give an age of $\sim$56--560~years. Our modelling suggests this could represent a mass ejection of $\sim$2$\times10^{22}~\mathrm{g}$ and an \ac{CME} energy of $\geq 2 \times 10^{36}~\mathrm{erg~M} / \mathrm{(}2 \times 10^{21}\mathrm{g)(v_{CME}/(}450\,\mathrm{km/s))}^{2}$. These estimates could be associated with super-flares from young rapidly-rotating stars \citep{2019ApJ...876...58N}. But would the radio continuum emission we see be expected from such a \ac{CME}? Presently, there is no example of a radio ring around a flare star (or any modelling that suggests there should be).

However, flare stars are known \mbox{X-ray} and radio emitters \citep{2018ApJ...856...39C}. For example, a stellar flare on HR~9024 \citep{2019NatAs...3..742A} has been observed using the High Energy Transmission Grating Spectrometer on {\it Chandra}. Doppler shifts of hot plasmas (up to $25 \times 10^6~\mathrm{K}$) from $\mathrm{S}_{\mathrm{xvi}}$, $\mathrm{S}_{\mathrm{xiv}}$ and $\mathrm{Mg}_{\mathrm{xii}}$ lines indicate motions with velocities of 50--500~km~s$^{-1}$. Additionally, a blueshift in the $\mathrm{O}_{\mathrm{viii}}$ line indicates an upward motion of relatively cool plasma with a velocity of $\upsilon=90\pm30~\mathrm{km~s^{-1}}$. This is thought to be a \ac{CME} with a mass of $\sim$10$^{21}\,\mathrm{g}$ and $5\times10^{34}\,\mathrm{ergs}$ as detailed in \citet{2019NatAs...3..742A}. This represents a major loss of stellar mass but still an order of magnitude smaller than our above estimates.

The fact that the {\it Gaia} star is clearly detected in the {\it XMM-Newton} image as a point source (see Table~\ref{tab3}) and its position on the rim of \lmcorc\ (see Fig.~\ref{fig:5new}) as well as its large proper motion makes this scenario feasible to explore. This may represent quasi-simultaneous flare/burst eruptions (temporally close together). There are several known coherent radio bursts from M-dwarfs \citep{2019ApJ...871..214V}. As a radio object, M-dwarfs could have a diverse morphology and the duration of these flare/bursts could range from seconds to hours. They all share strong (40~–-100~per~cent) circular polarisation \citep{2019ApJ...871..214V} which we do not see in any of our radio images nor is there a distinctive radio point source at the {\it Gaia} and/or {\it XMM-Newton} star position. \citet{2019ApJ...871..214V} also showed that no such events resemble solar Type~II bursts (often associated with \ac{CME}s), but they also could not rule out the occurrence of radio-quiet stellar \ac{CME}s. 

If this is indeed a nearby remnant of a stellar super-flare (or burst) then one would expect to see detectable proper motions due to the expansion of the ring. We estimate that over a 2-year period, assuming an expansion velocity of $\upsilon$=50--500~\kms, the distance of 58.5~pc and a shell diameter size of 0.057~pc the shell would expand between 0.4~arcsec to 5.6~arcsec which is well within the capabilities of the present generation of radio telescopes. However, it would be strange for the ring not to be concentric around that star as conservation of momentum would require it to move in the same way unless it moves through a dense medium.
Finally, if the circular shell material seen in radio is from a \ac{CME}, which has high density, we would expect to detect it in various optical lines.

\subsection{Planetary Nebula}

We rule out a \ac{PN} scenario as an explanation for \lmcorc. If it is at the \ac{LMC} distance, it would be an order of magnitude larger than any known \ac{PN} \citep{2009MNRAS.399..769F,2011MNRAS.412..223B,2016Ap&SS.361..108L,2017MNRAS.468.1794L}. If of Galactic origin, then prominent IR and optical emission would be seen in any (even shallow) image, which we do not see.

\section{What is \lmcorc?}

Multi-frequency radio analysis reveals an almost perfectly circular ring radio source \lmcorc\ in an unexpected location, in the direction between the \ac{LMC} and the plane of the \ac{MW}. At the distance of the \ac{LMC}, the source size of 47.2$\pm$1.0~pc is consistent with an intergalactic \ac{SNR} interpretation. Various other intrinsic properties including the spectral index ($\alpha=-0.4\pm0.1$) and morphology are also consistent with an identification as a bilateral \ac{SNR}. However, the lack of detectable radio polarisation is not consistent with such a scenario but also not totally unexpected. The progenitor of \ac{SNR} \lmcorc\ is more likely to be from the outer disk of the \ac{LMC}, but we cannot rule out a \ac{HVS} as an origin. At the same time, we favour a type~Ia \ac{SN} over a CCSN because of the location and stellar content of the area around the \lmcorc, as well as a very low (uniform) ambient density and the lack of morphological features expected for a CCSN of a fast-moving star. If \lmcorc\ is an \ac{SNR}, its most likely distance is 50~kpc and with the upper limit for its X-ray flux we obtain $n_{H}\lesssim$0.008~cm$^{-3}$, thus the \ac{SNR} would be younger than $\sim$7100~yr, implying it is in a late ED or early Sedov phase (Table~\ref{tab45}). This would make \lmcorc\ an ideal laboratory to study this evolutionary stage of \acp{SNR} in a low-density intergalactic environment.

While \lmcorc\ has the morphology of an \ac{ORC}, it is of larger angular size, has a flatter spectrum, and does not have a prominent central elliptical galaxy. 

Another scenario for \lmcorc\ would include a precessing \ac{AGN} jet, but the near-perfect symmetry with a complete ring structure, and the lack of a bright radio/optical central source, argue against it, although this could be a result of the \ac{AGN} being in a `switched-off' mode. 

A weak central radio source located 10.8~arcsec from the geometric centre is unlikely to be a central \ac{AGN} engine and even less likely a \ac{PWN}, as it coincides with a faint edge-on late-type or lenticular (S0) galaxy seen on \ac{SMASH} optical images. We also note four nearby bright radio sources towards the south-west edges of the \lmcorc\ ring -- all, most likely, unrelated background objects. 

We review other possible \ac{ORC} etiologies and discuss relevant models. These include its identity as a remnant of a stellar super-flare or ejected bubble shell from the nearby {\it Gaia}~EDR3~5278760380137682816 M-dwarf star, although an unfavourable direction of proper motion as well as an explanation for asymmetries to the parent star would be required. Finally, we rule out a \ac{PN} scenario.

In conclusion, our preferred interpretation for the nature of \lmcorc\ is an intergalactic \ac{SNR} that comes from a single-degenerate type~Ia \ac{SN} located in the far eastern outskirts of the \ac{LMC} which expands into a rarefied, intergalactic environment.


\section*{Acknowledgements}

The Australian SKA Pathfinder is part of the Australia Telescope National Facility which is managed by \ac{CSIRO}. Operation of \ac{ASKAP} is funded by the Australian Government with support from the National Collaborative Research Infrastructure Strategy. \ac{ASKAP} uses the resources of the Pawsey Supercomputing Centre. Establishment of \ac{ASKAP}, the Murchison Radio-astronomy Observatory and the Pawsey Supercomputing Centre are initiatives of the Australian Government, with support from the Government of Western Australia and the Science and Industry Endowment Fund. We acknowledge the Wajarri Yamatji people as the traditional owners of the Observatory site. 
The \ac{ATCA} is part of the Australia Telescope National Facility which is funded by the Australian Government for operation as a National Facility managed by \ac{CSIRO}.
This research has made use of the \ac{WISE}, operated by the Jet Propulsion Laboratory, California Institute of Technology, under contract with the National Aeronautics and Space Administration, and \ac{WISE} is also a joint project with the University of California, Los Angeles. Additionally, this research has made use of the \textsc{Simbad} database, operated at CDS, Strasbourg, France \citep{2000A&AS..143....9W}, \textsc{SAOImage DS9}, NASA's Astrophysics Data System Bibliographic Services, \textsc{VizieR} catalogue access tool, CDS, Strasbourg, France. The original description of the \textsc{VizieR} service was published in \cite{2000A&AS..143...23O}.

H.A. has benefited from grant CIIC 174/2021 of Universidad de Guanajuato, Mexico.
D.U. acknowledges Ministry of Education, Science and Technological Development of the Republic of Serbia through the contract No. 4451-03-9/2021-14/200104.
We thank B. Gaensler for constructive comments and suggestions. 
Robert Brose acknowledges funding from an Irish Research Council Starting Laureate Award (IRCLA/2017/83). 
We thank an anonymous referee for comments and suggestions that greatly improved our paper.


\section*{Data Availability}

The data that support the plots/images within this paper and other findings of this study are available from the corresponding author upon reasonable request. The \ac{ASKAP} data used in this article are available through the CSIRO ASKAP Science Data Archive (CASDA) and \ac{ATCA} data via the \ac{ATOA}.



\bibliographystyle{mnras}
\bibliography{orc_lmc} 

\begin{thebibliography}{}
\makeatletter
\relax
\def\mn@urlcharsother{\let\do\@makeother \do\$\do\&\do\#\do\^\do\_\do\%\do\~}
\def\mn@doi{\begingroup\mn@urlcharsother \@ifnextchar [ {\mn@doi@}
  {\mn@doi@[]}}
\def\mn@doi@[#1]#2{\def\@tempa{#1}\ifx\@tempa\@empty \href
  {http://dx.doi.org/#2} {doi:#2}\else \href {http://dx.doi.org/#2} {#1}\fi
  \endgroup}
\def\mn@eprint#1#2{\mn@eprint@#1:#2::\@nil}
\def\mn@eprint@arXiv#1{\href {http://arxiv.org/abs/#1} {{\tt arXiv:#1}}}
\def\mn@eprint@dblp#1{\href {http://dblp.uni-trier.de/rec/bibtex/#1.xml}
  {dblp:#1}}
\def\mn@eprint@#1:#2:#3:#4\@nil{\def\@tempa {#1}\def\@tempb {#2}\def\@tempc
  {#3}\ifx \@tempc \@empty \let \@tempc \@tempb \let \@tempb \@tempa \fi \ifx
  \@tempb \@empty \def\@tempb {arXiv}\fi \@ifundefined
  {mn@eprint@\@tempb}{\@tempb:\@tempc}{\expandafter \expandafter \csname
  mn@eprint@\@tempb\endcsname \expandafter{\@tempc}}}

\bibitem[\protect\citeauthoryear{{Acero}, {Ballet}  \& {Decourchelle}}{{Acero}
  et~al.}{2007}]{2007A&A...475..883A}
{Acero} F.,  {Ballet} J.,   {Decourchelle} A.,  2007, \mn@doi [Astron. \&
  Astrophys.] {10.1051/0004-6361:20077742}, \href
  {https://ui.adsabs.harvard.edu/abs/2007A&A...475..883A} {475, 883}

\bibitem[\protect\citeauthoryear{{Ajello} et~al.,}{{Ajello}
  et~al.}{2020}]{2020ApJ...892..105A}
{Ajello} M.,  et~al., 2020, \mn@doi [\apj] {10.3847/1538-4357/ab791e}, \href
  {https://ui.adsabs.harvard.edu/abs/2020ApJ...892..105A} {892, 105}

\bibitem[\protect\citeauthoryear{{Alsaberi} et~al.,}{{Alsaberi}
  et~al.}{2019}]{2019Ap&SS.364..204A}
{Alsaberi} R.~Z.~E.,  et~al., 2019, \mn@doi [\apss]
  {10.1007/s10509-019-3696-8}, \href
  {https://ui.adsabs.harvard.edu/abs/2019Ap&SS.364..204A} {364, 204}

\bibitem[\protect\citeauthoryear{{Anderson}, {Keohane}  \&
  {Rudnick}}{{Anderson} et~al.}{1995}]{1995ApJ...441..300A}
{Anderson} M.~C.,  {Keohane} J.~W.,   {Rudnick} L.,  1995, \mn@doi [\apj]
  {10.1086/175356}, \href
  {https://ui.adsabs.harvard.edu/abs/1995ApJ...441..300A} {441, 300}

\bibitem[\protect\citeauthoryear{{Argiroffi} et~al.,}{{Argiroffi}
  et~al.}{2019}]{2019NatAs...3..742A}
{Argiroffi} C.,  et~al., 2019, \mn@doi [Nature Astronomy]
  {10.1038/s41550-019-0781-4}, \href
  {https://ui.adsabs.harvard.edu/abs/2019NatAs...3..742A} {3, 742}

\bibitem[\protect\citeauthoryear{{Arnaud}}{{Arnaud}}{1996}]{Arnaud1996}
{Arnaud} K.~A.,  1996, in {Jacoby} G.~H.,  {Barnes} J.,  eds,  Astronomical
  Society of the Pacific Conference Series Vol. 101, Astronomical Data Analysis
  Software and Systems V. p.~17

\bibitem[\protect\citeauthoryear{{Balucinska-Church} \&
  {McCammon}}{{Balucinska-Church} \& {McCammon}}{1992}]{Bal1992}
{Balucinska-Church} M.,  {McCammon} D.,  1992, \mn@doi [\apj] {10.1086/172032},
  \href {http://cdsads.u-strasbg.fr/abs/1992ApJ...400..699B} {400, 699}

\bibitem[\protect\citeauthoryear{{Becker}, {Hurley-Walker}, {Weinberger},
  {Nicastro}, {Mayer}, {Merloni}  \& {Sanders}}{{Becker}
  et~al.}{2021}]{2021A&A...648A..30B}
{Becker} W.,  {Hurley-Walker} N.,  {Weinberger} C.,  {Nicastro} L.,  {Mayer}
  M.~G.~F.,  {Merloni} A.,   {Sanders} J.,  2021, \mn@doi [Astron. \&
  Astrophys.] {10.1051/0004-6361/202040156}, \href
  {https://ui.adsabs.harvard.edu/abs/2021A&A...648A..30B} {648, A30}

\bibitem[\protect\citeauthoryear{{Besla}, {Mart{\'\i}nez-Delgado}, {van der
  Marel}, {Beletsky}, {Seibert}, {Schlafly}, {Grebel}  \& {Neyer}}{{Besla}
  et~al.}{2016}]{2016ApJ...825...20B}
{Besla} G.,  {Mart{\'\i}nez-Delgado} D.,  {van der Marel} R.~P.,  {Beletsky}
  Y.,  {Seibert} M.,  {Schlafly} E.~F.,  {Grebel} E.~K.,   {Neyer} F.,  2016,
  \mn@doi [Astrophys. J.] {10.3847/0004-637X/825/1/20}, \href
  {https://ui.adsabs.harvard.edu/abs/2016ApJ...825...20B} {825, 20}

\bibitem[\protect\citeauthoryear{{Boji{\v{c}}i{\'c}}, {Parker}, {Filipovi{\'c}}
   \& {Frew}}{{Boji{\v{c}}i{\'c}} et~al.}{2011}]{2011MNRAS.412..223B}
{Boji{\v{c}}i{\'c}} I.~S.,  {Parker} Q.~A.,  {Filipovi{\'c}} M.~D.,   {Frew}
  D.~J.,  2011, \mn@doi [\mnras] {10.1111/j.1365-2966.2010.17900.x}, \href
  {https://ui.adsabs.harvard.edu/abs/2011MNRAS.412..223B} {412, 223}

\bibitem[\protect\citeauthoryear{{Boubert}, {Erkal}, {Evans}  \&
  {Izzard}}{{Boubert} et~al.}{2017}]{2017MNRAS.469.2151B}
{Boubert} D.,  {Erkal} D.,  {Evans} N.~W.,   {Izzard} R.~G.,  2017, \mn@doi
  [Mon. Not. R. Astron. Soc.] {10.1093/mnras/stx848}, \href
  {https://ui.adsabs.harvard.edu/abs/2017MNRAS.469.2151B} {469, 2151}

\bibitem[\protect\citeauthoryear{{Boubert}, {Erkal}  \& {Gualandris}}{{Boubert}
  et~al.}{2020}]{2020MNRAS.497.2930B}
{Boubert} D.,  {Erkal} D.,   {Gualandris} A.,  2020, \mn@doi [Mon. Not. R.
  Astron. Soc.] {10.1093/mnras/staa2211}, \href
  {https://ui.adsabs.harvard.edu/abs/2020MNRAS.497.2930B} {497, 2930}

\bibitem[\protect\citeauthoryear{{Bozzetto}, {Filipovi{\'c}},
  {Uro{\v{s}}evi{\'c}}, {Kothes}  \& {Crawford}}{{Bozzetto}
  et~al.}{2014}]{2014MNRAS.440.3220B}
{Bozzetto} L.~M.,  {Filipovi{\'c}} M.~D.,  {Uro{\v{s}}evi{\'c}} D.,  {Kothes}
  R.,   {Crawford} E.~J.,  2014, \mn@doi [\mnras] {10.1093/mnras/stu499}, \href
  {https://ui.adsabs.harvard.edu/abs/2014MNRAS.440.3220B} {440, 3220}

\bibitem[\protect\citeauthoryear{{Bozzetto} et~al.,}{{Bozzetto}
  et~al.}{2017}]{2017ApJS..230....2B}
{Bozzetto} L.~M.,  et~al., 2017, \mn@doi [Astrophys. J. Suppl.]
  {10.3847/1538-4365/aa653c}, \href
  {https://ui.adsabs.harvard.edu/abs/2017ApJS..230....2B} {230, 2}

\bibitem[\protect\citeauthoryear{{Brantseg}, {McEntaffer}, {Bozzetto},
  {Filipovic}  \& {Grieves}}{{Brantseg} et~al.}{2014}]{2014ApJ...780...50B}
{Brantseg} T.,  {McEntaffer} R.~L.,  {Bozzetto} L.~M.,  {Filipovic} M.,
  {Grieves} N.,  2014, \mn@doi [Astrophys. J.] {10.1088/0004-637X/780/1/50},
  \href {https://ui.adsabs.harvard.edu/abs/2014ApJ...780...50B} {780, 50}

\bibitem[\protect\citeauthoryear{{Brown}}{{Brown}}{2015}]{2015ARA&A..53...15B}
{Brown} W.~R.,  2015, \mn@doi [Ann. Rev. Astron. Astrophys.]
  {10.1146/annurev-astro-082214-122230}, \href
  {https://ui.adsabs.harvard.edu/abs/2015ARA&A..53...15B} {53, 15}

\bibitem[\protect\citeauthoryear{{Br{\"u}ggen} et~al.,}{{Br{\"u}ggen}
  et~al.}{2021}]{2021A&A...647A...3B}
{Br{\"u}ggen} M.,  et~al., 2021, \mn@doi [Astron. \& Astrophys.]
  {10.1051/0004-6361/202039533}, \href
  {https://ui.adsabs.harvard.edu/abs/2021A&A...647A...3B} {647, A3}

\bibitem[\protect\citeauthoryear{{Cendes}, {Gaensler}, {Ng}, {Zanardo},
  {Staveley-Smith}  \& {Tzioumis}}{{Cendes} et~al.}{2018}]{2018ApJ...867...65C}
{Cendes} Y.,  {Gaensler} B.~M.,  {Ng} C.~Y.,  {Zanardo} G.,  {Staveley-Smith}
  L.,   {Tzioumis} A.~K.,  2018, \mn@doi [\apj] {10.3847/1538-4357/aae261},
  \href {https://ui.adsabs.harvard.edu/abs/2018ApJ...867...65C} {867, 65}

\bibitem[\protect\citeauthoryear{{Choi}, {Dotter}, {Conroy}, {Cantiello},
  {Paxton}  \& {Johnson}}{{Choi} et~al.}{2016}]{2016ApJ...823..102C}
{Choi} J.,  {Dotter} A.,  {Conroy} C.,  {Cantiello} M.,  {Paxton} B.,
  {Johnson} B.~D.,  2016, \mn@doi [Astrophys. J.]
  {10.3847/0004-637X/823/2/102}, \href
  {https://ui.adsabs.harvard.edu/abs/2016ApJ...823..102C} {823, 102}

\bibitem[\protect\citeauthoryear{{Chomiuk} et~al.,}{{Chomiuk}
  et~al.}{2018}]{2018ATel11610....1C}
{Chomiuk} L.,  et~al., 2018, The Astronomer's Telegram, \href
  {https://ui.adsabs.harvard.edu/abs/2018ATel11610....1C} {11610, 1}

\bibitem[\protect\citeauthoryear{{Churazov}, {Br{\"u}ggen}, {Kaiser},
  {B{\"o}hringer}  \& {Forman}}{{Churazov} et~al.}{2001}]{2001ApJ...554..261C}
{Churazov} E.,  {Br{\"u}ggen} M.,  {Kaiser} C.~R.,  {B{\"o}hringer} H.,
  {Forman} W.,  2001, \mn@doi [Astrophys. J.] {10.1086/321357}, \href
  {https://ui.adsabs.harvard.edu/abs/2001ApJ...554..261C} {554, 261}

\bibitem[\protect\citeauthoryear{{Churazov}, {Khabibullin}, {Bykov}, {Chugai},
  {Sunyaev}  \& {Zinchenko}}{{Churazov} et~al.}{2021}]{2021MNRAS.507..971C}
{Churazov} E.~M.,  {Khabibullin} I.~I.,  {Bykov} A.~M.,  {Chugai} N.~N.,
  {Sunyaev} R.~A.,   {Zinchenko} I.~I.,  2021, \mn@doi [\mnras]
  {10.1093/mnras/stab2125}, \href
  {https://ui.adsabs.harvard.edu/abs/2021MNRAS.507..971C} {507, 971}

\bibitem[\protect\citeauthoryear{{Collier} et~al.,}{{Collier}
  et~al.}{2018}]{2018MNRAS.477..578C}
{Collier} J.~D.,  et~al., 2018, \mn@doi [\mnras] {10.1093/mnras/sty564}, \href
  {https://ui.adsabs.harvard.edu/abs/2018MNRAS.477..578C} {477, 578}

\bibitem[\protect\citeauthoryear{Conroy et~al.}{Conroy et~al.}{2021}]{conroy}
Conroy C.,  et~al., 2021, Nature, 592, 534

\bibitem[\protect\citeauthoryear{{Crawford} et~al.,}{{Crawford}
  et~al.}{2016}]{2016ApJS..227...23C}
{Crawford} T.~M.,  et~al., 2016, \mn@doi [\apjs] {10.3847/1538-4365/227/2/23},
  \href {https://ui.adsabs.harvard.edu/abs/2016ApJS..227...23C} {227, 23}

\bibitem[\protect\citeauthoryear{{Crosley} \& {Osten}}{{Crosley} \&
  {Osten}}{2018}]{2018ApJ...856...39C}
{Crosley} M.~K.,  {Osten} R.~A.,  2018, \mn@doi [Astrophys. J.]
  {10.3847/1538-4357/aaaec2}, \href
  {https://ui.adsabs.harvard.edu/abs/2018ApJ...856...39C} {856, 39}

\bibitem[\protect\citeauthoryear{{Cullinane} et~al.,}{{Cullinane}
  et~al.}{2020}]{2020MNRAS.497.3055C}
{Cullinane} L.~R.,  et~al., 2020, \mn@doi [Mon. Not. R. Astron. Soc.]
  {10.1093/mnras/staa2048}, \href
  {https://ui.adsabs.harvard.edu/abs/2020MNRAS.497.3055C} {497, 3055}

\bibitem[\protect\citeauthoryear{{De Horta} et~al.,}{{De Horta}
  et~al.}{2014}]{2014SerAJ.189...41D}
{De Horta} A.~Y.,  et~al., 2014, \mn@doi [Serbian Astronomical Journal]
  {10.2298/SAJ140605001H}, \href
  {https://ui.adsabs.harvard.edu/abs/2014SerAJ.189...41D} {189, 41}

\bibitem[\protect\citeauthoryear{{Dickey} \& {Lockman}}{{Dickey} \&
  {Lockman}}{1990}]{1990ARA&A..28..215D}
{Dickey} J.~M.,  {Lockman} F.~J.,  1990, \mn@doi [Ann. Rev. Astron. Astrophys.]
  {10.1146/annurev.aa.28.090190.001243}, \href
  {https://ui.adsabs.harvard.edu/abs/1990ARA&A..28..215D} {28, 215}

\bibitem[\protect\citeauthoryear{{Dickinson}}{{Dickinson}}{2018}]{2018Galax...6...56D}
{Dickinson} C.,  2018, \mn@doi [Galaxies] {10.3390/galaxies6020056}, \href
  {https://ui.adsabs.harvard.edu/abs/2018Galax...6...56D} {6, 56}

\bibitem[\protect\citeauthoryear{{Doi}, {Nakagawa}, {Isobe}, {Baba}, {Yano}  \&
  {Yamagishi}}{{Doi} et~al.}{2019}]{2019PASJ...71...26D}
{Doi} R.,  {Nakagawa} T.,  {Isobe} N.,  {Baba} S.,  {Yano} K.,   {Yamagishi}
  M.,  2019, \mn@doi [\pasj] {10.1093/pasj/psz019}, \href
  {https://ui.adsabs.harvard.edu/abs/2019PASJ...71...26D} {71, 26}

\bibitem[\protect\citeauthoryear{{Dotter}}{{Dotter}}{2016}]{2016ApJS..222....8D}
{Dotter} A.,  2016, \mn@doi [Astrophys. J. Suppl.] {10.3847/0067-0049/222/1/8},
  \href {https://ui.adsabs.harvard.edu/abs/2016ApJS..222....8D} {222, 8}

\bibitem[\protect\citeauthoryear{{Ducci}, {Kavanagh}, {Sasaki}  \&
  {Koribalski}}{{Ducci} et~al.}{2014}]{2014A&A...566A.115D}
{Ducci} L.,  {Kavanagh} P.~J.,  {Sasaki} M.,   {Koribalski} B.~S.,  2014,
  \mn@doi [Astron. \& Astrophys.] {10.1051/0004-6361/201423775}, \href
  {https://ui.adsabs.harvard.edu/abs/2014A&A...566A.115D} {566, A115}

\bibitem[\protect\citeauthoryear{{Ducci} et~al.,}{{Ducci}
  et~al.}{2020}]{2020ATel13545....1D}
{Ducci} L.,  et~al., 2020, The Astronomer's Telegram, \href
  {https://ui.adsabs.harvard.edu/abs/2020ATel13545....1D} {13545, 1}

\bibitem[\protect\citeauthoryear{{Edelmann}, {Napiwotzki}, {Heber},
  {Christlieb}  \& {Reimers}}{{Edelmann} et~al.}{2005}]{2005ApJ...634L.181E}
{Edelmann} H.,  {Napiwotzki} R.,  {Heber} U.,  {Christlieb} N.,   {Reimers} D.,
   2005, \mn@doi [Astrophys. J. Let.] {10.1086/498940}, \href
  {https://ui.adsabs.harvard.edu/abs/2005ApJ...634L.181E} {634, L181}

\bibitem[\protect\citeauthoryear{{Eldridge}, {Langer}  \& {Tout}}{{Eldridge}
  et~al.}{2011}]{2011MNRAS.414.3501E}
{Eldridge} J.~J.,  {Langer} N.,   {Tout} C.~A.,  2011, \mn@doi [Mon. Not. R.
  Astron. Soc.] {10.1111/j.1365-2966.2011.18650.x}, \href
  {https://ui.adsabs.harvard.edu/abs/2011MNRAS.414.3501E} {414, 3501}

\bibitem[\protect\citeauthoryear{{Erkal}, {Boubert}, {Gualandris}, {Evans}  \&
  {Antonini}}{{Erkal} et~al.}{2019}]{2019MNRAS.483.2007E}
{Erkal} D.,  {Boubert} D.,  {Gualandris} A.,  {Evans} N.~W.,   {Antonini} F.,
  2019, \mn@doi [Mon. Not. R. Astron. Soc.] {10.1093/mnras/sty2674}, \href
  {https://ui.adsabs.harvard.edu/abs/2019MNRAS.483.2007E} {483, 2007}

\bibitem[\protect\citeauthoryear{{Evans}, {Renzo}  \& {Rossi}}{{Evans}
  et~al.}{2020}]{2020MNRAS.497.5344E}
{Evans} F.~A.,  {Renzo} M.,   {Rossi} E.~M.,  2020, \mn@doi [Mon. Not. R.
  Astron. Soc.] {10.1093/mnras/staa2334}, \href
  {https://ui.adsabs.harvard.edu/abs/2020MNRAS.497.5344E} {497, 5344}

\bibitem[\protect\citeauthoryear{{Evans}, {Marchetti}, {Rossi}, {Baggen}  \&
  {Bloot}}{{Evans} et~al.}{2021}]{2021MNRAS.507.4997E}
{Evans} F.~A.,  {Marchetti} T.,  {Rossi} E.~M.,  {Baggen} J.~F.~W.,   {Bloot}
  S.,  2021, \mn@doi [\mnras] {10.1093/mnras/stab2271}, \href
  {https://ui.adsabs.harvard.edu/abs/2021MNRAS.507.4997E} {507, 4997}

\bibitem[\protect\citeauthoryear{{Filipovic}, {Haynes}, {White}  \&
  {Jones}}{{Filipovic} et~al.}{1998}]{1998A&AS..130..421F}
{Filipovic} M.~D.,  {Haynes} R.~F.,  {White} G.~L.,   {Jones} P.~A.,  1998,
  \mn@doi [\aaps] {10.1051/aas:1998417}, \href
  {https://ui.adsabs.harvard.edu/abs/1998A&AS..130..421F} {130, 421}

\bibitem[\protect\citeauthoryear{{Filipovi{\'c}} et~al.,}{{Filipovi{\'c}}
  et~al.}{2009}]{2009MNRAS.399..769F}
{Filipovi{\'c}} M.~D.,  et~al., 2009, \mn@doi [\mnras]
  {10.1111/j.1365-2966.2009.15307.x}, \href
  {https://ui.adsabs.harvard.edu/abs/2009MNRAS.399..769F} {399, 769}

\bibitem[\protect\citeauthoryear{{Filipovi{\'c}} et~al.,}{{Filipovi{\'c}}
  et~al.}{2021}]{2021MNRAS.507.2885F}
{Filipovi{\'c}} M.~D.,  et~al., 2021, \mn@doi [\mnras]
  {10.1093/mnras/stab2249}, \href
  {https://ui.adsabs.harvard.edu/abs/2021MNRAS.507.2885F} {507, 2885}

\bibitem[\protect\citeauthoryear{Filipović \& Tothill}{Filipović \&
  Tothill}{2021a}]{book2}
Filipović M.~D.,  Tothill N. F.~H.,  eds, 2021a, Multimessenger Astronomy in
  Practice.
2514-3433, IOP Publishing, \mn@doi{10.1088/2514-3433/ac2256}, \url
  {https://dx.doi.org/10.1088/2514-3433/ac2256}

\bibitem[\protect\citeauthoryear{Filipović \& Tothill}{Filipović \&
  Tothill}{2021b}]{10.1088/2514-3433/ac087e}
Filipović M.~D.,  Tothill N. F.~H.,  2021b, Principles of Multimessenger
  Astronomy.
2514-3433, IOP Publishing, \mn@doi{10.1088/2514-3433/ac087e}, \url
  {https://dx.doi.org/10.1088/2514-3433/ac087e}

\bibitem[\protect\citeauthoryear{{For} et~al.,}{{For}
  et~al.}{2018}]{2018MNRAS.480.2743F}
{For} B.-Q.,  et~al., 2018, \mn@doi [\mnras] {10.1093/mnras/sty1960}, \href
  {http://adsabs.harvard.edu/abs/2018MNRAS.480.2743F} {480, 2743}

\bibitem[\protect\citeauthoryear{{Fragione} \& {Gualandris}}{{Fragione} \&
  {Gualandris}}{2019}]{2019MNRAS.489.4543F}
{Fragione} G.,  {Gualandris} A.,  2019, \mn@doi [Mon. Not. R. Astron. Soc.]
  {10.1093/mnras/stz2451}, \href
  {https://ui.adsabs.harvard.edu/abs/2019MNRAS.489.4543F} {489, 4543}

\bibitem[\protect\citeauthoryear{{Freundlich} \& {Maoz}}{{Freundlich} \&
  {Maoz}}{2021}]{2021MNRAS.502.5882F}
{Freundlich} J.,  {Maoz} D.,  2021, \mn@doi [Mon. Not. R. Astron. Soc.]
  {10.1093/mnras/stab493}, \href
  {https://ui.adsabs.harvard.edu/abs/2021MNRAS.502.5882F} {502, 5882}

\bibitem[\protect\citeauthoryear{{Gaensler}}{{Gaensler}}{1998}]{1998ApJ...493..781G}
{Gaensler} B.~M.,  1998, \mn@doi [\apj] {10.1086/305146}, \href
  {https://ui.adsabs.harvard.edu/abs/1998ApJ...493..781G} {493, 781}

\bibitem[\protect\citeauthoryear{{Gaia Collaboration} et~al.,}{{Gaia
  Collaboration} et~al.}{2021}]{2021A&A...649A...1G}
{Gaia Collaboration} et~al., 2021, \mn@doi [\aap]
  {10.1051/0004-6361/202039657}, \href
  {https://ui.adsabs.harvard.edu/abs/2021A&A...649A...1G} {649, A1}

\bibitem[\protect\citeauthoryear{{Galvin} \& {Filipovic}}{{Galvin} \&
  {Filipovic}}{2014}]{2014SerAJ.189...15G}
{Galvin} T.~J.,  {Filipovic} M.~D.,  2014, \mn@doi [Serbian Astronomical
  Journal] {10.2298/SAJ140505002G}, \href
  {https://ui.adsabs.harvard.edu/abs/2014SerAJ.189...15G} {189, 15}

\bibitem[\protect\citeauthoryear{{Geier} et~al.,}{{Geier}
  et~al.}{2015}]{2015Sci...347.1126G}
{Geier} S.,  et~al., 2015, \mn@doi [Science] {10.1126/science.1259063}, \href
  {https://ui.adsabs.harvard.edu/abs/2015Sci...347.1126G} {347, 1126}

\bibitem[\protect\citeauthoryear{{Gooch}}{{Gooch}}{1995}]{1995ASPC...77..144G}
{Gooch} R.,  1995, in {Shaw} R.~A.,  {Payne} H.~E.,   {Hayes} J.~J.~E.,  eds,
  Astronomical Society of the Pacific Conference Series Vol. 77, Astronomical
  Data Analysis Software and Systems IV. p.~144

\bibitem[\protect\citeauthoryear{{Gualandris} \& {Portegies
  Zwart}}{{Gualandris} \& {Portegies Zwart}}{2007}]{2007MNRAS.376L..29G}
{Gualandris} A.,  {Portegies Zwart} S.,  2007, \mn@doi [Mon. Not. R. Astron.
  Soc.] {10.1111/j.1745-3933.2007.00280.x}, \href
  {https://ui.adsabs.harvard.edu/abs/2007MNRAS.376L..29G} {376, L29}

\bibitem[\protect\citeauthoryear{{HI4PI Collaboration} et~al.,}{{HI4PI
  Collaboration} et~al.}{2016}]{2016A&A...594A.116H}
{HI4PI Collaboration} et~al., 2016, \mn@doi [Astron. \& Astrophys.]
  {10.1051/0004-6361/201629178}, \href
  {https://ui.adsabs.harvard.edu/abs/2016A&A...594A.116H} {594, A116}

\bibitem[\protect\citeauthoryear{{Haberl}, {Sturm}, {Filipovi{\'c}}, {Pietsch}
  \& {Crawford}}{{Haberl} et~al.}{2012}]{2012A&A...537L...1H}
{Haberl} F.,  {Sturm} R.,  {Filipovi{\'c}} M.~D.,  {Pietsch} W.,   {Crawford}
  E.~J.,  2012, \mn@doi [\aap] {10.1051/0004-6361/201118369}, \href
  {https://ui.adsabs.harvard.edu/abs/2012A&A...537L...1H} {537, L1}

\bibitem[\protect\citeauthoryear{{Hakobyan} et~al.,}{{Hakobyan}
  et~al.}{2017}]{2017MNRAS.471.1390H}
{Hakobyan} A.~A.,  et~al., 2017, \mn@doi [Mon. Not. R. Astron. Soc.]
  {10.1093/mnras/stx1608}, \href
  {https://ui.adsabs.harvard.edu/abs/2017MNRAS.471.1390H} {471, 1390}

\bibitem[\protect\citeauthoryear{{Hancock}, {Trott}  \&
  {Hurley-Walker}}{{Hancock} et~al.}{2018}]{2018PASA...35...11H}
{Hancock} P.~J.,  {Trott} C.~M.,   {Hurley-Walker} N.,  2018, \mn@doi [\pasa]
  {10.1017/pasa.2018.3}, \href
  {https://ui.adsabs.harvard.edu/abs/2018PASA...35...11H} {35, e011}

\bibitem[\protect\citeauthoryear{{Harvey-Smith}, {Gaensler}, {Kothes},
  {Townsend}, {Heald}, {Ng}  \& {Green}}{{Harvey-Smith}
  et~al.}{2010}]{2010ApJ...712.1157H}
{Harvey-Smith} L.,  {Gaensler} B.~M.,  {Kothes} R.,  {Townsend} R.,  {Heald}
  G.~H.,  {Ng} C.~Y.,   {Green} A.~J.,  2010, \mn@doi [\apj]
  {10.1088/0004-637X/712/2/1157}, \href
  {https://ui.adsabs.harvard.edu/abs/2010ApJ...712.1157H} {712, 1157}

\bibitem[\protect\citeauthoryear{{Hills}}{{Hills}}{1988}]{1988Natur.331..687H}
{Hills} J.~G.,  1988, \mn@doi [Nature] {10.1038/331687a0}, \href
  {https://ui.adsabs.harvard.edu/abs/1988Natur.331..687H} {331, 687}

\bibitem[\protect\citeauthoryear{{Hirsch}, {Heber}, {O'Toole}  \&
  {Bresolin}}{{Hirsch} et~al.}{2005}]{2005A&A...444L..61H}
{Hirsch} H.~A.,  {Heber} U.,  {O'Toole} S.~J.,   {Bresolin} F.,  2005, \mn@doi
  [Astron. \& Astrophys.] {10.1051/0004-6361:200500212}, \href
  {https://ui.adsabs.harvard.edu/abs/2005A&A...444L..61H} {444, L61}

\bibitem[\protect\citeauthoryear{{Horton}, {Krause}  \& {Hardcastle}}{{Horton}
  et~al.}{2020}]{2020MNRAS.499.5765H}
{Horton} M.~A.,  {Krause} M. G.~H.,   {Hardcastle} M.~J.,  2020, \mn@doi [Mon.
  Not. R. Astron. Soc.] {10.1093/mnras/staa3020}, \href
  {https://ui.adsabs.harvard.edu/abs/2020MNRAS.499.5765H} {499, 5765}

\bibitem[\protect\citeauthoryear{{Hurley-Walker} et~al.,}{{Hurley-Walker}
  et~al.}{2019a}]{2019PASA...36...45H}
{Hurley-Walker} N.,  et~al., 2019a, \mn@doi [\pasa] {10.1017/pasa.2019.34},
  \href {https://ui.adsabs.harvard.edu/abs/2019PASA...36...45H} {36, e045}

\bibitem[\protect\citeauthoryear{{Hurley-Walker} et~al.,}{{Hurley-Walker}
  et~al.}{2019b}]{2019PASA...36...48H}
{Hurley-Walker} N.,  et~al., 2019b, \mn@doi [\pasa] {10.1017/pasa.2019.33},
  \href {https://ui.adsabs.harvard.edu/abs/2019PASA...36...48H} {36, e048}

\bibitem[\protect\citeauthoryear{{Irrgang}, {Kreuzer}  \& {Heber}}{{Irrgang}
  et~al.}{2018}]{2018A&A...620A..48I}
{Irrgang} A.,  {Kreuzer} S.,   {Heber} U.,  2018, \mn@doi [Astron. \&
  Astrophys.] {10.1051/0004-6361/201833874}, \href
  {https://ui.adsabs.harvard.edu/abs/2018A&A...620A..48I} {620, A48}

\bibitem[\protect\citeauthoryear{{Irrgang}, {Dimpel}, {Heber}  \&
  {Raddi}}{{Irrgang} et~al.}{2021}]{2021A&A...646L...4I}
{Irrgang} A.,  {Dimpel} M.,  {Heber} U.,   {Raddi} R.,  2021, \mn@doi [Astron.
  \& Astrophys.] {10.1051/0004-6361/202040178}, \href
  {https://ui.adsabs.harvard.edu/abs/2021A&A...646L...4I} {646, L4}

\bibitem[\protect\citeauthoryear{{Jarrett} et~al.,}{{Jarrett}
  et~al.}{2012}]{2012AJ....144...68J}
{Jarrett} T.~H.,  et~al., 2012, \mn@doi [Astron. J.]
  {10.1088/0004-6256/144/2/68}, \href
  {https://ui.adsabs.harvard.edu/abs/2012AJ....144...68J} {144, 68}

\bibitem[\protect\citeauthoryear{{Jarrett}, {Cluver}, {Brown}, {Dale}, {Tsai}
  \& {Masci}}{{Jarrett} et~al.}{2019}]{2019ApJS..245...25J}
{Jarrett} T.~H.,  {Cluver} M.~E.,  {Brown} M.~J.~I.,  {Dale} D.~A.,  {Tsai}
  C.~W.,   {Masci} F.,  2019, \mn@doi [Astrophys. J. Suppl.]
  {10.3847/1538-4365/ab521a}, \href
  {https://ui.adsabs.harvard.edu/abs/2019ApJS..245...25J} {245, 25}

\bibitem[\protect\citeauthoryear{{Kavanagh} et~al.,}{{Kavanagh}
  et~al.}{2016}]{2016A&A...586A...4K}
{Kavanagh} P.~J.,  et~al., 2016, \mn@doi [\aap] {10.1051/0004-6361/201527414},
  \href {https://ui.adsabs.harvard.edu/abs/2016A&A...586A...4K} {586, A4}

\bibitem[\protect\citeauthoryear{{Kavanagh}, {Sasaki}, {Filipovic}, {Points},
  {Bozzetto}, {Haberl}, {Maggi}  \& {Maitra}}{{Kavanagh}
  et~al.}{2021}]{2021arXiv211100446K}
{Kavanagh} P.~J.,  {Sasaki} M.,  {Filipovic} M.~D.,  {Points} S.~D.,
  {Bozzetto} L.~M.,  {Haberl} F.,  {Maggi} P.,   {Maitra} C.,  2021, arXiv
  e-prints, \href {https://ui.adsabs.harvard.edu/abs/2021arXiv211100446K} {p.
  arXiv:2111.00446}

\bibitem[\protect\citeauthoryear{{Khokhlov}, {Mueller}  \&
  {Hoeflich}}{{Khokhlov} et~al.}{1993}]{1993A&A...270..223K}
{Khokhlov} A.,  {Mueller} E.,   {Hoeflich} P.,  1993, \aap, \href
  {https://ui.adsabs.harvard.edu/abs/1993A&A...270..223K} {270, 223}

\bibitem[\protect\citeauthoryear{{Koposov} et~al.,}{{Koposov}
  et~al.}{2020}]{2020MNRAS.491.2465K}
{Koposov} S.~E.,  et~al., 2020, \mn@doi [Mon. Not. R. Astron. Soc.]
  {10.1093/mnras/stz3081}, \href
  {https://ui.adsabs.harvard.edu/abs/2020MNRAS.491.2465K} {491, 2465}

\bibitem[\protect\citeauthoryear{{Koribalski}, {Norris}, {Andernach},
  {Rudnick}, {Shabala}, {Filipovi{\'c}}  \& {Lenc}}{{Koribalski}
  et~al.}{2021}]{2021MNRAS.505L..11K}
{Koribalski} B.~S.,  {Norris} R.~P.,  {Andernach} H.,  {Rudnick} L.,  {Shabala}
  S.,  {Filipovi{\'c}} M.,   {Lenc} E.,  2021, \mn@doi [Mon. Not. R. Astron.
  Soc.] {10.1093/mnrasl/slab041}, \href
  {https://ui.adsabs.harvard.edu/abs/2021MNRAS.505L..11K} {505, L11}

\bibitem[\protect\citeauthoryear{{Kothes} \& {Reich}}{{Kothes} \&
  {Reich}}{2001}]{2001A&A...372..627K}
{Kothes} R.,  {Reich} W.,  2001, \mn@doi [\aap] {10.1051/0004-6361:20010407},
  \href {https://ui.adsabs.harvard.edu/abs/2001A&A...372..627K} {372, 627}

\bibitem[\protect\citeauthoryear{{Kothes}, {Furst}  \& {Reich}}{{Kothes}
  et~al.}{1998}]{1998A&A...331..661K}
{Kothes} R.,  {Furst} E.,   {Reich} W.,  1998, \aap, \href
  {https://ui.adsabs.harvard.edu/abs/1998A&A...331..661K} {331, 661}

\bibitem[\protect\citeauthoryear{{Kothes}, {Fedotov}, {Foster}  \&
  {Uyan{\i}ker}}{{Kothes} et~al.}{2006}]{2006A&A...457.1081K}
{Kothes} R.,  {Fedotov} K.,  {Foster} T.~J.,   {Uyan{\i}ker} B.,  2006, \mn@doi
  [\aap] {10.1051/0004-6361:20065062}, \href
  {https://ui.adsabs.harvard.edu/abs/2006A&A...457.1081K} {457, 1081}

\bibitem[\protect\citeauthoryear{{Kothes}, {Reich}, {Foster}  \&
  {Reich}}{{Kothes} et~al.}{2017}]{2017A&A...597A.116K}
{Kothes} R.,  {Reich} P.,  {Foster} T.~J.,   {Reich} W.,  2017, \mn@doi [\aap]
  {10.1051/0004-6361/201629848}, \href
  {https://ui.adsabs.harvard.edu/abs/2017A&A...597A.116K} {597, A116}

\bibitem[\protect\citeauthoryear{{Kothes}, {Reich}, {Safi-Harb}, {Guest},
  {Reich}  \& {F{\"u}rst}}{{Kothes} et~al.}{2020}]{2020MNRAS.496..723K}
{Kothes} R.,  {Reich} W.,  {Safi-Harb} S.,  {Guest} B.,  {Reich} P.,
  {F{\"u}rst} E.,  2020, \mn@doi [\mnras] {10.1093/mnras/staa1573}, \href
  {https://ui.adsabs.harvard.edu/abs/2020MNRAS.496..723K} {496, 723}

\bibitem[\protect\citeauthoryear{{Lacy} et~al.,}{{Lacy}
  et~al.}{2020}]{2020PASP..132c5001L}
{Lacy} M.,  et~al., 2020, \mn@doi [\pasp] {10.1088/1538-3873/ab63eb}, \href
  {https://ui.adsabs.harvard.edu/abs/2020PASP..132c5001L} {132, 035001}

\bibitem[\protect\citeauthoryear{{Leahy}}{{Leahy}}{2017}]{2017ApJ...837...36L}
{Leahy} D.~A.,  2017, \mn@doi [\apj] {10.3847/1538-4357/aa60c1}, \href
  {https://ui.adsabs.harvard.edu/abs/2017ApJ...837...36L} {837, 36}

\bibitem[\protect\citeauthoryear{{Leahy} \& {Ranasinghe}}{{Leahy} \&
  {Ranasinghe}}{2018}]{2018ApJ...866....9L}
{Leahy} D.~A.,  {Ranasinghe} S.,  2018, \mn@doi [\apj]
  {10.3847/1538-4357/aade48}, \href
  {https://ui.adsabs.harvard.edu/abs/2018ApJ...866....9L} {866, 9}

\bibitem[\protect\citeauthoryear{{Leahy}, {Wang}, {Lawton}, {Ranasinghe}  \&
  {Filipovi{\'c}}}{{Leahy} et~al.}{2019}]{2019AJ....158..149L}
{Leahy} D.,  {Wang} Y.,  {Lawton} B.,  {Ranasinghe} S.,   {Filipovi{\'c}} M.,
  2019, \mn@doi [Astron. J.] {10.3847/1538-3881/ab3d2c}, \href
  {https://ui.adsabs.harvard.edu/abs/2019AJ....158..149L} {158, 149}

\bibitem[\protect\citeauthoryear{{Leahy}, {Ranasinghe}  \& {Gelowitz}}{{Leahy}
  et~al.}{2020}]{2020ApJS..248...16L}
{Leahy} D.~A.,  {Ranasinghe} S.,   {Gelowitz} M.,  2020, \mn@doi [Astrophys. J.
  Suppl.] {10.3847/1538-4365/ab8bd9}, \href
  {https://ui.adsabs.harvard.edu/abs/2020ApJS..248...16L} {248, 16}

\bibitem[\protect\citeauthoryear{{Leverenz}, {Filipovi{\'c}},
  {Boji{\v{c}}i{\'c}}, {Crawford}, {Collier}, {Grieve}, {Dra{\v{s}}kovi{\'c}}
  \& {Reid}}{{Leverenz} et~al.}{2016}]{2016Ap&SS.361..108L}
{Leverenz} H.,  {Filipovi{\'c}} M.~D.,  {Boji{\v{c}}i{\'c}} I.~S.,  {Crawford}
  E.~J.,  {Collier} J.~D.,  {Grieve} K.,  {Dra{\v{s}}kovi{\'c}} D.,   {Reid}
  W.~A.,  2016, \mn@doi [\apss] {10.1007/s10509-016-2686-3}, \href
  {https://ui.adsabs.harvard.edu/abs/2016Ap&SS.361..108L} {361, 108}

\bibitem[\protect\citeauthoryear{{Leverenz}, {Filipovi{\'c}}, {Vukoti{\'c}},
  {Uro{\v{s}}evi{\'c}}  \& {Grieve}}{{Leverenz}
  et~al.}{2017}]{2017MNRAS.468.1794L}
{Leverenz} H.,  {Filipovi{\'c}} M.~D.,  {Vukoti{\'c}} B.,  {Uro{\v{s}}evi{\'c}}
  D.,   {Grieve} K.,  2017, \mn@doi [\mnras] {10.1093/mnras/stx555}, \href
  {https://ui.adsabs.harvard.edu/abs/2017MNRAS.468.1794L} {468, 1794}

\bibitem[\protect\citeauthoryear{{Lopez}, {Ramirez-Ruiz}, {Huppenkothen},
  {Badenes}  \& {Pooley}}{{Lopez} et~al.}{2011}]{2011ApJ...732..114L}
{Lopez} L.~A.,  {Ramirez-Ruiz} E.,  {Huppenkothen} D.,  {Badenes} C.,
  {Pooley} D.~A.,  2011, \mn@doi [Astrophys. J.] {10.1088/0004-637X/732/2/114},
  \href {https://ui.adsabs.harvard.edu/abs/2011ApJ...732..114L} {732, 114}

\bibitem[\protect\citeauthoryear{{Luken} et~al.,}{{Luken}
  et~al.}{2020}]{2020MNRAS.492.2606L}
{Luken} K.~J.,  et~al., 2020, \mn@doi [Mon. Not. R. Astron. Soc.]
  {10.1093/mnras/stz3439}, \href
  {https://ui.adsabs.harvard.edu/abs/2020MNRAS.492.2606L} {492, 2606}

\bibitem[\protect\citeauthoryear{{Maggi} et~al.,}{{Maggi}
  et~al.}{2016}]{2016A&A...585A.162M}
{Maggi} P.,  et~al., 2016, \mn@doi [Astron. \& Astrophys.]
  {10.1051/0004-6361/201526932}, \href
  {https://ui.adsabs.harvard.edu/abs/2016A&A...585A.162M} {585, A162}

\bibitem[\protect\citeauthoryear{{Maggi} et~al.,}{{Maggi}
  et~al.}{2019}]{2019A&A...631A.127M}
{Maggi} P.,  et~al., 2019, \mn@doi [Astron. \& Astrophys.]
  {10.1051/0004-6361/201936583}, \href
  {https://ui.adsabs.harvard.edu/abs/2019A&A...631A.127M} {631, A127}

\bibitem[\protect\citeauthoryear{{Maitra}, {Haberl}, {Maggi}, {Kavanagh},
  {Vasilopoulos}, {Sasaki}, {Filipovi{\'c}}  \& {Udalski}}{{Maitra}
  et~al.}{2021}]{2021MNRAS.504..326M}
{Maitra} C.,  {Haberl} F.,  {Maggi} P.,  {Kavanagh} P.~J.,  {Vasilopoulos} G.,
  {Sasaki} M.,  {Filipovi{\'c}} M.~D.,   {Udalski} A.,  2021, \mn@doi [Mon.
  Not. R. Astron. Soc.] {10.1093/mnras/stab716}, \href
  {https://ui.adsabs.harvard.edu/abs/2021MNRAS.504..326M} {504, 326}

\bibitem[\protect\citeauthoryear{{Maoz} \& {Mannucci}}{{Maoz} \&
  {Mannucci}}{2012}]{2012PASA...29..447M}
{Maoz} D.,  {Mannucci} F.,  2012, \mn@doi [Pub. Astron. Soc. Aust.]
  {10.1071/AS11052}, \href
  {https://ui.adsabs.harvard.edu/abs/2012PASA...29..447M} {29, 447}

\bibitem[\protect\citeauthoryear{{Marocco} et~al.,}{{Marocco}
  et~al.}{2021}]{2021ApJS..253....8M}
{Marocco} F.,  et~al., 2021, \mn@doi [Astrophys. J. Suppl.]
  {10.3847/1538-4365/abd805}, \href
  {https://ui.adsabs.harvard.edu/abs/2021ApJS..253....8M} {253, 8}

\bibitem[\protect\citeauthoryear{{McKee} \& {Truelove}}{{McKee} \&
  {Truelove}}{1995}]{1995PhR...256..157M}
{McKee} C.~F.,  {Truelove} J.~K.,  1995, \mn@doi [\physrep]
  {10.1016/0370-1573(94)00106-D}, \href
  {https://ui.adsabs.harvard.edu/abs/1995PhR...256..157M} {256, 157}

\bibitem[\protect\citeauthoryear{{Meyer}, {Langer}, {Mackey}, {Vel{\'a}zquez}
  \& {Gusdorf}}{{Meyer} et~al.}{2015}]{2015MNRAS.450.3080M}
{Meyer} D.~M.~A.,  {Langer} N.,  {Mackey} J.,  {Vel{\'a}zquez} P.~F.,
  {Gusdorf} A.,  2015, \mn@doi [\mnras] {10.1093/mnras/stv898}, \href
  {https://ui.adsabs.harvard.edu/abs/2015MNRAS.450.3080M} {450, 3080}

\bibitem[\protect\citeauthoryear{{Meyer}, {Pohl}, {Petrov}  \&
  {Oskinova}}{{Meyer} et~al.}{2021}]{2021MNRAS.502.5340M}
{Meyer} D.~M.~A.,  {Pohl} M.,  {Petrov} M.,   {Oskinova} L.,  2021, \mn@doi
  [\mnras] {10.1093/mnras/stab452}, \href
  {https://ui.adsabs.harvard.edu/abs/2021MNRAS.502.5340M} {502, 5340}

\bibitem[\protect\citeauthoryear{{Millar}, {White}  \& {Filipovic}}{{Millar}
  et~al.}{2012}]{2012SerAJ.184...19M}
{Millar} W.~C.,  {White} G.~L.,   {Filipovic} M.~D.,  2012, \mn@doi [Serbian
  Astronomical Journal] {10.2298/SAJ1284019M}, \href
  {https://ui.adsabs.harvard.edu/abs/2012SerAJ.184...19M} {184, 19}

\bibitem[\protect\citeauthoryear{{Namekata} et~al.,}{{Namekata}
  et~al.}{2021}]{2021NatAs.tmp..246N}
{Namekata} K.,  et~al., 2021, \mn@doi [Nature Astronomy]
  {10.1038/s41550-021-01532-8}, \href
  {https://ui.adsabs.harvard.edu/abs/2021NatAs.tmp..246N} {}

\bibitem[\protect\citeauthoryear{{Neunteufel}}{{Neunteufel}}{2020}]{2020A&A...641A..52N}
{Neunteufel} P.,  2020, \mn@doi [Astron. \& Astrophys.]
  {10.1051/0004-6361/202037792}, \href
  {https://ui.adsabs.harvard.edu/abs/2020A&A...641A..52N} {641, A52}

\bibitem[\protect\citeauthoryear{{Nidever} et~al.,}{{Nidever}
  et~al.}{2021}]{2021AJ....161...74N}
{Nidever} D.~L.,  et~al., 2021, \mn@doi [Astron. J.]
  {10.3847/1538-3881/abceb7}, \href
  {https://ui.adsabs.harvard.edu/abs/2021AJ....161...74N} {161, 74}

\bibitem[\protect\citeauthoryear{{Nikolaev}, {Drake}, {Keller}, {Cook},
  {Dalal}, {Griest}, {Welch}  \& {Kanbur}}{{Nikolaev}
  et~al.}{2004}]{2004ApJ...601..260N}
{Nikolaev} S.,  {Drake} A.~J.,  {Keller} S.~C.,  {Cook} K.~H.,  {Dalal} N.,
  {Griest} K.,  {Welch} D.~L.,   {Kanbur} S.~M.,  2004, \mn@doi [\apj]
  {10.1086/380439}, \href
  {https://ui.adsabs.harvard.edu/abs/2004ApJ...601..260N} {601, 260}

\bibitem[\protect\citeauthoryear{{Norris}, {Crawford}  \& {Macgregor}}{{Norris}
  et~al.}{2021a}]{2021Galax...9...83N}
{Norris} R.~P.,  {Crawford} E.,   {Macgregor} P.,  2021a, \mn@doi [Galaxies]
  {10.3390/galaxies9040083}, \href
  {https://ui.adsabs.harvard.edu/abs/2021Galax...9...83N} {9, 83}

\bibitem[\protect\citeauthoryear{{Norris} et~al.,}{{Norris}
  et~al.}{2021b}]{2021PASA...38....3N}
{Norris} R.~P.,  et~al., 2021b, \mn@doi [Pub. Astron. Soc. Aust.]
  {10.1017/pasa.2020.52}, \href
  {https://ui.adsabs.harvard.edu/abs/2021PASA...38....3N} {38, e003}

\bibitem[\protect\citeauthoryear{Norris et~al.,}{Norris et~al.}{2021c}]{emups}
Norris R.~P.,  et~al., 2021c, \mn@doi [Publications of the Astronomical Society
  of Australia] {10.1017/pasa.2021.42}, 38, e046

\bibitem[\protect\citeauthoryear{{Notsu} et~al.,}{{Notsu}
  et~al.}{2019}]{2019ApJ...876...58N}
{Notsu} Y.,  et~al., 2019, \mn@doi [Astrophys. J.] {10.3847/1538-4357/ab14e6},
  \href {https://ui.adsabs.harvard.edu/abs/2019ApJ...876...58N} {876, 58}

\bibitem[\protect\citeauthoryear{O'Dea \& Owen}{O'Dea \&
  Owen}{1986}]{1986ApJ...301..841O}
O'Dea C.~P.,  Owen F.~N.,  1986, Astrophys. J., 301, 841

\bibitem[\protect\citeauthoryear{{Ochsenbein}, {Bauer}  \&
  {Marcout}}{{Ochsenbein} et~al.}{2000}]{2000A&AS..143...23O}
{Ochsenbein} F.,  {Bauer} P.,   {Marcout} J.,  2000, \mn@doi [\aaps]
  {10.1051/aas:2000169}, \href
  {https://ui.adsabs.harvard.edu/abs/2000A&AS..143...23O} {143, 23}

\bibitem[\protect\citeauthoryear{{Onken} et~al.,}{{Onken}
  et~al.}{2019}]{2019PASA...36...33O}
{Onken} C.~A.,  et~al., 2019, \mn@doi [\pasa] {10.1017/pasa.2019.27}, \href
  {https://ui.adsabs.harvard.edu/abs/2019PASA...36...33O} {36, e033}

\bibitem[\protect\citeauthoryear{{Paiano}, {Falomo}, {Treves}  \&
  {Scarpa}}{{Paiano} et~al.}{2020}]{2020MNRAS.497...94P}
{Paiano} S.,  {Falomo} R.,  {Treves} A.,   {Scarpa} R.,  2020, \mn@doi [Mon.
  Not. R. Astron. Soc.] {10.1093/mnras/staa1840}, \href
  {https://ui.adsabs.harvard.edu/abs/2020MNRAS.497...94P} {497, 94}

\bibitem[\protect\citeauthoryear{{Pakmor}, {Kromer}, {Taubenberger}  \&
  {Springel}}{{Pakmor} et~al.}{2013}]{2013ApJ...770L...8P}
{Pakmor} R.,  {Kromer} M.,  {Taubenberger} S.,   {Springel} V.,  2013, \mn@doi
  [\apjl] {10.1088/2041-8205/770/1/L8}, \href
  {https://ui.adsabs.harvard.edu/abs/2013ApJ...770L...8P} {770, L8}

\bibitem[\protect\citeauthoryear{{Pavlovic}, {Dobardzic}, {Vukotic}  \&
  {Urosevic}}{{Pavlovic} et~al.}{2014}]{2014SerAJ.189...25P}
{Pavlovic} M.~Z.,  {Dobardzic} A.,  {Vukotic} B.,   {Urosevic} D.,  2014,
  \mn@doi [Serbian Astronomical Journal] {10.2298/SAJ1489025P}, \href
  {https://ui.adsabs.harvard.edu/abs/2014SerAJ.189...25P} {189, 25}

\bibitem[\protect\citeauthoryear{{Pavlovi{\'c}}, {Uro{\v{s}}evi{\'c}},
  {Arbutina}, {Orlando}, {Maxted}  \& {Filipovi{\'c}}}{{Pavlovi{\'c}}
  et~al.}{2018}]{2018ApJ...852...84P}
{Pavlovi{\'c}} M.~Z.,  {Uro{\v{s}}evi{\'c}} D.,  {Arbutina} B.,  {Orlando} S.,
  {Maxted} N.,   {Filipovi{\'c}} M.~D.,  2018, \mn@doi [Astrophys. J.]
  {10.3847/1538-4357/aaa1e6}, \href
  {https://ui.adsabs.harvard.edu/abs/2018ApJ...852...84P} {852, 84}

\bibitem[\protect\citeauthoryear{{Pejcha} \& {Prieto}}{{Pejcha} \&
  {Prieto}}{2015}]{2015ApJ...806..225P}
{Pejcha} O.,  {Prieto} J.~L.,  2015, \mn@doi [\apj]
  {10.1088/0004-637X/806/2/225}, \href
  {https://ui.adsabs.harvard.edu/abs/2015ApJ...806..225P} {806, 225}

\bibitem[\protect\citeauthoryear{{Pennock} et~al.,}{{Pennock}
  et~al.}{2021}]{2021MNRAS.506.3540P}
{Pennock} C.~M.,  et~al., 2021, \mn@doi [Mon. Not. R. Astron. Soc.]
  {10.1093/mnras/stab1858}, \href
  {https://ui.adsabs.harvard.edu/abs/2021MNRAS.506.3540P} {506, 3540}

\bibitem[\protect\citeauthoryear{{Petruk}, {Kuzyo}, {Orlando}, {Pohl}  \&
  {Brose}}{{Petruk} et~al.}{2021}]{2021MNRAS.505..755P}
{Petruk} O.,  {Kuzyo} T.,  {Orlando} S.,  {Pohl} M.,   {Brose} R.,  2021,
  \mn@doi [\mnras] {10.1093/mnras/stab1319}, \href
  {https://ui.adsabs.harvard.edu/abs/2021MNRAS.505..755P} {505, 755}

\bibitem[\protect\citeauthoryear{{Piatti} \& {Geisler}}{{Piatti} \&
  {Geisler}}{2013}]{2013AJ....145...17P}
{Piatti} A.~E.,  {Geisler} D.,  2013, \mn@doi [Astron. J.]
  {10.1088/0004-6256/145/1/17}, \href
  {https://ui.adsabs.harvard.edu/abs/2013AJ....145...17P} {145, 17}

\bibitem[\protect\citeauthoryear{{Pietrzy{\'n}ski} et~al.,}{{Pietrzy{\'n}ski}
  et~al.}{2019}]{2019Natur.567..200P}
{Pietrzy{\'n}ski} G.,  et~al., 2019, \mn@doi [\nat]
  {10.1038/s41586-019-0999-4}, \href
  {https://ui.adsabs.harvard.edu/abs/2019Natur.567..200P} {567, 200}

\bibitem[\protect\citeauthoryear{{Platais} et~al.,}{{Platais}
  et~al.}{2018}]{2018AJ....156...98P}
{Platais} I.,  et~al., 2018, \mn@doi [Astron. J.] {10.3847/1538-3881/aad280},
  \href {https://ui.adsabs.harvard.edu/abs/2018AJ....156...98P} {156, 98}

\bibitem[\protect\citeauthoryear{{Przybilla}, {Nieva}, {Heber}, {Firnstein},
  {Butler}, {Napiwotzki}  \& {Edelmann}}{{Przybilla}
  et~al.}{2008}]{2008A&A...480L..37P}
{Przybilla} N.,  {Nieva} M.~F.,  {Heber} U.,  {Firnstein} M.,  {Butler} K.,
  {Napiwotzki} R.,   {Edelmann} H.,  2008, \mn@doi [Astron. \& Astrophys.]
  {10.1051/0004-6361:200809391}, \href
  {https://ui.adsabs.harvard.edu/abs/2008A&A...480L..37P} {480, L37}

\bibitem[\protect\citeauthoryear{{Raddi} et~al.,}{{Raddi}
  et~al.}{2019}]{2019MNRAS.489.1489R}
{Raddi} R.,  et~al., 2019, \mn@doi [Mon. Not. R. Astron. Soc.]
  {10.1093/mnras/stz1618}, \href
  {https://ui.adsabs.harvard.edu/abs/2019MNRAS.489.1489R} {489, 1489}

\bibitem[\protect\citeauthoryear{Ranasinghe \& Leahy}{Ranasinghe \&
  Leahy}{2019}]{Ranasinghe:2019quc}
Ranasinghe S.,  Leahy D.,  2019, \mn@doi [JHEP Grav. Cosmol.]
  {10.4236/jhepgc.2019.53046}, 5, 907

\bibitem[\protect\citeauthoryear{{Reynolds}, {Gaensler}  \&
  {Bocchino}}{{Reynolds} et~al.}{2012}]{2012SSRv..166..231R}
{Reynolds} S.~P.,  {Gaensler} B.~M.,   {Bocchino} F.,  2012, \mn@doi [Space
  Science Reviews] {10.1007/s11214-011-9775-y}, \href
  {https://ui.adsabs.harvard.edu/abs/2012SSRv..166..231R} {166, 231}

\bibitem[\protect\citeauthoryear{{Reynoso}, {Hughes}  \& {Moffett}}{{Reynoso}
  et~al.}{2013}]{2013AJ....145..104R}
{Reynoso} E.~M.,  {Hughes} J.~P.,   {Moffett} D.~A.,  2013, \mn@doi [\aj]
  {10.1088/0004-6256/145/4/104}, \href
  {https://ui.adsabs.harvard.edu/abs/2013AJ....145..104R} {145, 104}

\bibitem[\protect\citeauthoryear{{Roper} et~al.,}{{Roper}
  et~al.}{2018}]{2018MNRAS.479.1800R}
{Roper} Q.,  et~al., 2018, \mn@doi [\mnras] {10.1093/mnras/sty1196}, \href
  {https://ui.adsabs.harvard.edu/abs/2018MNRAS.479.1800R} {479, 1800}

\bibitem[\protect\citeauthoryear{{Rudnick}}{{Rudnick}}{2002}]{2002PASP..114..427R}
{Rudnick} L.,  2002, \mn@doi [Pub. Astron. Soc. Pac.] {10.1086/342499}, \href
  {https://ui.adsabs.harvard.edu/abs/2002PASP..114..427R} {114, 427}

\bibitem[\protect\citeauthoryear{{Sadeh}, {Abdalla}  \& {Lahav}}{{Sadeh}
  et~al.}{2016}]{annz_bib}
{Sadeh} I.,  {Abdalla} F.~B.,   {Lahav} O.,  2016, \mn@doi [\pasp]
  {10.1088/1538-3873/128/968/104502}, \href
  {https://ui.adsabs.harvard.edu/abs/2016PASP..128j4502S} {128, 104502}

\bibitem[\protect\citeauthoryear{{Sault}, {Teuben}  \& {Wright}}{{Sault}
  et~al.}{1995}]{1995ASPC...77..433S}
{Sault} R.~J.,  {Teuben} P.~J.,   {Wright} M.~C.~H.,  1995, in {Shaw} R.~A.,
  {Payne} H.~E.,   {Hayes} J.~J.~E.,  eds,  Astronomical Society of the Pacific
  Conference Series Vol. 77, Astronomical Data Analysis Software and Systems
  IV. p.~433 (\mn@eprint {arXiv} {astro-ph/0612759})

\bibitem[\protect\citeauthoryear{{Schlafly} \& {Finkbeiner}}{{Schlafly} \&
  {Finkbeiner}}{2011}]{2011ApJ...737..103S}
{Schlafly} E.~F.,  {Finkbeiner} D.~P.,  2011, \mn@doi [\apj]
  {10.1088/0004-637X/737/2/103}, \href
  {https://ui.adsabs.harvard.edu/abs/2011ApJ...737..103S} {737, 103}

\bibitem[\protect\citeauthoryear{{Schlafly}, {Meisner}  \& {Green}}{{Schlafly}
  et~al.}{2019}]{2019ApJS..240...30S}
{Schlafly} E.~F.,  {Meisner} A.~M.,   {Green} G.~M.,  2019, \mn@doi [Astrophys.
  J. Suppl.] {10.3847/1538-4365/aafbea}, \href
  {https://ui.adsabs.harvard.edu/abs/2019ApJS..240...30S} {240, 30}

\bibitem[\protect\citeauthoryear{{Shabala}, {Jurlin}, {Morganti}, {Brienza},
  {Hardcastle}, {Godfrey}, {Krause}  \& {Turner}}{{Shabala}
  et~al.}{2020}]{2020MNRAS.496.1706S}
{Shabala} S.~S.,  {Jurlin} N.,  {Morganti} R.,  {Brienza} M.,  {Hardcastle}
  M.~J.,  {Godfrey} L. E.~H.,  {Krause} M. G.~H.,   {Turner} R.~J.,  2020,
  \mn@doi [Mon. Not. R. Astron. Soc.] {10.1093/mnras/staa1172}, \href
  {https://ui.adsabs.harvard.edu/abs/2020MNRAS.496.1706S} {496, 1706}

\bibitem[\protect\citeauthoryear{{Shen} et~al.,}{{Shen}
  et~al.}{2018}]{2018ApJ...865...15S}
{Shen} K.~J.,  et~al., 2018, \mn@doi [Astrophys. J.]
  {10.3847/1538-4357/aad55b}, \href
  {https://ui.adsabs.harvard.edu/abs/2018ApJ...865...15S} {865, 15}

\bibitem[\protect\citeauthoryear{{Str{\"u}der} et~al.,}{{Str{\"u}der}
  et~al.}{2001}]{2001A&A...365L..18S}
{Str{\"u}der} L.,  et~al., 2001, \mn@doi [\aap] {10.1051/0004-6361:20000066},
  \href {https://ui.adsabs.harvard.edu/abs/2001A&A...365L..18S} {365, L18}

\bibitem[\protect\citeauthoryear{{Sun}, {Reich}, {Wang}, {Han}  \&
  {Reich}}{{Sun} et~al.}{2011}]{2011A&A...535A..64S}
{Sun} X.~H.,  {Reich} W.,  {Wang} C.,  {Han} J.~L.,   {Reich} P.,  2011,
  \mn@doi [\aap] {10.1051/0004-6361/201117679}, \href
  {https://ui.adsabs.harvard.edu/abs/2011A&A...535A..64S} {535, A64}

\bibitem[\protect\citeauthoryear{{Truelove} \& {McKee}}{{Truelove} \&
  {McKee}}{1999}]{1999ApJS..120..299T}
{Truelove} J.~K.,  {McKee} C.~F.,  1999, \mn@doi [Astrophys. J. Suppl.]
  {10.1086/313176}, \href
  {https://ui.adsabs.harvard.edu/abs/1999ApJS..120..299T} {120, 299}

\bibitem[\protect\citeauthoryear{{Turner} et~al.,}{{Turner}
  et~al.}{2001}]{2001A&A...365L..27T}
{Turner} M.~J.~L.,  et~al., 2001, \mn@doi [\aap] {10.1051/0004-6361:20000087},
  \href {https://ui.adsabs.harvard.edu/abs/2001A&A...365L..27T} {365, L27}

\bibitem[\protect\citeauthoryear{{Turtle}, {Pugh}, {Kenderdine}  \&
  {Pauliny-Toth}}{{Turtle} et~al.}{1962}]{1962MNRAS.124..297T}
{Turtle} A.~J.,  {Pugh} J.~F.,  {Kenderdine} S.,   {Pauliny-Toth} I.~I.~K.,
  1962, \mn@doi [\mnras] {10.1093/mnras/124.4.297}, \href
  {https://ui.adsabs.harvard.edu/abs/1962MNRAS.124..297T} {124, 297}

\bibitem[\protect\citeauthoryear{{Uro{\v{s}}evi{\'c}}}{{Uro{\v{s}}evi{\'c}}}{2020}]{2020NatAs...4..910U}
{Uro{\v{s}}evi{\'c}} D.,  2020, \mn@doi [Nature Astronomy]
  {10.1038/s41550-020-01228-5}, \href
  {https://ui.adsabs.harvard.edu/abs/2020NatAs...4..910U} {4, 910}

\bibitem[\protect\citeauthoryear{{Villadsen} \& {Hallinan}}{{Villadsen} \&
  {Hallinan}}{2019}]{2019ApJ...871..214V}
{Villadsen} J.,  {Hallinan} G.,  2019, \mn@doi [Astrophys. J.]
  {10.3847/1538-4357/aaf88e}, \href
  {https://ui.adsabs.harvard.edu/abs/2019ApJ...871..214V} {871, 214}

\bibitem[\protect\citeauthoryear{{Vukoti{\'c}}, {{\'C}iprijanovi{\'c}},
  {Vu{\v{c}}eti{\'c}}, {Oni{\'c}}  \& {Uro{\v{s}}evi{\'c}}}{{Vukoti{\'c}}
  et~al.}{2019}]{2019SerAJ.199...23S}
{Vukoti{\'c}} B.,  {{\'C}iprijanovi{\'c}} A.,  {Vu{\v{c}}eti{\'c}} M.~M.,
  {Oni{\'c}} D.,   {Uro{\v{s}}evi{\'c}} D.,  2019, \mn@doi [Serb. Astron. J.]
  {10.2298/SAJ1999023V}, \href
  {https://ui.adsabs.harvard.edu/abs/2019SerAJ.199...23S} {199, 23}

\bibitem[\protect\citeauthoryear{{Wan}, {Guglielmo}, {Lewis}, {Mackey}  \&
  {Ibata}}{{Wan} et~al.}{2020}]{2020MNRAS.492..782W}
{Wan} Z.,  {Guglielmo} M.,  {Lewis} G.~F.,  {Mackey} D.,   {Ibata} R.~A.,
  2020, \mn@doi [Mon. Not. R. Astron. Soc.] {10.1093/mnras/stz3493}, \href
  {https://ui.adsabs.harvard.edu/abs/2020MNRAS.492..782W} {492, 782}

\bibitem[\protect\citeauthoryear{{Webbink}}{{Webbink}}{1984}]{1984ApJ...277..355W}
{Webbink} R.~F.,  1984, \mn@doi [Astrophys. J.] {10.1086/161701}, \href
  {https://ui.adsabs.harvard.edu/abs/1984ApJ...277..355W} {277, 355}

\bibitem[\protect\citeauthoryear{{Wenger} et~al.,}{{Wenger}
  et~al.}{2000}]{2000A&AS..143....9W}
{Wenger} M.,  et~al., 2000, \mn@doi [\aaps] {10.1051/aas:2000332}, \href
  {https://ui.adsabs.harvard.edu/abs/2000A&AS..143....9W} {143, 9}

\bibitem[\protect\citeauthoryear{{West}, {Safi-Harb}, {Jaffe}, {Kothes},
  {Landecker}  \& {Foster}}{{West} et~al.}{2016}]{2016A&A...587A.148W}
{West} J.~L.,  {Safi-Harb} S.,  {Jaffe} T.,  {Kothes} R.,  {Landecker} T.~L.,
  {Foster} T.,  2016, \mn@doi [\aap] {10.1051/0004-6361/201527001}, \href
  {https://ui.adsabs.harvard.edu/abs/2016A&A...587A.148W} {587, A148}

\bibitem[\protect\citeauthoryear{{West}, {Safi-Harb}  \& {Ferrand}}{{West}
  et~al.}{2017}]{2017A&A...597A.121W}
{West} J.~L.,  {Safi-Harb} S.,   {Ferrand} G.,  2017, \mn@doi [\aap]
  {10.1051/0004-6361/201628079}, \href
  {https://ui.adsabs.harvard.edu/abs/2017A&A...597A.121W} {597, A121}

\bibitem[\protect\citeauthoryear{{West}, {Landecker}, {Gaensler}, {Jaffe}  \&
  {Hill}}{{West} et~al.}{2021}]{2021ApJ...923...58W}
{West} J.~L.,  {Landecker} T.~L.,  {Gaensler} B.~M.,  {Jaffe} T.,   {Hill}
  A.~S.,  2021, \mn@doi [\apj] {10.3847/1538-4357/ac2ba2}, \href
  {https://ui.adsabs.harvard.edu/abs/2021ApJ...923...58W} {923, 58}

\bibitem[\protect\citeauthoryear{{Whelan} \& {Iben}}{{Whelan} \&
  {Iben}}{1973}]{1973ApJ...186.1007W}
{Whelan} J.,  {Iben} Icko J.,  1973, \mn@doi [Astrophys. J.] {10.1086/152565},
  \href {https://ui.adsabs.harvard.edu/abs/1973ApJ...186.1007W} {186, 1007}

\bibitem[\protect\citeauthoryear{{Wilms}, {Allen}  \& {McCray}}{{Wilms}
  et~al.}{2000}]{Wilms2000}
{Wilms} J.,  {Allen} A.,   {McCray} R.,  2000, \mn@doi [\apj] {10.1086/317016},
  \href {http://cdsads.u-strasbg.fr/abs/2000ApJ...542..914W} {542, 914}

\bibitem[\protect\citeauthoryear{{Xiao}, {F{\"u}rst}, {Reich}  \& {Han}}{{Xiao}
  et~al.}{2008}]{2008A&A...482..783X}
{Xiao} L.,  {F{\"u}rst} E.,  {Reich} W.,   {Han} J.~L.,  2008, \mn@doi [\aap]
  {10.1051/0004-6361:20078461}, \href
  {https://ui.adsabs.harvard.edu/abs/2008A&A...482..783X} {482, 783}

\bibitem[\protect\citeauthoryear{{Yew} et~al.,}{{Yew}
  et~al.}{2021}]{2021MNRAS.500.2336Y}
{Yew} M.,  et~al., 2021, \mn@doi [Mon. Not. R. Astron. Soc.]
  {10.1093/mnras/staa3382}, \href
  {https://ui.adsabs.harvard.edu/abs/2021MNRAS.500.2336Y} {500, 2336}

\bibitem[\protect\citeauthoryear{{Zinn}, {Grunden}  \& {Bomans}}{{Zinn}
  et~al.}{2011}]{2011A&A...536A.103Z}
{Zinn} P.~C.,  {Grunden} P.,   {Bomans} D.~J.,  2011, \mn@doi [Astron. \&
  Astrophys.] {10.1051/0004-6361/201117631}, \href
  {https://ui.adsabs.harvard.edu/abs/2011A&A...536A.103Z} {536, A103}

\makeatother
\end{thebibliography}

\section*{}
Please note: Oxford University Press is not responsible for the content or functionality of any supporting materials supplied by the authors. Any queries (other than missing material) should be directed to the corresponding author for the article.

\section*{}
{\it 
$^{1}$School of Science, Western Sydney University, Locked Bag 1797, Penrith South DC, NSW 2751, Australia \\
$^{2}$CSIRO Space \& Astronomy, PO Box 76, Epping, NSW 1710, Australia \\
$^{3}$Minnesota Institute for Astrophysics, School of Physics and Astronomy, University of Minnesota, 116 Church Street SE, Minneapolis, MN 55455, USA\\
$^{4}$Department of Physics and Astronomy, University of Calgary, University of Calgary, Calgary, Alberta, T2N 1N4, Canada\\
$^{5}$Institut f\"ur Astronomie und Astrophysik, Kepler Center for Astro and Particle Physics, Universit\"at T\"ubingen, Sand 1, 72076 T\"ubingen, Germany\\
$^{6}$ISDC Data Center for Astrophysics, Universit\'e de Gen\`eve, 16 chemin d’\'Ecogia, 1290 Versoix, Switzerland\\
$^{7}$Dominion Radio Astrophysical Observatory, Herzberg Astronomy and Astrophysics, National Research Council Canada, PO Box 248, Penticton BC V2A 6J9, Canada \\
$^{8}$Departamento de Astronom\'ia, DCNE, Universidad de Guanajuato, Cj\'on. de Jalisco s/n, Col. Valenciana, Guanajuato, CP 36023, Gto., Mexico\\
$^{9}$Dublin Institute for Advanced Studies, Astronomy \& Astrophysics Section, 31 Fitzwilliam Place, D02 XF86 Dublin 2, Ireland \\
$^{10}$University of Potsdam, Institute of Physics and Astronomy, 14476 Potsdam, Germany \\
$^{11}$The Inter-University Institute for Data Intensive Astronomy (IDIA), Department of Astronomy, University of Cape Town, Rondebosch 7701, South Africa\\
$^{12}$Research School of Astronomy and Astrophysics, Australian National University, Canberra 2611, ACT, Australia\\
$^{13}$International Centre for Radio Astronomy Research, Curtin University, Bentley, WA 6102, Australia\\
$^{14}$Max-Planck-Institut f\"{u}r extraterrestrische Physik, Gie{\ss}enbachstra{\ss}e 1, D-85748 Garching, Germany\\
$^{15}$Remeis Observatory and ECAP, Universit\"{a}t Erlangen-N\"{u}rnberg, Sternwartstra{\ss}e 7, D-96049 Bamberg, Germany\\
$^{16}$Australian Astronomical Optics, AAO-Macquarie, Faculty of Science and Engineering, Macquarie University, 105 Delhi Rd, North Ryde, NSW 2113, Australia \\
$^{17}$INAF -- Osservatorio Astrofisico di Catania, via Santa Sofia 78, I-95123 Catania, Italia \\
$^{18}$Department of Astronomy, University of Cape Town, Private Bag X3, Rondebosch 7701, South Africa\\
$^{19}$School of Cosmic Physics, Dublin Institute for Advanced Studies, 31 Fitzwilliam Place, Dublin 2, Ireland\\
$^{20}$Observatoire Astronomique de Strasbourg, Universit\'e de Strasbourg, CNRS, 11 rue de l'Universit\'e, F-67000 Strasbourg, France\\
$^{21}$Lennard-Jones Laboratories, Keele University, Staffordshire, ST5 5BG, UK\\
$^{22}$Cerro Tololo Inter-American Observatory/NSF's NOIRLab, Casilla 603, La Serena, Chile\\
$^{23}$School of Physical Sciences, The University of Adelaide, Adelaide 5005, Australia\\
$^{24}$Department of Physics and Astronomy, University of Manitoba, Winnipeg, MB R3T 2N2, Canada\\
$^{25}$National Astronomical Observatory of Japan, Mitaka, Tokyo 181-8588, Japan\\
$^{26}$School of Natural Sciences, University of Tasmania, Private Bag 37, Hobart 7001, Australia\\
$^{27}$Department of Astronomy, Faculty of Mathematics, University of Belgrade, Studentski trg 16, 11000 Belgrade, Serbia\\
$^{28}$Isaac Newton Institute of Chile, Yugoslavia Branch\\
$^{29}$Dunlap Institute for Astronomy and Astrophysics, University of Toronto, Toronto, ON M5S 3H4, Canada\\
$^{30}$Sydney Institute for Astronomy, School of Physics, The University of Sydney, Sydney, New South Wales, Australia
}



\appendix 

\renewcommand{\thetable}{A1}
\begin{table*}
 \setcounter{table}{4}
\centering
\caption{\lmcorc\ as an \ac{SNR} of type~Ia at various distances (radius) and various \ac{ISM} densities ($n_{\rm H}=10^{-1} - 10^{-4}$~cm$^{-3}$). Based on the model of \citet{2018ApJ...866....9L}, we assume an \ac{SN} explosion energy of $0.5\times10^{51}$~erg and type~Ia \ac{SN} ejecta mass of 1.4~M$_{\odot}$. Column~6 lists the emission-measure-weighted shocked gas electron temperature, Column~7 the emission measure (EM), Column~8 the X-ray luminosity (L$_{X}$) and Column~9 the swept-up mass (M$_{sw}$). For $n_{\rm H}=0.008$~cm$^{-3}$ the reverse shock reaches the centre at 10800~years for all cases.}
\begin{tabular}{cccccccccc}
\hline
  (1)    & (2)    & (3)          & (4)    & (5)         & (6)           & (7)           & (8)           & (9)  & (10) \\
Distance & Radius &  $n_{\rm H}$ & Age    & Shock $\upsilon$& Shock T$_e$& EM           & L$_{X}$  & M$_{sw}$     & Comment  \\
  (kpc)  &   (pc) &  (cm$^{-3}$) & (years)&   (\kms)    & (K)           & (cm$^{-3}$)   &(erg~s$^{-1}$) & ($M_{\odot}$) & \\
\hline
  50     & 23.8   &   0.1       & 25000   &  400        & 4.5$\times10^6$ & 4.4$\times10^{58}$ &  2.2$\times10^{35}$ & 175 & Sedov phase, \\
         &        &             &         &             &               &               &            &   & reverse shock reaches centre at 4630~yr \\  
  50     & 23.8   &   0.01      & 8600    & 1250        & 1.0$\times10^7$ & 4.5$\times10^{56}$ & 4.9$\times10^{33}$ & 17.5  & Late transition from ED phase, \\
         &        &             &         &             &               &               &     &           & reverse shock reaches centre at 9980~yr \\  
  50     & 23.8   &   0.001     & 4100    & 3400        & 1.7$\times10^7$  & 4.8$\times10^{54}$ & 8.8$\times10^{31}$ & 1.75 & Early transition from ED phase, \\
         &        &             &         &             &               &               &         &       & reverse shock reaches centre at 21500~yr \\  
  50     & 23.8   &   0.0001    & 2300    & 6050        & 3.4$\times10^7$  & 1.0$\times10^{53}$ & 2.4$\times10^{30}$ & 0.175 & ED phase, \\
         &        &             &         &             &               &               &            &    & reverse shock reaches centre at 46300~yr \\  
\hline\hline
  5      & 2.4    &   0.008     & 113    & 11780        & 1.2$\times10^8$ & 5.6$\times10^{53}$ & 1.8$\times10^{31}$ & 0.014 & ED phase \\
  10     & 4.7    &   0.008     & 378    & 7019         & 5.0$\times10^7$ & 4.5$\times10^{54}$ & 1.3$\times10^{32}$ & 0.114 & ED phase \\
  20     & 9.5    &   0.008     & 1390   & 4010         & 3.6$\times10^7$ & 2.7$\times10^{55}$ & 8.1$\times10^{32}$ & 0.915 & ED phase \\
  30     & 14.3   &   0.008     & 2830   & 2960         & 2.5$\times10^7$ & 6.8$\times10^{55}$ & 1.2$\times10^{33}$ & 3.12 & Early transition from ED phase \\
  40     & 19.0   &   0.008     & 4900   & 1950         & 1.7$\times10^7$ & 1.5$\times10^{56}$ & 2.0$\times10^{33}$ & 7.32 & Mid transition from ED phase \\
  50     & 23.8   &   0.008     & 7900   & 1390         & 1.3$\times10^7$ & 2.9$\times10^{56}$ & 3.2$\times10^{33}$ & 14.0 & Late transition from ED phase \\
  60     & 28.5   &   0.008     & 11900  & 1060         & 1.1$\times10^7$ & 4.9$\times10^{56}$ & 4.8$\times10^{33}$ & 24.7 & Sedov phase \\
  70     & 33.3   &   0.008     & 17200  & 840          & 9.8$\times10^6$ & 7.8$\times10^{56}$ & 6.8$\times10^{33}$ & 39.4 & Sedov phase  \\
\hline
\end{tabular}
\label{tab67}
\end{table*}
\section{\lmcorc\ and the unified \ac{SNR} evolution model}
 \label{app:1}
 
In addition to \citet{1995PhR...256..157M}, we also explore the `unified \ac{SNR} evolution models' \citep{2018ApJ...866....9L,1999ApJS..120..299T}, using the software SNRpy from \citet{2019AJ....158..149L}, a \ac{SN} explosion energy of $0.5\times10^{51}$~erg \citep{2017ApJ...837...36L,2020ApJS..248...16L} and type~Ia \ac{SN} ejecta mass of 1.4~M$_{\odot}$. For ejecta with a power-law envelope with index 7 and depending on the above range of the local density ($10^{-1} - 10^{-4}$~cm$^{-3}$), the \ac{SNR} would reach a radius of 23.8~pc at an age of 2300 to 25000~yrs and shock velocities of 6050~\kms\ to 400~\kms\ (see Table~\ref{tab67}). If as old as 25000~yrs, it would still be in the Sedov phase. However, a more likely scenario has an ambient density of $\sim$0.001 to 0.01~cm$^{-3}$ which would argue for an \ac{SNR} of age 4000 to 9000~yrs that is still in early (4000~yr) to late (9000~yr) transition from the ejecta-dominated (ED) phase to the Sedov phase. 

As additional test, we placed \lmcorc\ at various distances from 5~kpc to 70~kpc, which changed the radius from 2.4 to 33.3~pc (Table~\ref{tab67}). For those distances, we initially set the \ac{ISM} density to 0.01~cm$^{-3}$ which is not extreme for an \ac{SNR}. From the 47 Galactic \acp{SNR} modelled by \citet{2020ApJS..248...16L}, four have density $<0.01$~cm$^{-3}$. The models for a density of 0.01~cm$^{-3}$ predict somewhat brighter \acp{SNR} than the limit given by the X-ray flux upper limits. By adjusting the \ac{ISM} density to range from 0.0094~cm$^{-3}$ (for 20~kpc distance) to 0.0083~cm$^{-3}$ (for 70~kpc distance), the unified \ac{SNR} evolution model \citep{2019AJ....158..149L,1999ApJS..120..299T} fluxes are equal to the upper limits. Thus, we set the density to 0.008~cm$^{-3}$ to be below the X-ray upper limits which produces estimated ages from 113~years to 17200~years (see Table~\ref{tab67}). These are faint in X-rays, consistent with the X-ray limits (see next paragraph). However, for the nearest and youngest of these (ages 113 and 378~yr), the SNe would have been almost certainly observed historically because they are bright type~Ia at high Galactic latitude where there is no significant extinction. For distances $\geq$20~kpc, the age is large enough that they could have escaped the historical record.
This, together with paucity of progenitors in the 20--40~kpc distance range (see Section~\ref{sec:progenitor}), would also suggest that the positioning of \lmcorc\ as an \ac{SNR} is more likely to be at the \ac{LMC} distance of $\sim$50~kpc.

\bsp	
\label{lastpage}
\end{document}